\documentclass[iop,author-year]{emulateapj}
\usepackage{url}
\usepackage{graphicx}
\usepackage{color}
\usepackage{amsmath}
\usepackage{setspace}
\usepackage{threeparttable}
\usepackage{epstopdf}
\usepackage{graphics}
\usepackage{subfigure}
\usepackage{url}             
\usepackage{rotating}
\usepackage{lipsum}
\usepackage{float}
\usepackage{wrapfig}
\usepackage{longtable}
\usepackage{natbib}
\usepackage{hyperref}

\renewcommand{\thefootnote}{\ifcase\value{footnote}\or * \or
\# \or $\dagger$ \or ** \or \# \or * \or \# \or $\dagger$ \or ($\infty$)\fi}

\begin{document}

\title{Multi-wavelength polarimetry and spectral study of M87 jet during 2002-2008 \footnotemark[*]}

\author{Sayali S. Avachat \altaffilmark{1}, Eric S. Perlman \altaffilmark{1}, Steven C. Adams\altaffilmark{2},
Mihai Cara\altaffilmark{3}, Frazer Owen \altaffilmark{4}, William B. Sparks \altaffilmark{3}, Markos Georganopoulos \altaffilmark{5}}

\altaffiltext{1}{Department of Physics and Space Sciences, 150 W. University Blvd., Florida Institute of Technology, Melbourne, FL 32901, USA}
\altaffiltext{2}{Department of Physics and Astronomy, University of Georgia, Athens, GA, 30605, USA}
\altaffiltext{3}{Space Telescope Science Institute, 3700 San Martin Drive, Baltimore, MD 21218, USA}
\altaffiltext{4}{National Radio Astronomy Observatory, Array Operations Center, P.O. Box O, 1003 Lopezville Road, Socorro, NM 87801-0387, USA}
\altaffiltext{5}{Department of Physics, University of Maryland–Baltimore County, 1000 Hilltop Circle, Baltimore, MD 21250, USA}
\footnotetext[*]{ Based on the observations made with \textit{Karl G. Jansky Very Large Array (VLA)}, operated by National Radio Astronomy Observatory (NRAO), and Hubble Sapce Telescope (\textit{HST}), obtained at the Space Telescope Science Institute (STScI), which is operated by the Association of Universities for Research in Astronomy, Inc.}

\begin{abstract}
We present a multi-wavelength polarimetric and spectral study of M87 jet obtained at sub-
arcsecond resolution between 2002 and 2008. The observations include multi-band archival VLA
polarimetry data sets along with the HST imaging polarimetry. These observations have better angular resolution than previous work by factors of 2-3 and in addition, allow us to explore the time domain. These observations envelope the huge flare in HST-1 located at 0.$\arcsec$86 from the nucleus \citep{2007ApJ...663L..65C, 2009ApJ...699..305H, 2009AJ....137.3864M, 2011ApJ...743..119P}. The increased resolution enables us to view more structure in each knot, showing several resolved sub-components. We also see apparent helical structure in the polarization vectors in several knots, with polarization vectors turning either clockwise or counterclockwise near the flux maxima in various places as well as show filamentary undulations. Some of these characteristics are correlated with flux and polarization maxima while others are not. We also examine the total flux and fractional polarization and look for changes in both radio and optical since the observations of \citet{1999AJ....117.2185P} and test them against various models based on shocks and instabilities in the jet. Our results are broadly consistent with previous spine-sheath models and recollimation shock models, however, they require additional combinations of features to explain the observed complexity, e.g. shearing of magnetic field lines near the jet surface and compression of the toroidal component near shocks. In particular, in many regions we find apparently helical features both in total flux and polarization. We discuss the physical interpretation of these features.
\end{abstract}


\section{Introduction}
\indent M87 hosts one of the nearest (d=16 Mpc, translating to a scale of $\approx$78 pc per arcsecond) relativistic jets. The kpc scale jet is under observation in X-rays with \textit{Chandra}, optical-ultraviolet with \textit{HST}, and in radio with \textit{VLA} and \textit{VLBA}. During the last decade, a major flare was seen in knot HST-1, located 0.$''$86 from M87's nucleus. This flare, which featured an increase in optical and X-ray flux of more than a factor of 100, was observed extensively in the optical \citep{2009AJ....137.3864M, 2011ApJ...743..119P}, X-rays \citep{2006ApJ...640..211H}. \citet{2007ApJ...663L..65C} suggest that HST-1 was also the site of a TeV flare observed around the same time by the H.E.S.S. experiment, however there are other views about the origin of the TeV emission. While \citet{2011ApJ...743..177H} think that both the nucleus as well as HST-1 can be sources of TeV emission, \citet{2005ApJ...634L..33G} suggest that the 2005 TeV flare originated from the nucleus. The current facilities do not have enough angular resolution in TeV to comment on the origin of these flares and the time resolution of the observations is insufficient to discriminate as well \citep{2012ApJ...746..151A}. \\
\indent The jet morphology at all wavelengths appear broadly similar \citep{1996ApJ...473..254S, 2005ApJ...627..140P}. The observed differences can be accounted for by highly polarized synchrotron radiation at all wavelengths and a nearly constant radio-optical spectral index throughout the jet \citep{2001ApJ...551..206P}. The jet has typical fractional polarization 10\% to 20\% in most regions \citep{1990ApJ...362..449O,1999AJ....117.2185P}. Radio polarization maps on large scale show large Faraday rotations in the direction of the 2 kpc radio lobes ranging from 350 rad m$^{-2}$ in the jet to 8000 rad m$^{-2}$ in the eastern radio lobes. In a more recent study, \citet{2016ApJ...823...86A} report rotation measures of a few hundreds of rad m$^{-2}$ over the most of the jet region along with some higher values of $\sim$1000 rad m$^{2}$ in knot C, the values which are in agreement with \citet{1990ApJ...362..449O}. They fit RM with two Gaussian, one for higher values in knot C and another one for the rest of the jet, with similar standard deviation, $\sigma_{RM}$ $\sim$ 120-180 rad m$^{2}$, (see their Fig.3). The observed polarization and high rotation measure suggests that the rotation is taking place in a Faraday screen in front of the radio emitting plasma. \citet{2016ApJ...823...86A}'s results suggest this screen is much closer to the jet vicinity and most likely associated with the sheath of the jet.\\
\indent Polarimetry can reveal the configuration of the magnetic field in the emitting region, and is thus a very useful diagnostic for jets. Many knot regions show high polarization ($\approx40\%-50\%$, close to the theoretical maximum for optically thin synchrotron emission) suggesting a highly ordered magnetic field. Previous radio and optical polarization images show the magnetic field is mostly parallel to the direction the of jet except in the shock-like knot regions, HST-1, knots A and C, where it becomes perpendicular to the jet axis (\citet{1999AJ....117.2185P}, hereafter P99). \\
\indent \citet{2001ApJ...551..206P} observed changes in the spectral indices of other knots in the jet, particularly D and F, which when combined with the MFPA vector morphology at 0.$''$2 resolution, suggest high energy synchrotron emitting particles may represent a very different population than those that emit in radio. P99 proposed a ``stratified" jet model to explain the differences seen in the radio and optical flux and polarization morphology. The model suggests that the radio and optical electrons may originate from different locations within the jet (P99, Fig. 7). According to their model, the observed radio emission is coming from the outer layer or ``strata" of the jet, shown by dotted lines in the figure; whereas the optical emission is coming from the central region close to the axis of the jet shown by solid lines. \\
\indent We describe the details of the polarimetry observations used for this study and the error analysis carried out in $\S$ \ref{sec:observations}. In $\S$ \ref{sec:compare_pol}, we discuss the general trends in flux and polarization structure along the jet, and compare the observed features of the individual knots with the previous studies. In $\S$ \ref{sec:variability}, we analyze the flux and polarization variability seen over the period of observations. Finally, we discuss our findings in the $\S$ \ref{sec:discussion} and conclude our discussion in $\S$ \ref{sec:concl}.

\begin{deluxetable*}{cccccccccc}
\scriptsize
\tablewidth{0pt}
\tablecaption{\textit{VLA} and \textit{HST} Polarimetry Observations}
\tablehead{\multicolumn{2}{c}{Project ID} & \multicolumn{2}{c}{Telescope Configuration} & \multicolumn{2}{c}{Energy Band} & \multicolumn{2}{c}{Date of observation}\\
\hline \\
\colhead{\textit{VLA}} & \colhead{\textit{HST}} & \colhead{\textit{VLA}} & \colhead{\textit{HST}} & \colhead{\textit{VLA}} & \colhead{\textit{HST}} & \colhead{\textit{VLA}} & \colhead{\textit{HST $ ^\dagger$}}}
\startdata
AH295 & 9705 & VLA:C:1 & ACS/HRC & X, Q & F606W & 19-Oct-02 & Dec 07 2002~ (1)\\
(J. Biretta) & (E. Perlman) & \nodata & \nodata & \nodata & \nodata & \nodata & Dec 10 2002~ (2)\\
AH822  & 9829 & VLA:A:1 & ACS/HRC & X,U,K & F606W & 02-Jun-03 & Nov 29 2003~ (3)\\
(D.E.Harris) & 10133 & \nodata & \nodata & \nodata & \nodata & 03-Jun-03 & Nov 28 2004~ (4)\\
\nodata & (J. Biretta) & \nodata & \nodata & \nodata& \nodata & 24-Aug-03$^ \ast$ & Dec 26 2004~ (5)\\
\nodata & \nodata & VLA:B:1  & \nodata & X,U,K,Q  & \nodata & 16-Nov-03  & Feb 09 2005~ (6) \\
 \\
AH862  & \nodata & VLA:A:1  & ACS/HRC &  X,U,K  & F606W & 15-Nov-04 & Mar 27 2005~ (7)\\
(D.E.Harris) & \nodata & \nodata  & \nodata &  \nodata  & \nodata & 31-Dec-04 & May 09 2005 ~(8)\\
\nodata  & \nodata & VLA:B:1  & \nodata & X,U,K,Q & \nodata &  03-May-05 & Jun 22 2005 ~(9) \\
\\
AH885  & \nodata & VLA:A:1  &  ACS/HRC & X,U,K  & F606W & 15-Feb-06 & Aug 01 2005~(10)\\
(D.E.Harris)  & 10617 & VLA:B:1   & \nodata & X  & \nodata & 07-May-06 & Nov 29 2005~(11)\\
\nodata  & (J. Biretta) & \nodata  & \nodata & X,U,K  & \nodata & 08-May-06 & Dec 26 2005~(12)\\
\nodata  & \nodata & \nodata  & \nodata & X,U,K,Q  & \nodata & 31-Jul-06 & Feb 08 2006~(13)\\
\nodata  & \nodata & \nodata  & \nodata & \nodata  & \nodata & 01-Aug-06$^ \ast$ & Mar 30 2006~(14) \\
\\
AC843  & \nodata & VLA:A:1  & ACS/HRC & X,U,K  & F606W &  11-Jun-07 & May 23 2006~(15)\\
(D.E.Harris)  & 10910 & \nodata  & \nodata & \nodata  & \nodata &  12-Jun-07 & Nov 28 2006~(16)\\
\nodata  & (J. Biretta) & \nodata  & \nodata & \nodata  & \nodata & 10-Aug-07$^ \ast$  & Dec 30 2006~(17)\\
\nodata  & 11216 & \nodata  & \nodata & \nodata  & \nodata & 11-Aug-07$^ \ast$ & Nov 25 2007~(18) \\
\nodata  & (J. Biretta) & VLA:B:1  & \nodata & X,U,K,Q  & \nodata & 19-Jan-08  & \nodata &\\
\enddata
\\
 \tablenotetext{$^\ast$}{\textit{VLA} observations on these dates were not used due to the bad weather.}
 \tablenotetext{$^\dagger$}{\textit{HST} observations sequence numbers are taken from \citet{2011ApJ...743..119P}.}
\label{tbl-obs_data}
\end{deluxetable*}

\begin{deluxetable*}{cccccc}
\tablewidth{0pt}
\tablecaption{Radio Flux and Polarization Data}
\tablehead{\colhead{Region}& \colhead{X$^{\dagger}$} & \colhead{Y$^{\dagger}$} & \colhead{Flux Density (mJy)}& \colhead{Polarization (\%)}& \colhead{Position Angle (deg)}}
\startdata
Nucleus	&	769-781	&	429-441	&	4358.4	$\pm$	0.9	&	1.3	$\pm$	0.2	&	3	$\pm$	1	\\
HST-1	&	803-813	&	445-455	&	89.8	$\pm$	0.8	&	8.1	$\pm$	0.2	&	-7	$\pm$	4	\\
D-E	&	869-897	&	469-479	&	62.4	$\pm$	1.2	&	23.6	$\pm$	0.2	&	20	$\pm$	3	\\
D-M	&	905-923	&	473-489	&	28.4	$\pm$	1.2	&	32.1	$\pm$	0.2	&	18	$\pm$	5	\\
D-W	&	919-931	&	485-497	&	19.9	$\pm$	0.9	&	19.5	$\pm$	0.2	&	13	$\pm$	9	\\
E	&	973-1033	&	499-535	&	78.0	$\pm$	3.3	&	11.4	$\pm$	0.6	&	28	$\pm$	13	\\
F	&	1075-1133	&	537-585	&	137.8	$\pm$	3.7	&	17.9	$\pm$	0.7	&	17	$\pm$	6	\\
I	&	1175-1213	&	571-607	&	99.3	$\pm$	2.6	&	8.6	$\pm$	0.5	&	35	$\pm$	10	\\
A-shock	&	1223-1251	&	585-629	&	600.1	$\pm$	2.5	&	22.7	$\pm$	0.5	&	41	$\pm$	1	\\
A	&	1221-1295	&	581-643	&	1291.4	$\pm$	4.7	&	10.3	$\pm$	0.9	&	-36	$\pm$	1	\\
B1	&	1313-1341	&	617-657	&	337.8	$\pm$	2.4	&	26.4	$\pm$	0.4	&	32	$\pm$	1	\\
B2	&	1367-1397	&	625-679	&	149.5	$\pm$	2.8	&	31.5	$\pm$	0.5	&	-31	$\pm$	2	\\
C1	&	1431-1471	&	687-729	&	320.1	$\pm$	2.9	&	24.4	$\pm$	0.5	&	30	$\pm$	1	\\
C2	&	1479-1505	&	681-741	&	50.4	$\pm$	2.8	&	56.7	$\pm$	0.5	&	33	$\pm$	3	\\
G1	&	1475-1531	&	753-771	&	64.0	$\pm$	2.3	&	28.5	$\pm$	0.4	&	-11	$\pm$	5	\\
G2	&	1527-1559	&	725-759	&	105.4	$\pm$	2.3	&	37.1	$\pm$	0.4	&	-39	$\pm$	2	

\enddata
\tablenotetext{$^\dagger$}{Box coordinates (X, Y) are in pixels. The jet is at $\sim$ 20.5$^{\circ}$ north from the x axis, with a scale of 0.''025 pixel$^{-1}$.}
 \label{tbl-rad-poln-flux_data}
\end{deluxetable*}

\begin{deluxetable*}{cccccc}
\tablewidth{0pt}
\tablecaption{Optical Flux and Polarization Data}
\tablehead{\colhead{Region}& \colhead{X$^{\dagger}$} & \colhead{Y$^{\dagger}$} & \colhead{Flux Density ($\mu$Jy)}& \colhead{Polarization (\%)}& \colhead{Position Angle (deg)}}
\startdata
Nucleus	&	769-781	&	429-441	&	630.9	$\pm$	0.1	&	3.1	$\pm$	1.1	&	-17	$\pm$	3	\\
HST-1	&	803-813	&	445-455	&	562.7	$\pm$	0.0	&	27.1	$\pm$	1.0	&	-15	$\pm$	3	\\
D-E	&	869-897	&	469-479	&	42.0	$\pm$	0.1	&	5.1	$\pm$	6.7	&	-33	$\pm$	3	\\
D-M	&	905-923	&	473-489	&	14.3	$\pm$	0.1	&	20.5	$\pm$	4.0	&	-17	$\pm$	3	\\
D-W	&	919-931	&	485-497	&	11.3	$\pm$	0.1	&	26.5	$\pm$	3.3	&	-31	$\pm$	3	\\
E	&	973-1033	&	499-535	&	43.2	$\pm$	0.2	&	10.8	$\pm$	6.4	&	-27	$\pm$	3	\\
F	&	1075-1133	&	537-585	&	94.3	$\pm$	0.2	&	12.8	$\pm$	5.6	&	-38	$\pm$	3	\\
I	&	1175-1213	&	571-607	&	36.9	$\pm$	0.2	&	22.0	$\pm$	3.0	&	-26	$\pm$	3	\\
A-shock	&	1223-1251	&	585-629	&	33.4	$\pm$	0.2	&	182.0	$\pm$	2.1	&	-3	$\pm$	3	\\
A	&	1221-1295	&	581-643	&	839.6	$\pm$	0.3	&	20.0	$\pm$	1.0	&	19	$\pm$	3	\\
B1	&	1313-1341	&	617-657	&	203.2	$\pm$	0.2	&	15.9	$\pm$	1.2	&	-14	$\pm$	3	\\
B2	&	1367-1397	&	625-679	&	149.0	$\pm$	0.2	&	22.6	$\pm$	1.5	&	10	$\pm$	3	\\
C1	&	1431-1471	&	687-729	&	215.6	$\pm$	0.2	&	8.0	$\pm$	2.7	&	-9	$\pm$	3	\\
C2	&	1479-1505	&	681-741	&	52.0	$\pm$	0.2	&	15.6	$\pm$	4.1	&	-33	$\pm$	3	\\
G1	&	1475-1531	&	753-771	&	29.8	$\pm$	0.1	&	21.9	$\pm$	3.2	&	18	$\pm$	3	\\
G2	&	1527-1559	&	725-759	&	15.8	$\pm$	0.2	&	26.9	$\pm$	9.5	&	7	$\pm$	3	
\enddata
\tablenotetext{$^\dagger$}{Box coordinates (X, Y) are in pixels. The jet is at $\sim$ 20.5$^{\circ}$ north from the +X axis, with a scale of 0.''025 pixel$^{-1}$.}
 \label{tbl-opt-poln-flux_data}
\end{deluxetable*}

\section{Observations and data reduction} \label{sec:observations}
\indent We use {\it VLA} and {\it HST} polarimetry observations for our analysis. The details of the observing runs are summarized in Table \ref{tbl-obs_data}. We describe the details of the observations and data reduction steps in the following sub-sections.

\subsection{\textit{VLA} observations}
\indent M87 was under intensive observations during its flare. During 2002-2008, M87 was observed by the {\it VLA} every 5-6 months at 8, 15, 22, and 43 GHz in A and B configurations.
We extracted the {\it VLA} data from the {\it NRAO}\footnote{The National Radio Astronomy Observatory (NRAO) is a facility of the National Science Foundation operated under cooperative agreement by Associated Universities, Inc (\url{https://archive.nrao.edu}).} data archive. \\
\indent Data reduction was carried out using standard reduction techniques in the {\it Astronomical Image Processing System} ({\it AIPS}). Data were first calibrated in {\it AIPS} task CALIB, using 3C286 as primary flux calibrator, 3C138 as a polarization position angle (PA) calibrator and 1224+035 as instrument polarization calibrator, for phase only and amplitude \& phase corrections. Subsequent runs of CALIB, using M87 as a self-calibrator, were performed to minimize the calibration errors.  Antenna $'$D$'$ terms were corrected using task PCAL and CLCOR with 1224+035 as polarization calibrator. Multi-source data sets were then separated into single source data sets using task SPLIT. After another run of self-calibration on single source data sets, Stokes I, Q and U images were obtained using task IMAGR. The final images, thus obtained, have a beam size of $\approx$0.$\arcsec$09 at 22 GHz and $\approx$0.$\arcsec$14 at 15GHz, in the B array configuration. The resolution of these images is comparable to that of optical images. \\
\indent Next we used the procedure DOFARS to correct the polarization values using the rotation measure (RM), as described in \citet{2005A&A...441.1217B}. This procedure reads the Q and U polarization cubes as inputs to run the task FARS which evaluates the brightness distribution as a function of Faraday rotation using the measured brightness and given set of (wavelength)$^{2}$. FARS outputs the rotation measure (RM) cubes, {\bf which were used as input } for the task AFARS. This task produces a map of positions of maximum rotation measure and flux densities. Next we used the output images of AFARS to correct and find the error maps of Stokes Q and U images, using task RFARS. The magnetic field position angle (MFPA) and fractional polarization maps were then obtained using RM corrected Stokes Q and U images in task COMB. \\
\indent Task PCNTR was used to plot the total flux contour maps of MFPA and fractional polarization. The radio maps at all epochs were convolved using task CONVL to the same resolution as 15 GHz image. The errors in PA and fractional polarization were found by propagating errors in Stokes Q and U images. We then compared them to the optical polarimetry images detailed in \citet{2011ApJ...743..119P}. Our K band (22 GHz) data have the highest signal to noise (S/N) of all the \textit{VLA} data and the U band (15 GHz) have the most coverage over time domain. In this paper, we use only these data for the comparison of the jet's radio and optical polarization structure. Optical images at all wavelengths were smoothed to the same resolution as 15 GHz radio images and were resampled at 0.$\arcsec$025/pixel to compare both the bands at the same scale.

\subsection{\textit{HST} observations}
\indent Before the onset of the HST-1 flare in 2001-2002, M87 was a regular target of \textit{HST} since 1994, with observations occurring roughly every year. With more intense monitoring between December 2002 and November 2007, the M87 jet was observed at 4-5 week intervals \citep{2009AJ....137.3864M, 2009ApJ...699..305H}. All the observed epochs are listed in Table 1. The polarimetry was done in F606W and F330W bands.
The F606W polarimetry observations are used in this paper. The numbers in the bracket in front of dates are the original sequence number of observations used in \citet{2011ApJ...743..119P}. \\
\indent The High-Resolution Channel (HRC) of the Advanced Camera for Surveys (ACS) was used for polarimetry observations of 17 of these epochs in F606W band. The ACS HRC is a single-chip CCD camera, with a plate scale of 0.$\arcsec$028$\times$ 0.$\arcsec$025 pixel$^{-1}$, corresponding to a field of view of about 28$\times$25$\arcsec$ and yielding diffraction limited resolution of $\approx$0.$\arcsec$06 for the F606W observations. These observations were reduced using methods of re-calibration following the ACS and WFPC2 Instrument Handbooks. \\
\indent To prepare the observations for photometry and polarimetry, all the epochs of ACS/HRC F606W were combined to create a composite (``master'') image. This improved the signal to noise ratio of the background galaxy. The image was further modelled for galaxy emission using ellipse STSDAS, which was subtracted and split into three images, one corresponding to each polarizer on \textit{HST}. For more details of individual \textit{HST} observations and data reduction procedures the reader may refer to \citet{2011ApJ...743..119P}.

\subsection{Error Analysis} \label{sec:err_ana}
\indent To evaluate the relevance of the flux and polarization images and their measurements in terms of different dynamic ranges of radio and optical data, we performed the error analysis on our radio images. The errors in flux, polarization and PAs are the propagated statistical errors and systematic errors for the \textit{VLA} instrument. The statistical error in flux (Stokes I, Q and U) is calculated using the off source rms noise, the number of pixels (N) in a box and beam area. Note that the radio images are in units of flux density (Jy) per beam.  \\
\begin{equation}
error \ = \ rms \ast \sqrt{\frac{N}{beam \ area}}
\end{equation} \\
As can be seen, the error is smaller for the larger regions as the error is reduced as 1/$\sqrt{N}$. \\ \indent Polarized flux and PA images were made in $\it{AIPS}$ using the following equations:
\begin{equation}
P \ = \ \frac{\sqrt{Q^{2}+U^{2}}}{I}\\
\end{equation}
\begin{equation}
PA \ = \ \frac{1}{2}tan^{-1}\Big(\frac{U}{Q}\Big) \label{eq:3.2}
\end{equation}
\indent To calculate the errors we used the standard error propagation formulae. To find the error in polarized flux, we assumed that the errors in Stokes Q and U add up in quadrature as follows -
\begin{equation}
\sigma_{P} \ = \ \sqrt{\sigma^{2}_{Q}+\sigma^{2}_{U}}
\end{equation}
To find the error in PA, we used the following formula -
\begin{equation}
\sigma_{PA} \ = \ \frac{1}{2}\frac{\sqrt{(Q^{2}\ast \sigma^{2}_{U})+(U^{2}\ast \sigma^{2}_{Q})}}{Q^{2}+U^{2}}
\end{equation}
\indent Table \ref{tbl-rad-poln-flux_data} and \ref{tbl-opt-poln-flux_data} show the flux and polarimetry information for all the identified knots in M87's jet in radio and optical bands respectively. The table lists average flux, average polarization and average PA of magnetic field vectors in all the knots. The regions in Stokes Q and U used to obtain these values were obtained by putting boxes around each knot. We list the (X, Y) coordinates of these boxes in respective tables. \\
\indent Optical data was treated differently using the methods of debiasing. After the Stokes I, Q and U images were obtained, we accounted for the well known Rician bias in P \citep{Serkowski} using a python code adapted from the STECF IRAF package \citep{2000ASPC..216..671H}. This code debiases the P image, following \citet{1974ApJ...194..249W}, and calculates the error in the polarization PA accounting for the non-gaussian nature of distribution (see also \citet{1993A&A...274..968N}). In the calculation, pixels with SNR $\textless$ 0.1 are excluded outright. Also since the debiasing is done with the ``most probable value'' estimator, pixels where the values of P was negative or was above the Stokes I value (i.e. P $\textgreater$ 100\%) were blanked. Interested readers may refer to \citet{2013ApJ...773..186C, 2006ApJ...651..735P, 2011ApJ...743..119P} for further details on the application of this method. \\
\indent The use of ACS/HRC polarizers with different orientation angles and the PA\_V3 angle, which is the angle between the North and V3 axis of the telescope, rendered it necessary to correct the PA of the final image to obtain the real magnetic field PAs. The PA equation (eq.~\ref{eq:3.2}) then gets modified to -
\begin{equation}
PA \ = \ \frac{1}{2}tan^{-1}\Big(\frac{U}{Q}\Big) \ + \ PA\_V3 \ + \ \chi
\end{equation}
where $\chi$ is the angle of the instrument in the focal plane. 
This converts the Stokes Q and U from the instrumental frame to the sky frame. A script used to do these corrections will also produce the polarization (P), fractional polarization (FP) and PA images for each epoch and combine them all in one ``master" image. We use this combined image in our analysis. To find the significant levels of polarized flux images, we make the vectors plots using the $\it{AIPS}$ task PCNTR and setting the parameter PCUT = 3$\sigma$, where $\sigma$ is the background noise in the Stokes Q and U images. \\
\indent We show the polarization vector plots using fractional polarization images in the left hand side panels and those using polarized flux images on the right hand side panels of figures \ref{fig:nuc-hst1} through \ref{fig:knot-cg}, except in Fig.~\ref{fig:knot-efi} we do not show the fractional polarization plot because of the loss of details due to poor SNR. The polarized flux images were obtained using the total flux and polarized flux. To display the magnetic field PA vectors, we used the PA image obtained from Stokes Q and U images. The orientation of the vectors represent the direction of local magnetic field, whereas their lengths represent the degree of polarization. The images show the MFPA vectors which are above the 3$\sigma$ level in the polarized flux. The regions where we do not see any vectors within these figures, we believe, are the regions of lower polarization (P $\textless$ 3$\sigma$) or of depolarization (P $\approx$ 0). \\
\indent We label the subcomponents within the jet by visually inspecting the fractional polarization images on the left. The radio as well as optical images show a highly resolved flux and polarization structure of the jet, which was not seen in the old \textit{VLA} observations presented in P99. By comparing our fractional polarization and polarized flux images, we can identify new sub-structures based on the detected total flux, and the regions of significant polarized flux and hence claim that these new sub-structures are real, especially near the nucleus and HST-1, although we cannot comment if these are newly emerged or are just a result of better resolution.

\section{Radio and Optical Polarimetry}\label{sec:compare_pol}
\indent In this section we present a comprehensive discussion of the comparison of radio and optical polarimetry. In Figures \ref{fig:full_jet} and \ref{fig:flux_poln_profile}, we present the flux and polarization features of the jet in terms of general similarities and differences. In figures~\ref{fig:nuc-hst1} through \ref{fig:knot-cg} we show detailed maps of the radio (top, 22 GHz) and optical (F606W, bottom) flux and polarization images. In these images, the left side panel shows the fractional polarization maps while the right side panel shows polarized flux maps. As mentioned in \S~\ref{sec:err_ana}, these maps were plotted using a cut-off at 3$\sigma$ level of polarized flux. \\
\indent For convenience of discussion, we divide the jet in three parts, namely inner jet, intermediate jet and outer jet, and explain the polarization morphology for each in the following subsections.

\subsection{General Trends along the Jet}\label{sec:trends}

\begin{figure*}
    \begin{sideways}
        \includegraphics[width=1.2\textwidth]{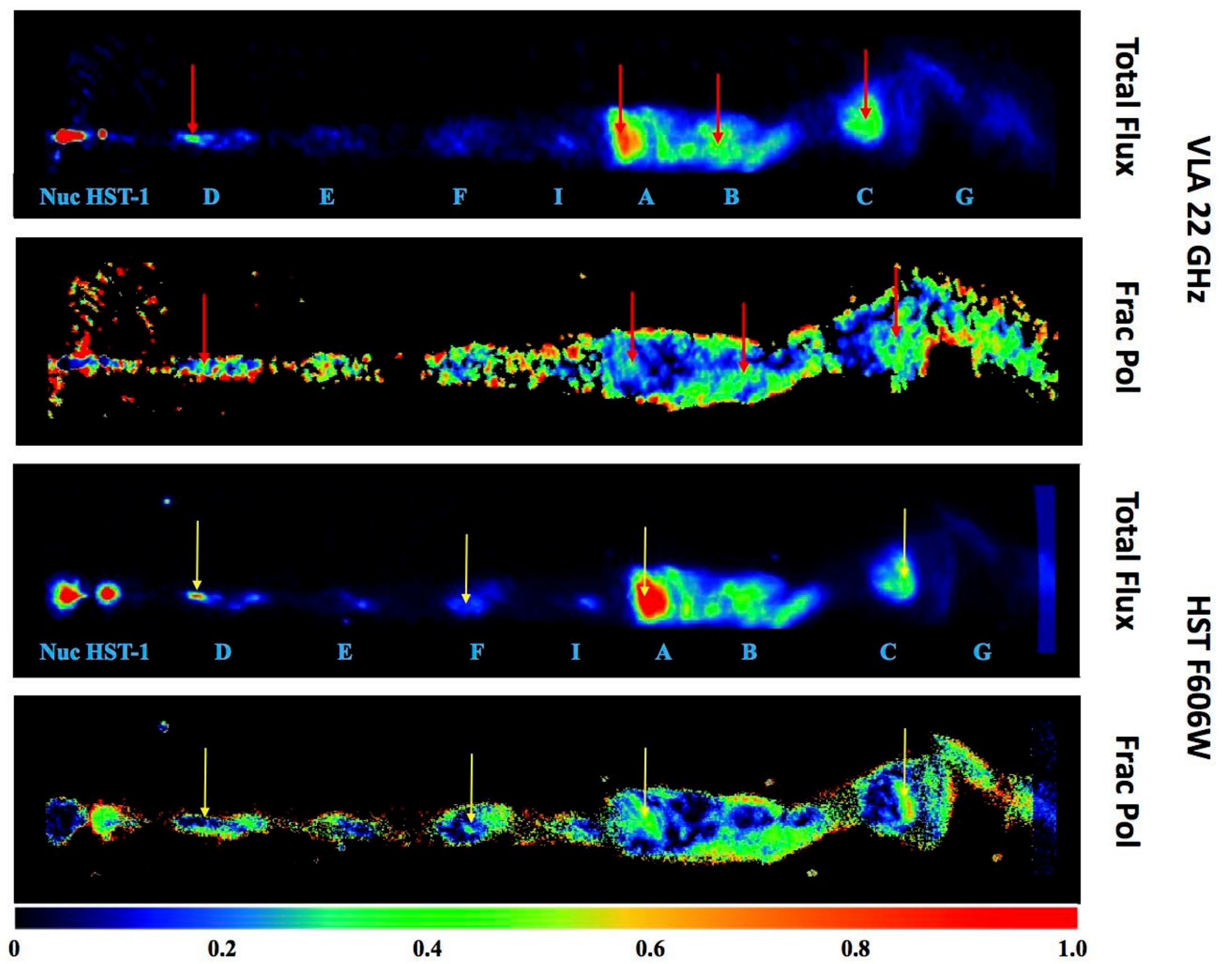} 
    \end{sideways}
  \caption[Flux and polarization of M87 jet in radio and optical bands.]{Color scale images showing of total flux and fractional polarization of M87 jet in the radio (top two panels) and optical (bottom two panels). All maps are rotated so that the jet axis lies along the x-axis. The false-color panel at the bottom represents the degree of polarization in both bands. The red arrows on top two panels indicate the locations of flux and polarization maxima, respectively. Please refer to the text for discussion.}\label{fig:full_jet}

\end{figure*}

\begin{figure*}
    \begin{center}
        \begin{tabular}{@{}cc@{}}
          \includegraphics[width=0.5\textwidth]{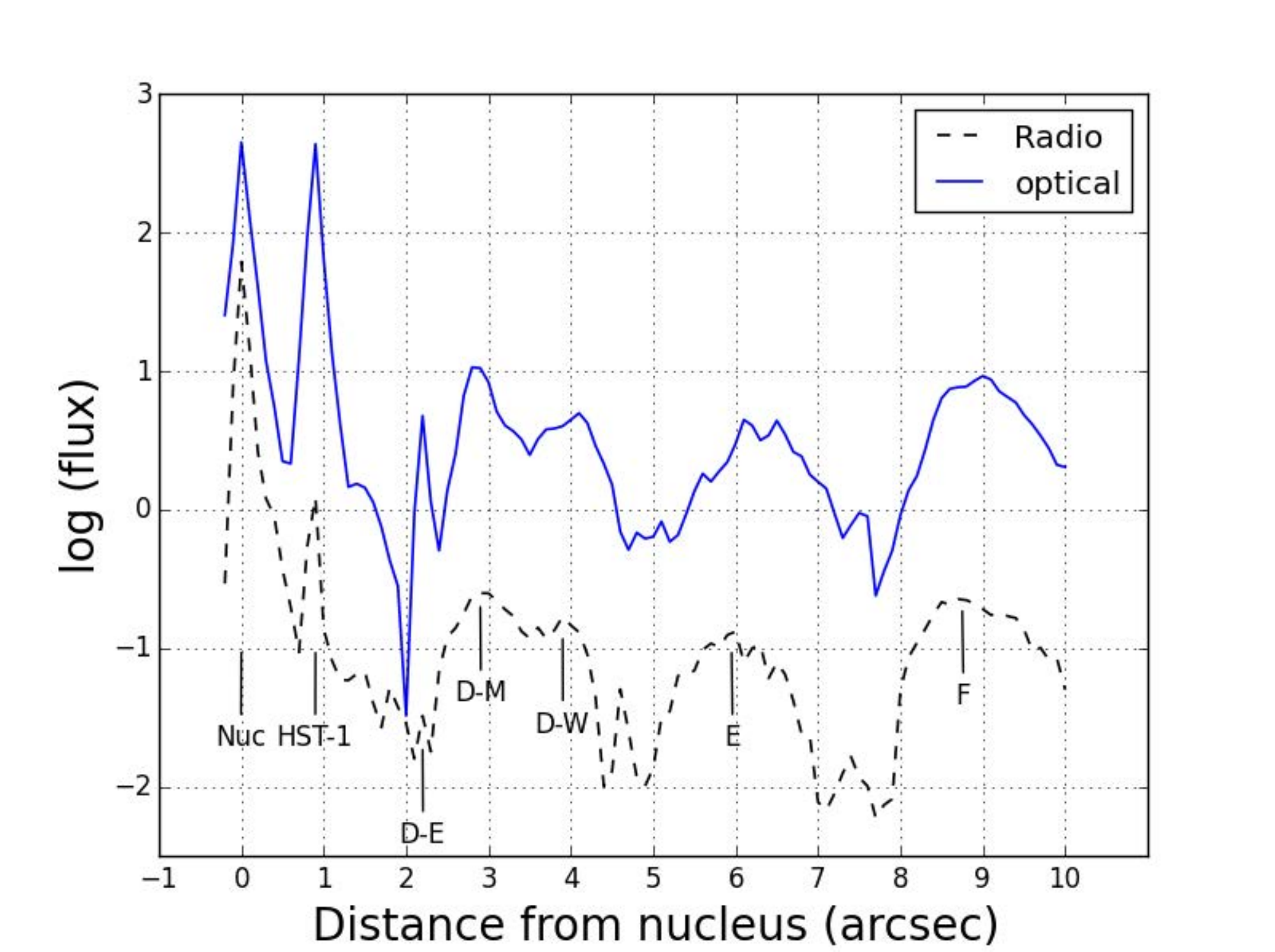}
          \includegraphics[width=0.5\textwidth]{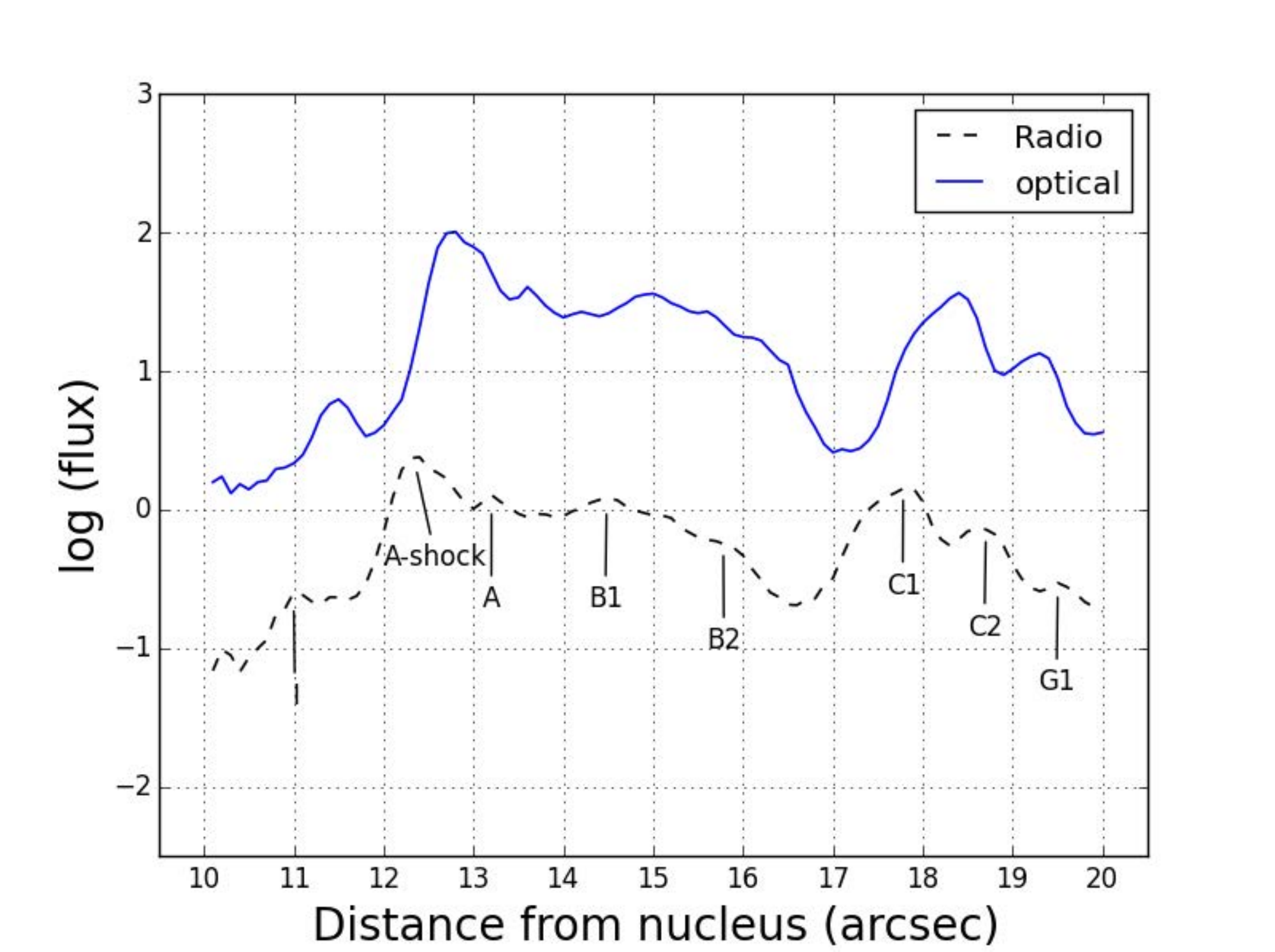}\\
          \includegraphics[width=0.5\textwidth]{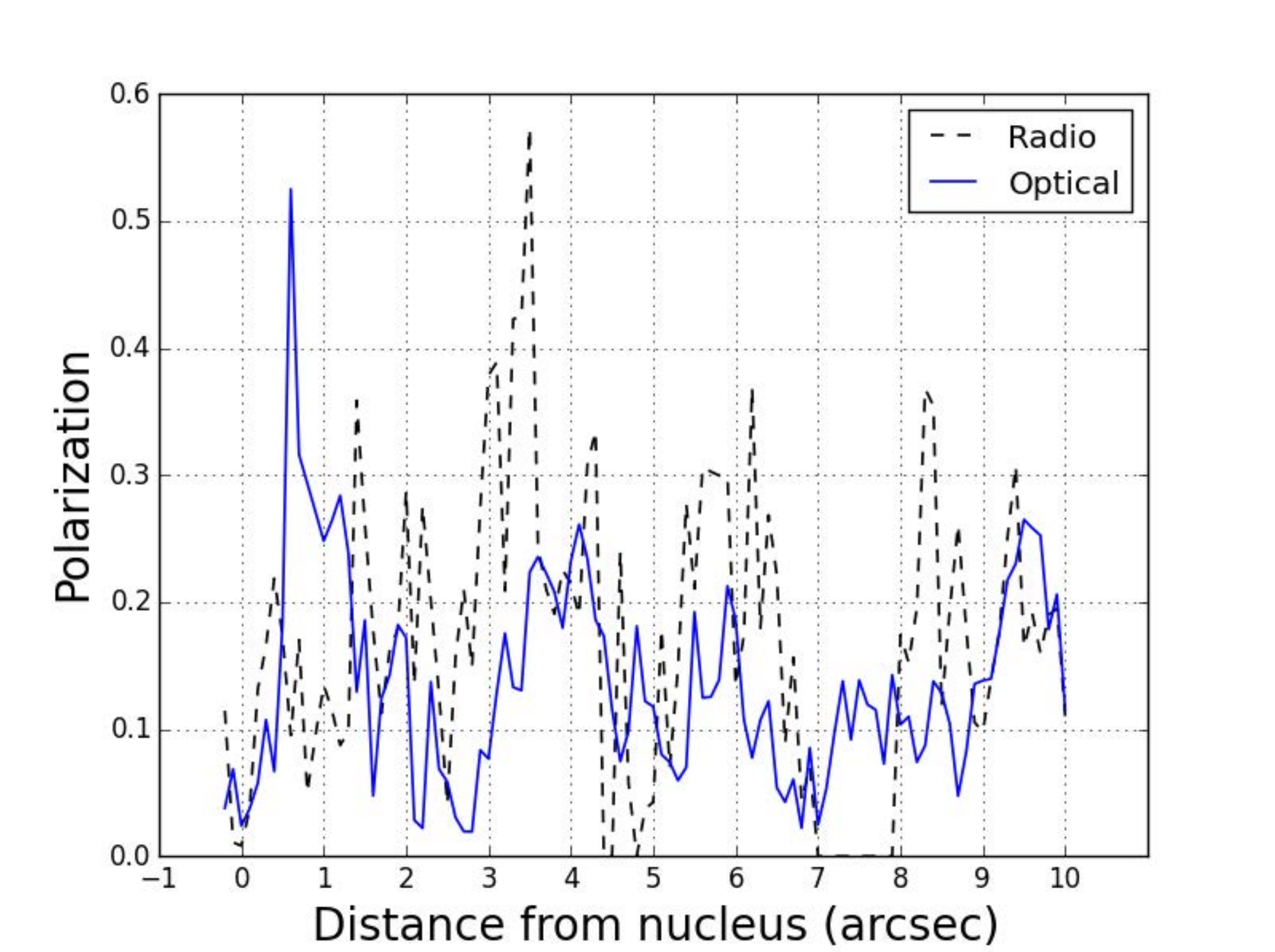}
          \includegraphics[width=0.5\textwidth]{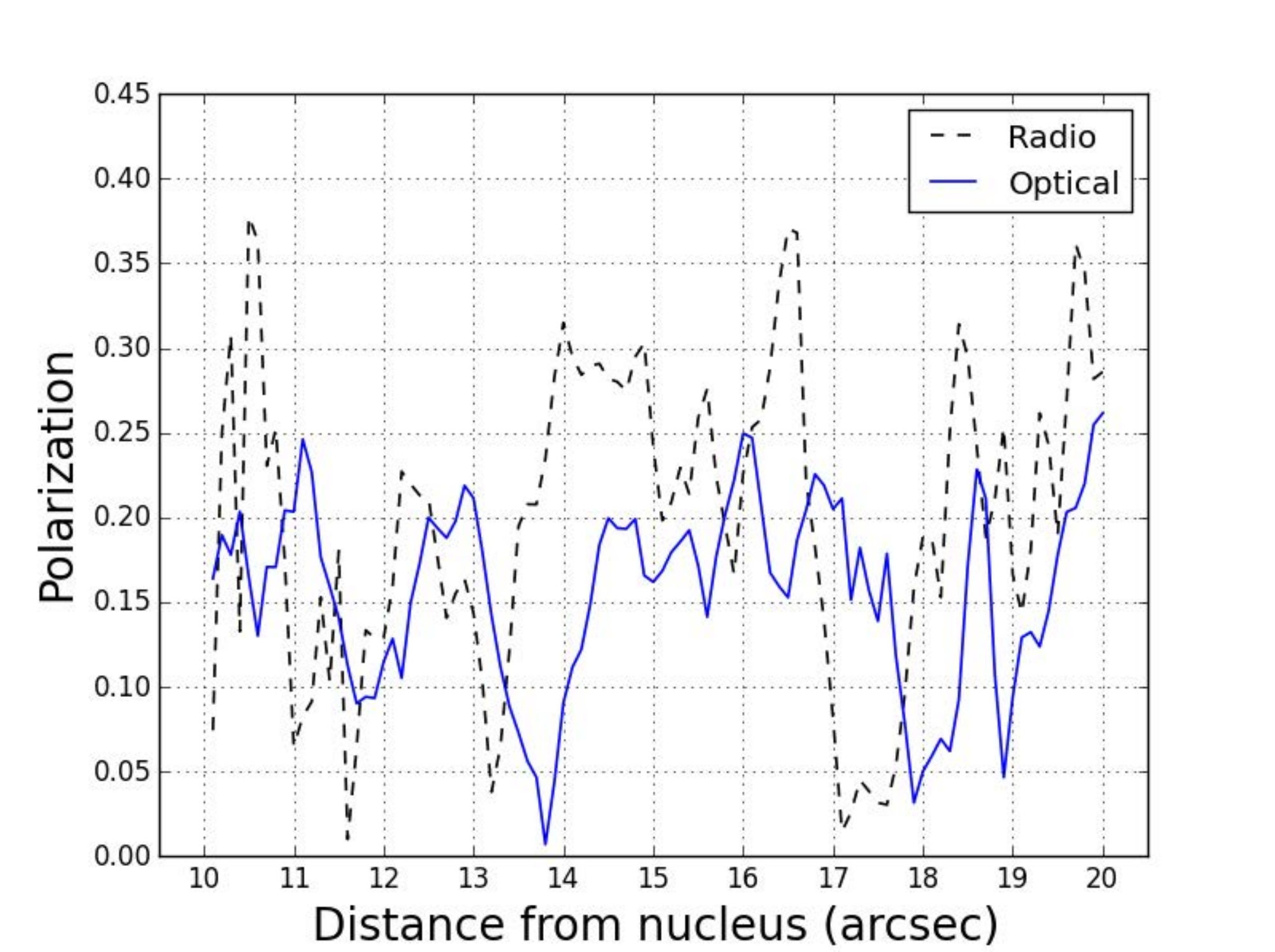}\\
          \includegraphics[width=0.5\textwidth]{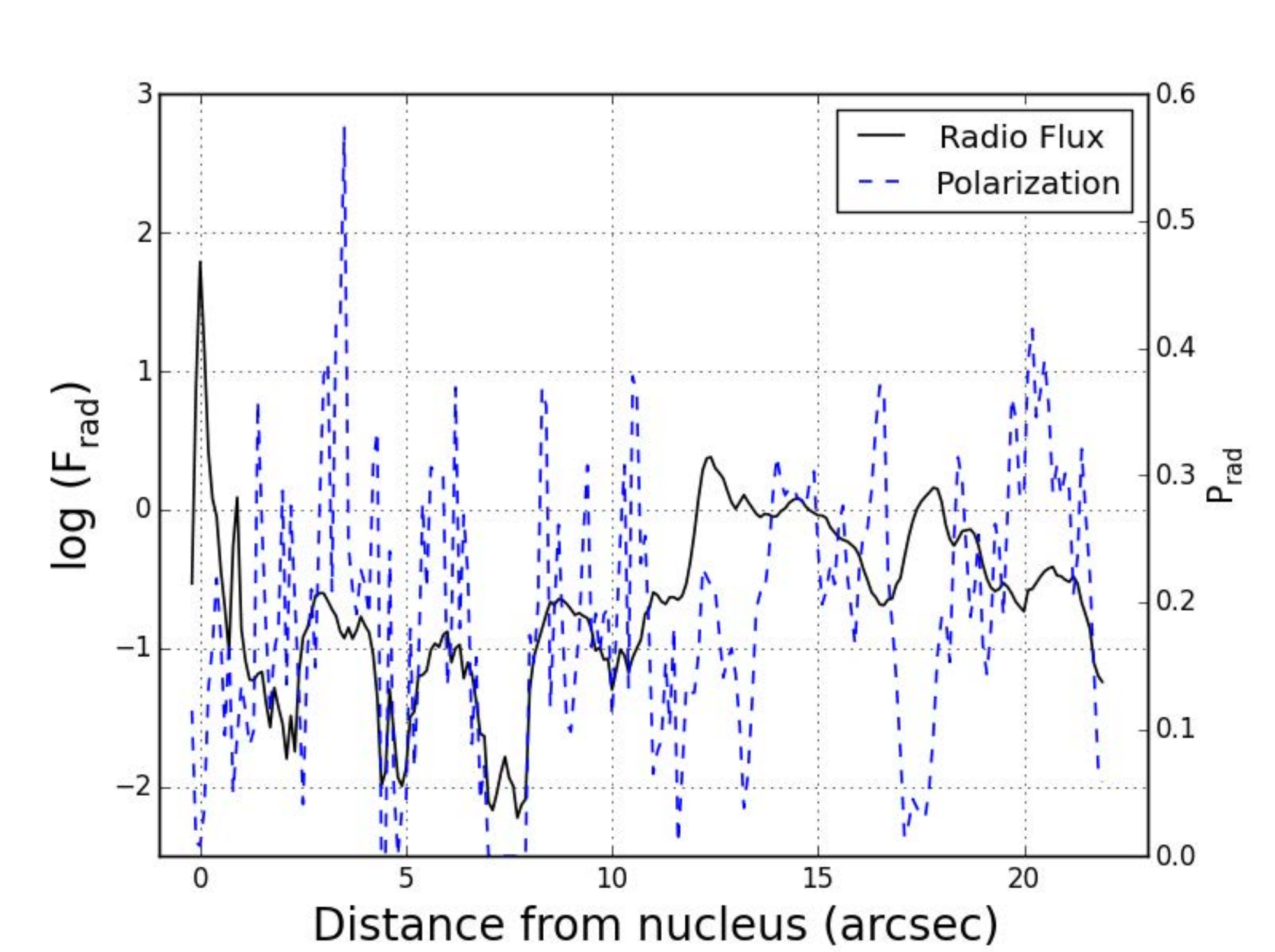}
          \includegraphics[width=0.5\textwidth]{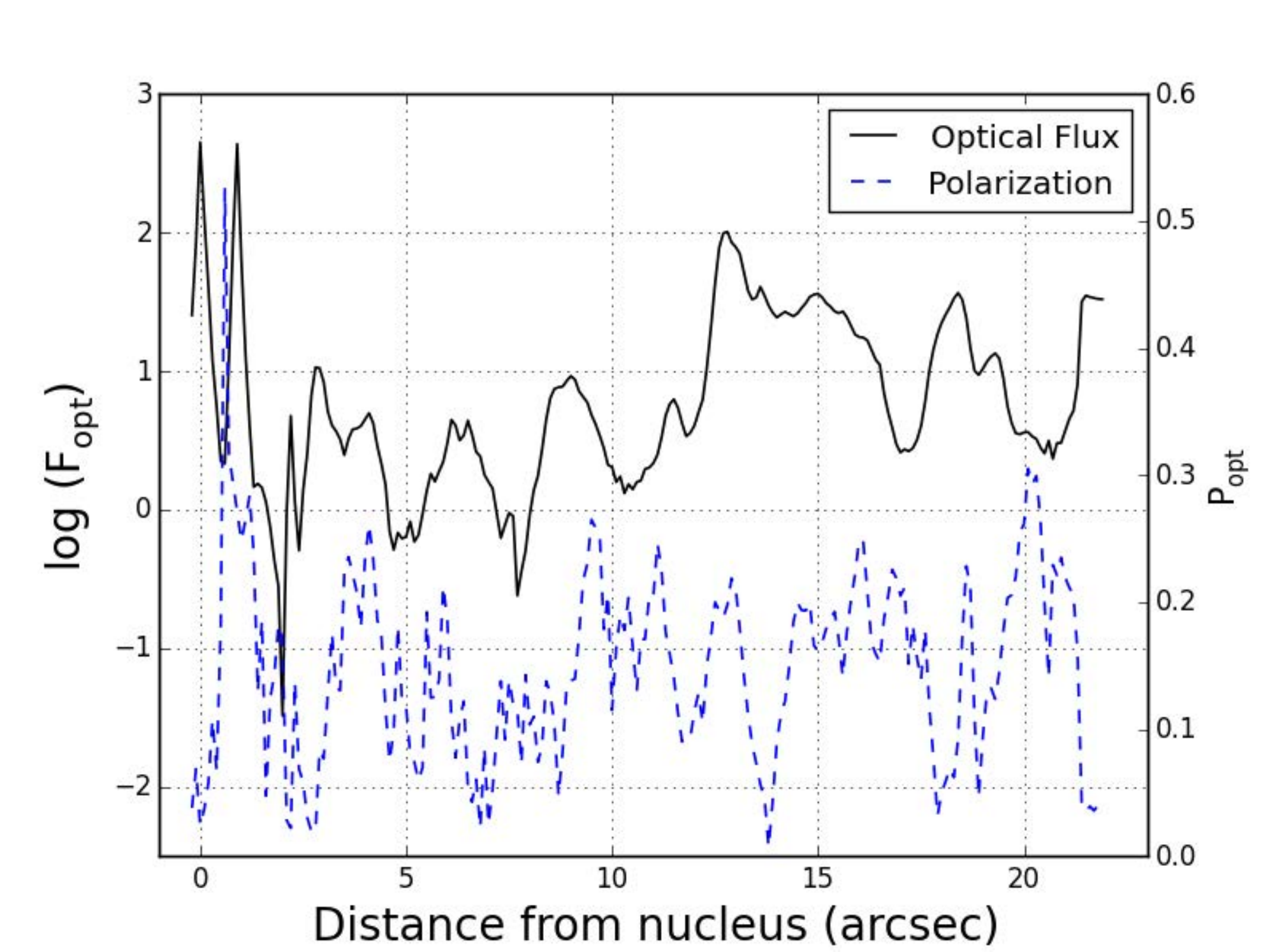}
        \end{tabular}
    \end{center}
    \caption{Flux and polarization profiles of radio and optical. Top two show the flux profile, middle two show the polarization profile in two bands of the inner and intermediate jet (0-10$\arcsec$ from the nucleus) on the right and outer jet (10$\arcsec$-20$\arcsec$ from the nucleus). In bottom two panels, we show the flux and polarization profiles in individual bands to point out the apparent correlation between the flux and polarization maxima and minima.} \label{fig:flux_poln_profile}
\end{figure*}

\indent The {\it VLA} and {\it HST} images show a wealth of information about the polarization and magnetic field structure of the M87 jet. The total flux and polarization images have many common general characteristics which are observed in both the bands. Fig~\ref{fig:full_jet} shows false color flux and fractional polarization images in radio (top two panels) and optical (bottom two panels). The jet shows some striking similarities in terms of the total flux structure in the radio and the optical. In general the radio jet shows a broader jet with more diffuse emission from near the surface as well as from inter-knot regions, as compared to the optical trend previously discussed in \citet{1996ApJ...473..254S}. \\
\indent In Fig.~\ref{fig:flux_poln_profile} we have plotted the flux and polarization profiles of radio and optical (top and middle two panels) which quantifies the above differences in the locations of flux and polarization maxima. The flux profiles show slight differences in the locations of flux maxima in both the bands. These differences are more prominent in the outer jet i.e. 10$\arcsec$-20$\arcsec$ from the nucleus. The optical flux maxima of knots are in several cases observed to be slightly downstream as compared to the radio by $\sim$ 0.5 - 1$\arcsec$. We also see similar differences in case of locations of polarization maxima (or minima) in both the bands. There are several places within the jet, where the maxima of radio polarization falls in the same place as optical polarization minima, e.g in D-East, E and F in the inner jet and I, upstream and downstream ends of A, B2 and C2 in the outer jet. The close inspection of the bottom two panels show that the flux and polarization do not necessarily follow each other, however; the locations of polarization maxima are shifted downstream by about $\sim$0.$\arcsec$25-0.$\arcsec$5, especially in the radio band, although; the difference is not that significant in case of the optical.\\
\indent Both the bands show a much more resolved fractional polarization structure as compared to the similar previous studies. The fractional polarization of radio is much more uniform as compared to the optical. As we will discuss later in this paper, we can clearly see this trend in the polarized flux images of individual knots as well. The optical images show regions of very high polarization and very low polarizations or depolarization close to each other, especially in the outer jet knots, A, B and C, whereas, this difference in degree of polarization is much less prominent in radio.\\

\indent The fractional polarization seems to be significantly higher in the regions of possible shocks in the jet, for example, knot HST-1, A-shock and the downstream end of knot C. The diffuse emission in all the inter-knot regions seem to have higher polarization as well. In the radio band, the flux and polarization maximum often do not appear to be at the same location. In fact, the polarization maxima are located $\approx$0.$''$25-0.$''$5 of the flux maxima in some individual knots, e.g. in knot D-East and A-shock, the polarization maxima is 0.25$''$ downstream of total flux maxima. We have pointed out flux and polarization maxima by red arrows in the radio maps (top two) in Fig.~\ref{fig:full_jet} and very clearly seen the bottom two panels of Fig.~\ref{fig:flux_poln_profile}, where we see the polarization maxima clearly shifted downstream of flux maxima in both the bands. This trend is much more clear in case of optical knots F, A and C, shown by yellow arrows in bottom two panels and is also discussed in respective sub-sections. \\
\indent One other important but not so obvious trend in Fig~\ref{fig:full_jet} is the apparent helical structure of the jet. This is seen in general in both radio and optical images. We see this much more clearly in the diagrams of MFPA vectors of individual knots (Fig.~\ref{fig:nuc-hst1} through ~\ref{fig:knot-cg}). We discuss individual knots in detail in terms of the common trends discussed above as well as their flux and polarization structure next.

\subsection{Inner jet}
\begin{figure*}
    \begin{center}
        \begin{tabular}{@{}cc@{}}
            \includegraphics[width=0.5\textwidth]{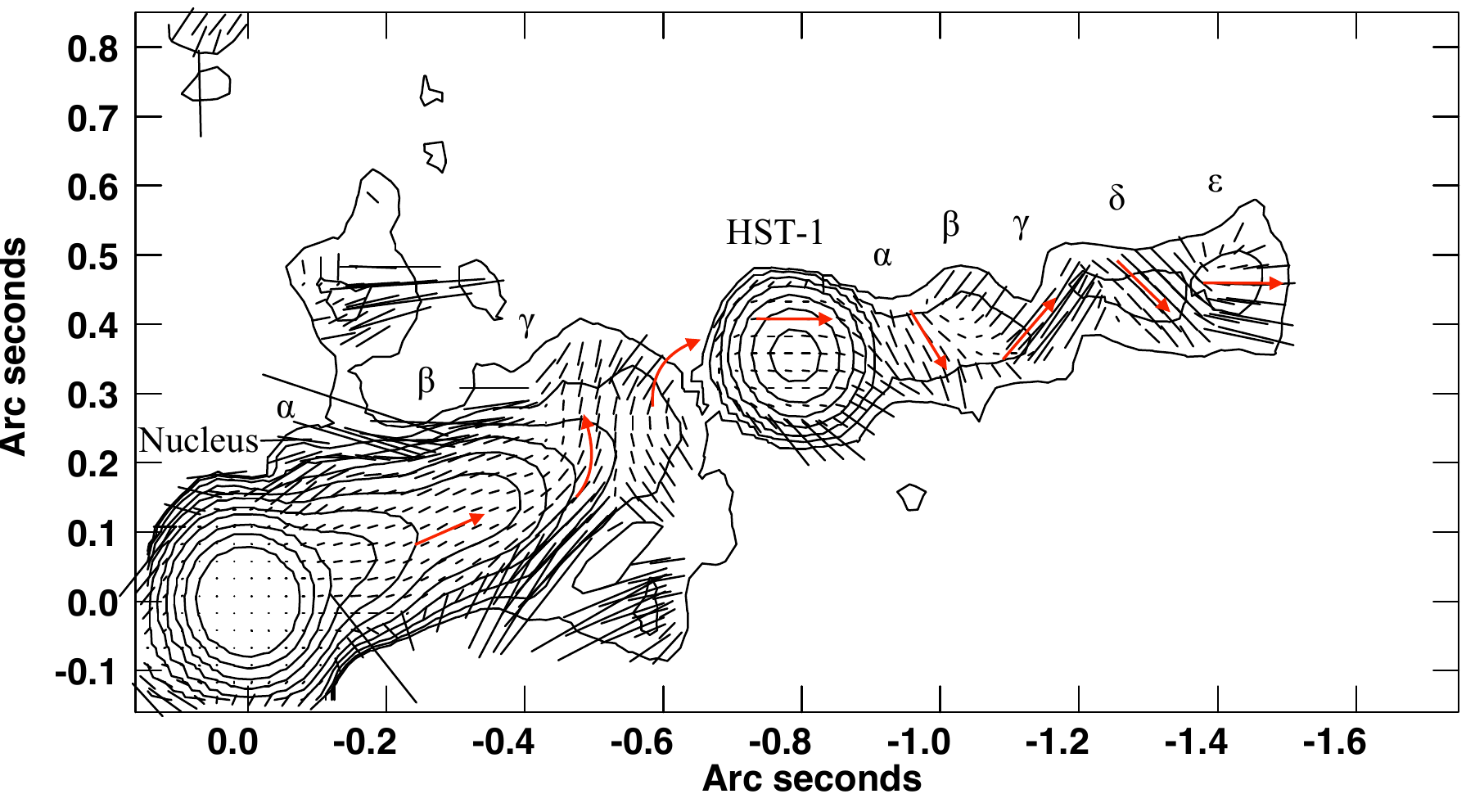}
            \includegraphics[width=0.51\textwidth]{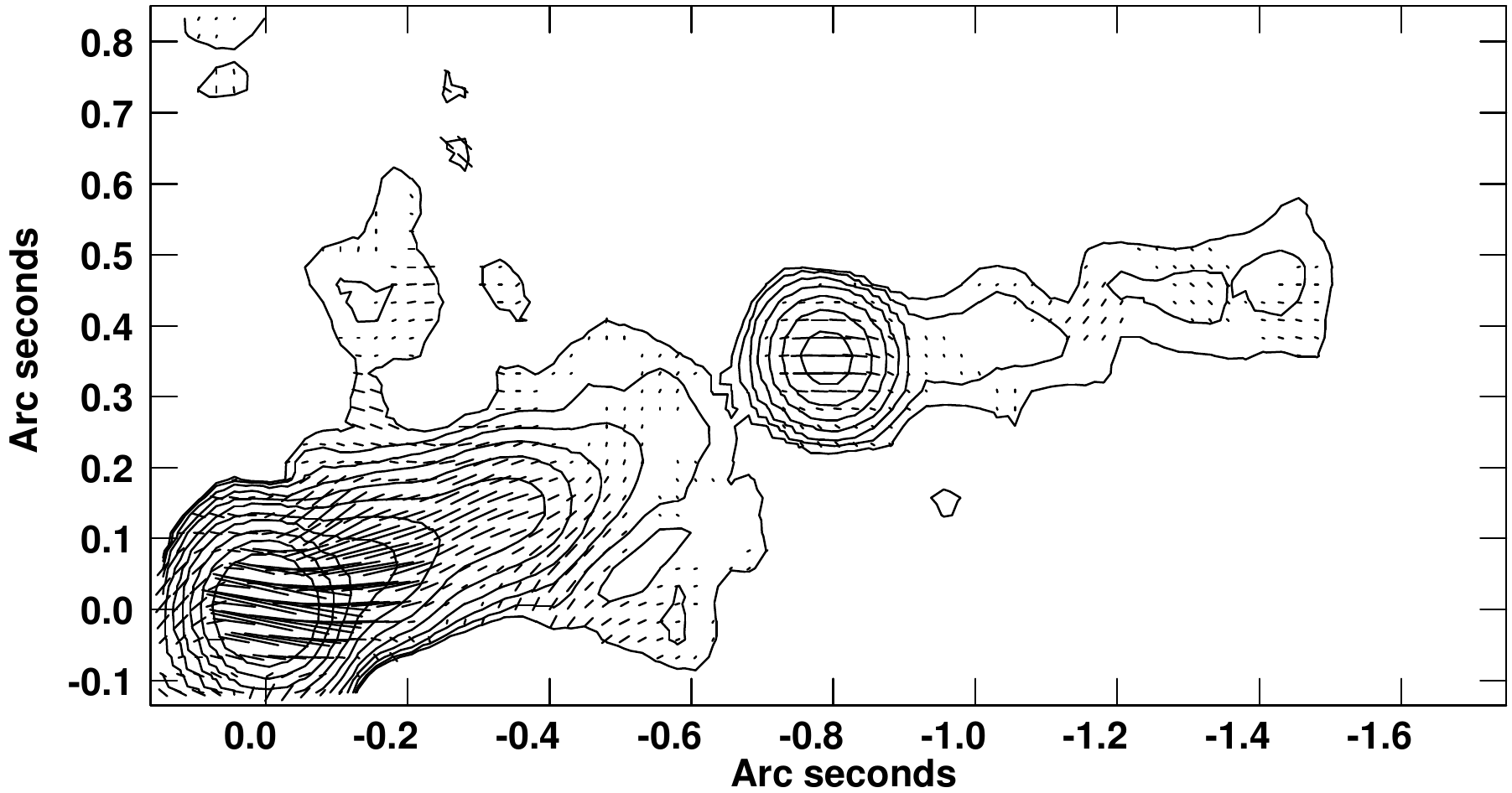}\\
            \includegraphics[width=0.5\textwidth]{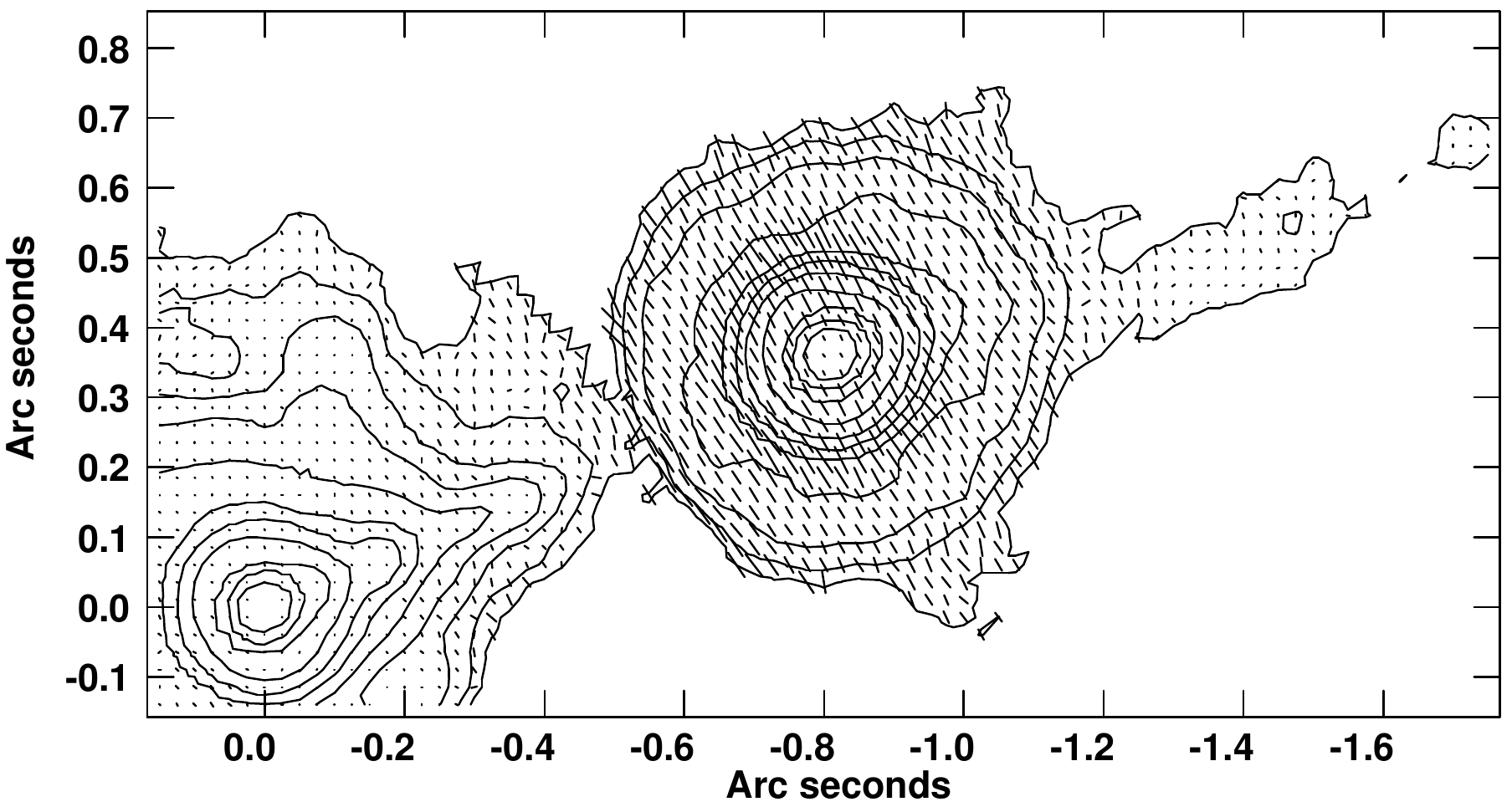}
            \includegraphics[width=0.51\textwidth]{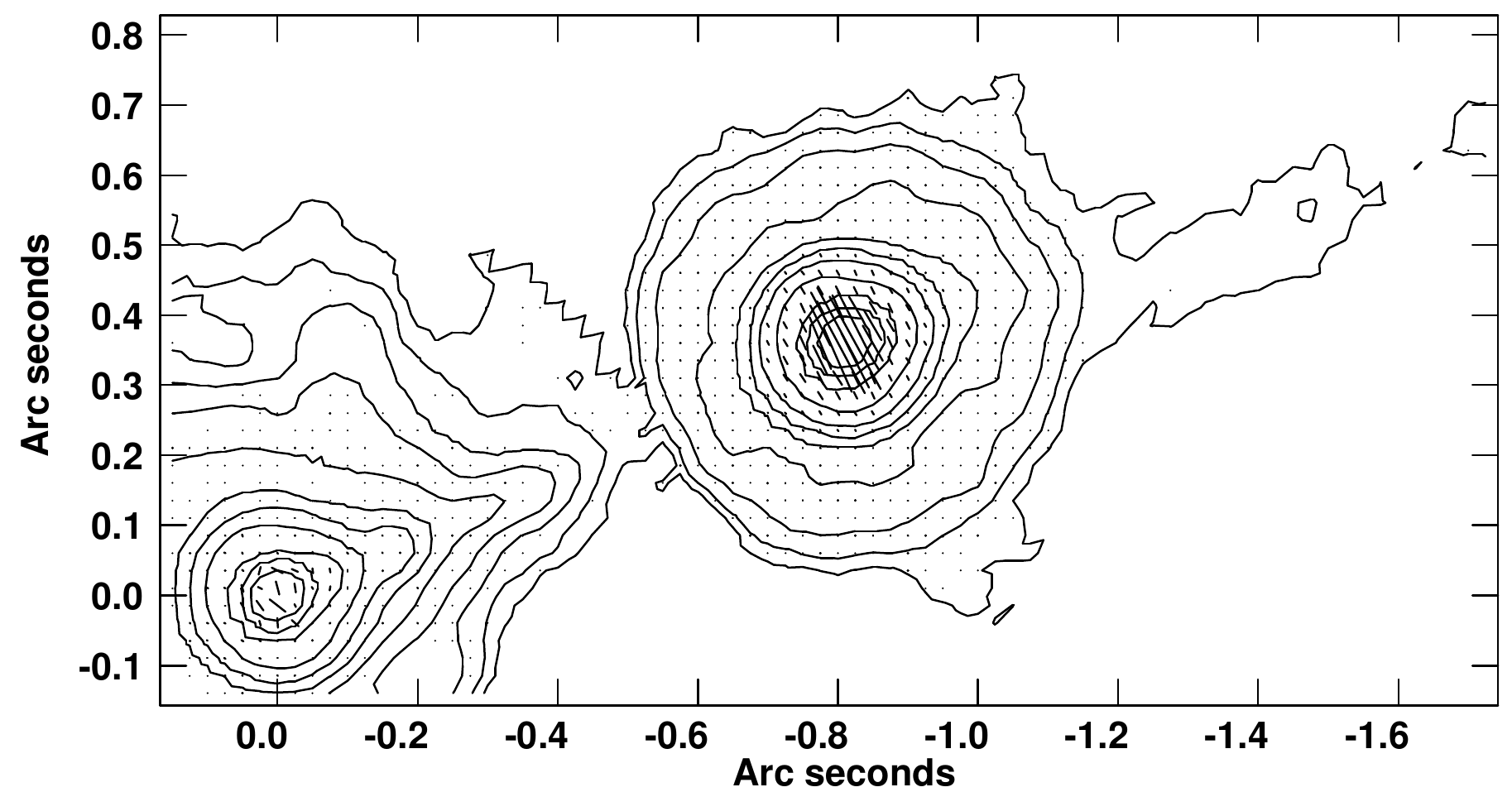}
        \end{tabular}
    \end{center}
    \caption[Radio and optical flux and polarization images of inner jet, the nucleus and HST-1.]{Nucleus and knot HST-1. Comparison of the radio and optical polarimetry. Top: 22 GHz, Bottom: F606W; Left panel: Fractional polarization, Right panel: Polarized flux. The images show combined epochs between 2002 and 2008 for both the wavebands. The contours represent the flux overlaid by the MFPA vectors. The contour levels are at (1, 2, 4, 6, 8, 12, 16, 32, 64, 128, 256, 512, 1014)$\ast$ 0.95 mJy beam$^{-1}$ in radio and 2e-2 $\mu$Jy in optical. The length of the vectors represent the amount of percentage polarization in the region. The red arrows in the radio polarization image show an apparent helical pattern traced by the MFPA vectors in the form of a wrapping within the jet boundaries. We show similar apparent wrapping patterns traced in other images.} \label{fig:nuc-hst1}
\end{figure*}

\indent The inner jet consists of the nucleus, and knots HST-1 and D along with the faint inter-knot emission. The inner jet extends out to about 4$''$ from the nucleus. Figures~\ref{fig:nuc-hst1}, top and bottom, show the stacked image of the nucleus and knot HST-1 in radio and optical bands. On the left panel we show the fractional polarization images and on the right panel are polarized flux images. \\
\indent Similarly, the total flux and polarization images of knot D are shown in Fig.~\ref{fig:knotD}, top and bottom. We discuss the radio and optical total flux and polarization structure of the inner jet in the following section, starting with the nucleus and HST-1.

\subsubsection{\textit{Nucleus}}
\indent The nucleus itself shows a well resolved polarization and flux structure which was not observed in any previous similar study. The region downstream of the nucleus shows a bright extended feature with several sub-components. The faint emission downstream of the nucleus can be distinguished in 3 distinct regions:
Nucleus-$\alpha$ at (0.1, 0.05)\footnote{Distances measured from the nucleus in ($\Delta \rm$RA, $\Delta \rm$Dec). Note that we leave out the `-' sign in the RA for brevity.}arcsec, Nucleus-$\beta$ at (0.3, 0.1) arcsec and Nucleus-$\gamma$ at (0.5, 0.2) arcsec, decreasing in brightness slightly in that order. These unresolved components, similar to the ones seen in \citet{2007ApJ...663L..65C} in their {\it VLBA} observations very close to the nucleus, were not seen in prior {\it VLA} observations. These could be either standing features in the jet or could be moving downstream and feeding the matter into the upstream region of knot HST-1. We do not have enough resolution to comment on their exact nature. \\
\indent The most significant difference between the previous radio polarimetry images and our images is in terms of well resolved sub-components just downstream of the nucleus out to $\sim$0.$''$6. We can distinctly identify at least three regions downstream of the nucleus, which are bright in the flux and show differences in the polarization morphology (Fig.~\ref{fig:nuc-hst1} (top, left)). The innermost diffuse regions, Nucleus-$\alpha$ and Nucleus-$\beta$ are not well resolved in our images. The observed MFPA in this region is parallel to the jet direction. Moving out to about 0.$''$5 from the nucleus, the jet is broadened and the vectors are seen to rotate counterclockwise by about 45$^{\circ}$ in region Nucleus-$\gamma$. Toward the downstream end of $\gamma$, the vectors turn back toward the jet center and ultimately become parallel again in the center of HST-1. Although the nucleus is less polarized, these unresolved regions downstream are found to have higher polarization of about 10-20\%, which is of the order of the knot HST-1's polarization. This nuclear structure was not observed in the previous images of P99.\\
\indent The sub-components show a complex polarization morphology. The nucleus is only weakly polarized (below 5\%) in the center, compared to $\approx$10-20\% near the edges, which is consistent with the general trend of the radio jet (see \S3.1). The high polarization region in the nucleus also does not coincide with the high flux region. In fact, the polarization is seen to be lowest at the flux maximum of the nucleus. The MFPA, in the region of the nucleus and in the region downstream of it (up to $\sim$0.$''$5), are predominantly parallel to the jet flow direction. The vectors are observed to rotate counterclockwise by about 5-10$^{\circ}$ near southern edge and clockwise near the northern edge by approximately same amount. In optical, the nuclear flux and polarization do not show any organized correlation either.\\
\indent Fig.~\ref{fig:nuc-hst1} (bottom) shows the combined F606W observations of the inner optical jet of M87 with a very bright nucleus and an equally bright knot HST-1. In making the optical images, the galaxy subtraction was stopped at $0.''4$ from the nucleus, which significantly affected the intensity of the nucleus. The innermost isophote, the wings of the point source and galaxy in part, of the optical jet is distorted due to the galaxy subtraction and do not have the shape as expected. As a result we do not see as much structure in optical as in radio near the nucleus. We do see a faint trailing emission corresponding to Nucleus-$\gamma$ at radio, which merges into the broad emission of knot HST-1 beyond about 0.$''$5 from the nucleus.\\
\indent The radio and optical polarized flux images show a similar morphology in general. We can identify the region corresponding to the nucleus-$\beta$ and $\gamma$ as seen in our radio images. Both the regions show low polarization and the MFPA is mostly parallel in the center, however this region in optical have a very low SNR near the edges where we see that the MFPA becomes more random, a feature seen in radio images as well. The MFPA becomes perpendicular at the downstream end of this region, near knot HST-1, although, we do not really as much structure in optical as we can in radio.

\subsubsection{\textit{Knot HST-1}}
\indent Fig.~\ref{fig:nuc-hst1} (top, left) shows the flux contours, MFPA vectors and fractional polarization of the most interesting region of the jet, knot HST-1. In this combined radio image, HST-1 is seen as bright as the nucleus. Similar to the extended structure observed and discussed for the nucleus region, we see fainter emission beyond knot HST-1 spread out from 0.$''$9 to 1.$''$6. Within this emission, we can distinguish at-least 5 regions, based on flux brightness and polarization morphology. Fig~\ref{fig:nuc-hst1} (top, left) shows the sub-components of HST-1's extended emission. The first of these, at (0.9, 0.36) arcsec from the nucleus is a faint and fairly diffuse region, called HST-1$\alpha$ connecting the flux maxima of HST-1 and another bright region labelled HST-1$\beta$ at (1.01, 0.38) arcsec from the nucleus. It is followed by less bright region HST-1$\gamma$ at about (1.15, 0.4) arcsec from the nucleus. HST-1$\delta$ and HST-1$\epsilon$, (1.3, 0.43) and (1.4, 0.45) arcsec respectively, from the nucleus, are very similar in brightness. However, not all these components are identified in polarized flux image on the top right. Out of the 5 sub-components described above, we can identify a very low polarization region of HST-1$\alpha$, somewhat higher polarization region HST-1$\gamma$, and moderately polarized HST-1$\delta$ and southern edge of HST-1$\epsilon$. \\
\indent The center of HST-1 and the extended sub-structure show high radio polarization (typically around 20\%, but variable in the flux maximum region, as discussed in \S 4), which is evidence of a highly ordered magnetic field. For these images, the MFPA vectors lie mostly along the jet direction in the center of the knot. They are seen to rotate counter-clockwise in going further out from the center of HST-1, with a complex pattern of undulations downstream.\\
\indent The radio polarization of HST1-$\alpha$ is slightly higher than at the flux maximum of knot HST-1. The polarization vectors in this region are seen to lie oblique to the jet flow direction and turn slightly southward. We do not see significant polarized flux emission at the region corresponding to HST-1$\beta$, hence we will not discuss it further. In HST-1$\gamma$, the MFPA vectors turned northward and the fractional polarization reaches a local maximum ($\sim$50\%).  Further downstream, twin sub-components HST-1$\delta$ and southern part of HST-1$\epsilon$ are very similar in fractional polarization. HST-1$\delta$ displays vectors rotated downward as compared to the MFPA vectors at HST-1$\gamma$ while they again turn upward and become almost parallel to the jet direction in southern part of HST-1$\epsilon$. These patterns  trace an envelope of helical structure, shown by the guiding arrows on the Fig~\ref{fig:nuc-hst1} (top, left). This peculiar MFPA structure may indicate a helical magnetic field structure in the jet. We discuss this  in \S\ref{sec:helical}. \\
\indent The optical polarization structure of HST-1 is shown in Fig~\ref{fig:nuc-hst1} (bottom left) with the polarization vectors on the flux contours. HST-1 is highly polarized in the optical as compared to the nucleus with the fractional polarization ranging from 20\% to 45\%. Unlike in radio, the optical MFPA vectors are predominantly perpendicular to the direction of jet flow at the locations of flux maxima. The extended emission seen downstream of HST-1 in radio is not seen in optical. Out of the five sub-components identified in radio fractional polarization image, we can identify outer only three faint components corresponding to HST-1$\gamma$, HST-1$\delta$ and HST-1$\epsilon$. Due to the faintness of the emission we do not see a clear flux maxima in any of these sub-components. The fainter emission downstream, corresponding to HST-1$\epsilon$, has lower polarization. The vectors in this region are oriented in a peculiar circular pattern which can be explained in terms of the wrappings due to the helical magnetic field in the region.

\begin{figure*}
\centering
  \begin{tabular}{@{}cc@{}}
  \includegraphics[width=0.5\textwidth]{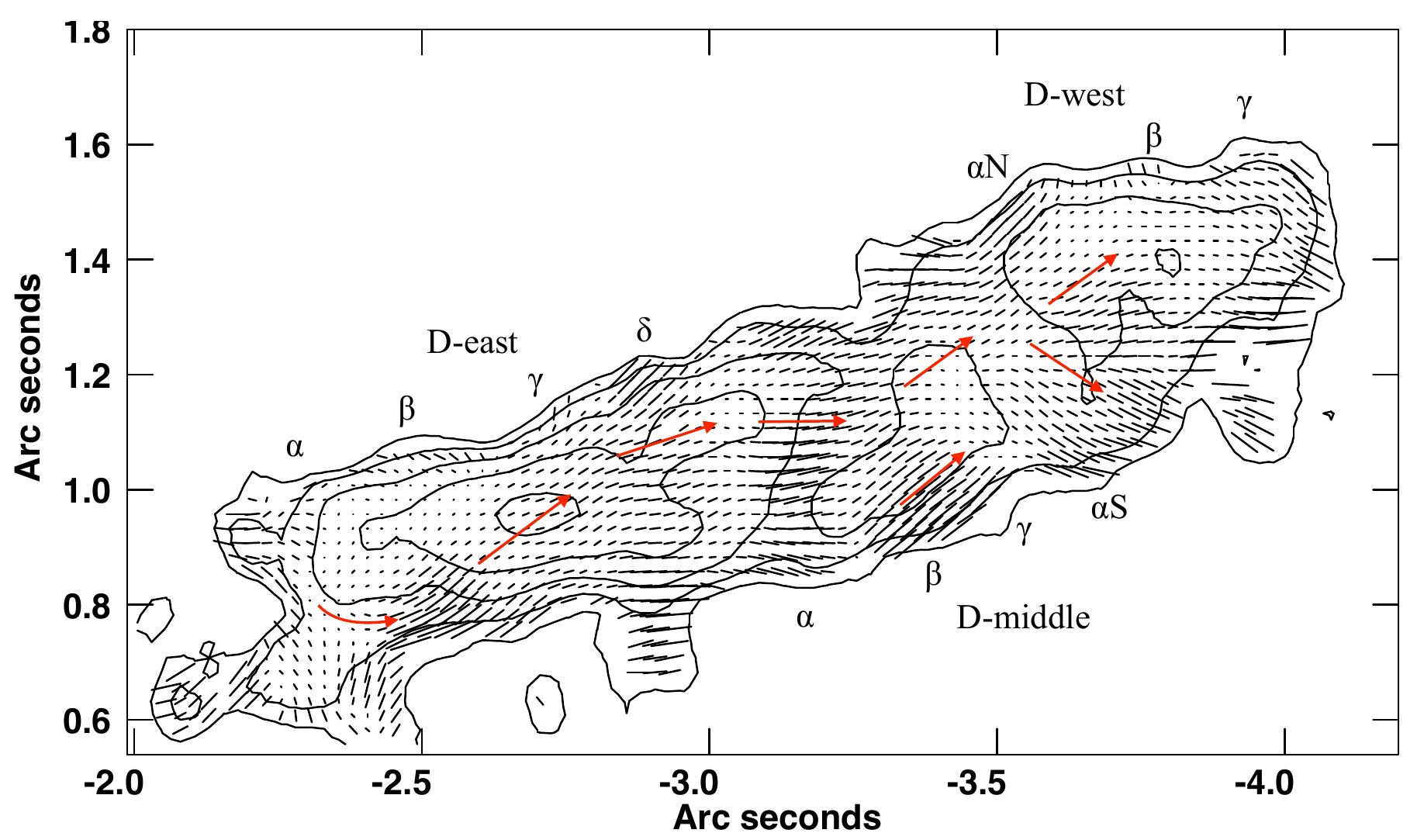} 
  \includegraphics[width=0.5\textwidth]{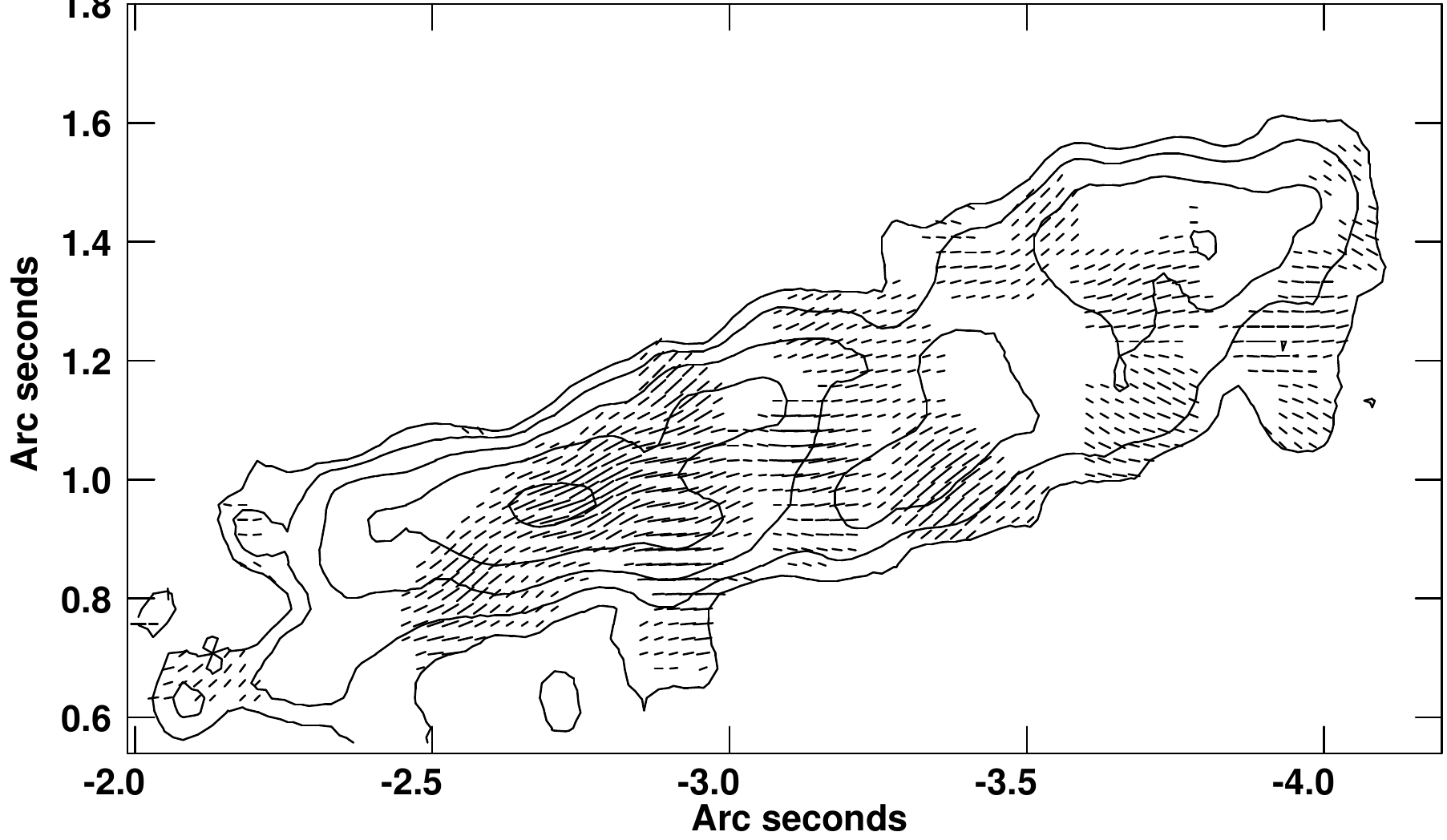} \\
    \includegraphics[width=0.5\textwidth]{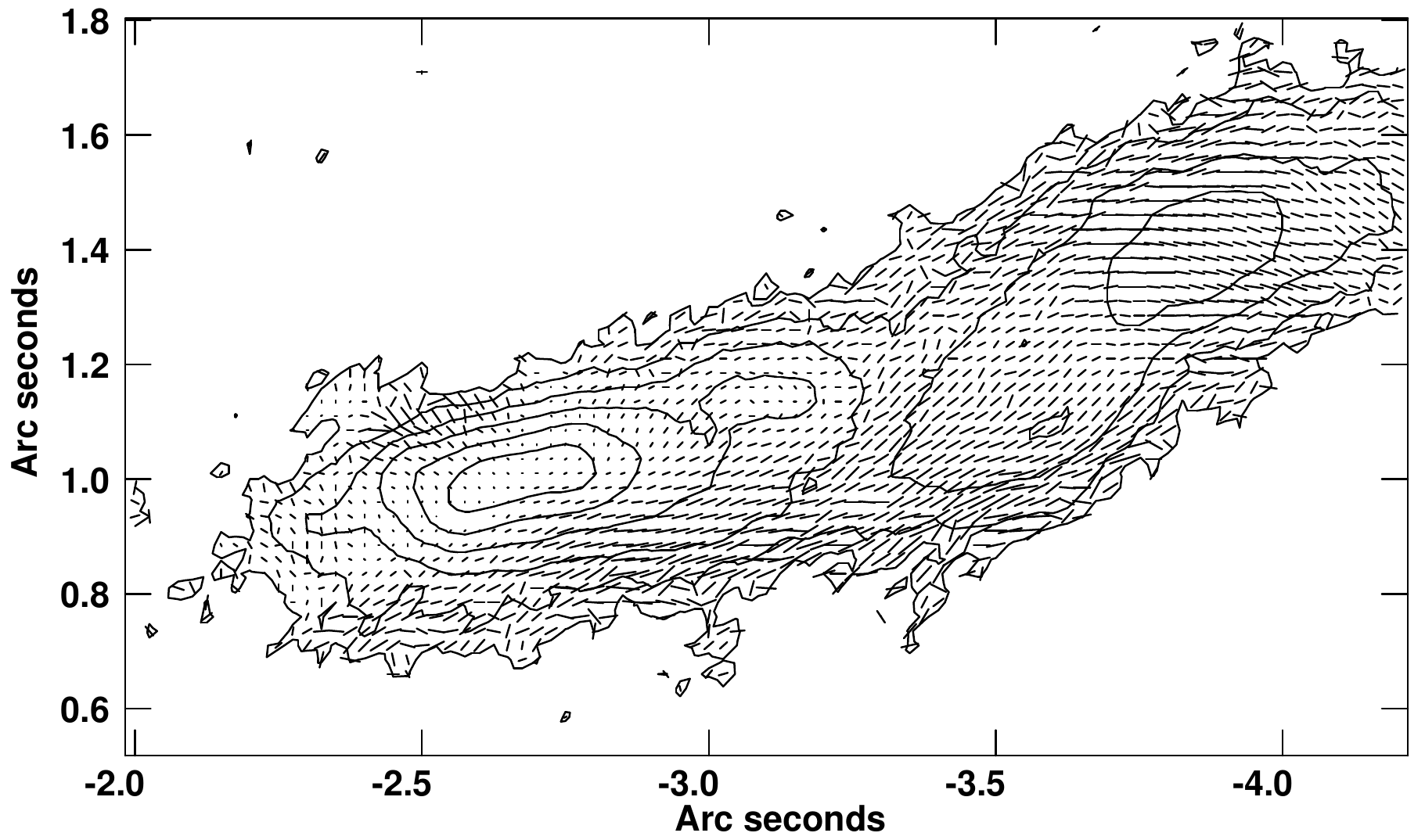} 
  \includegraphics[width=0.5\textwidth]{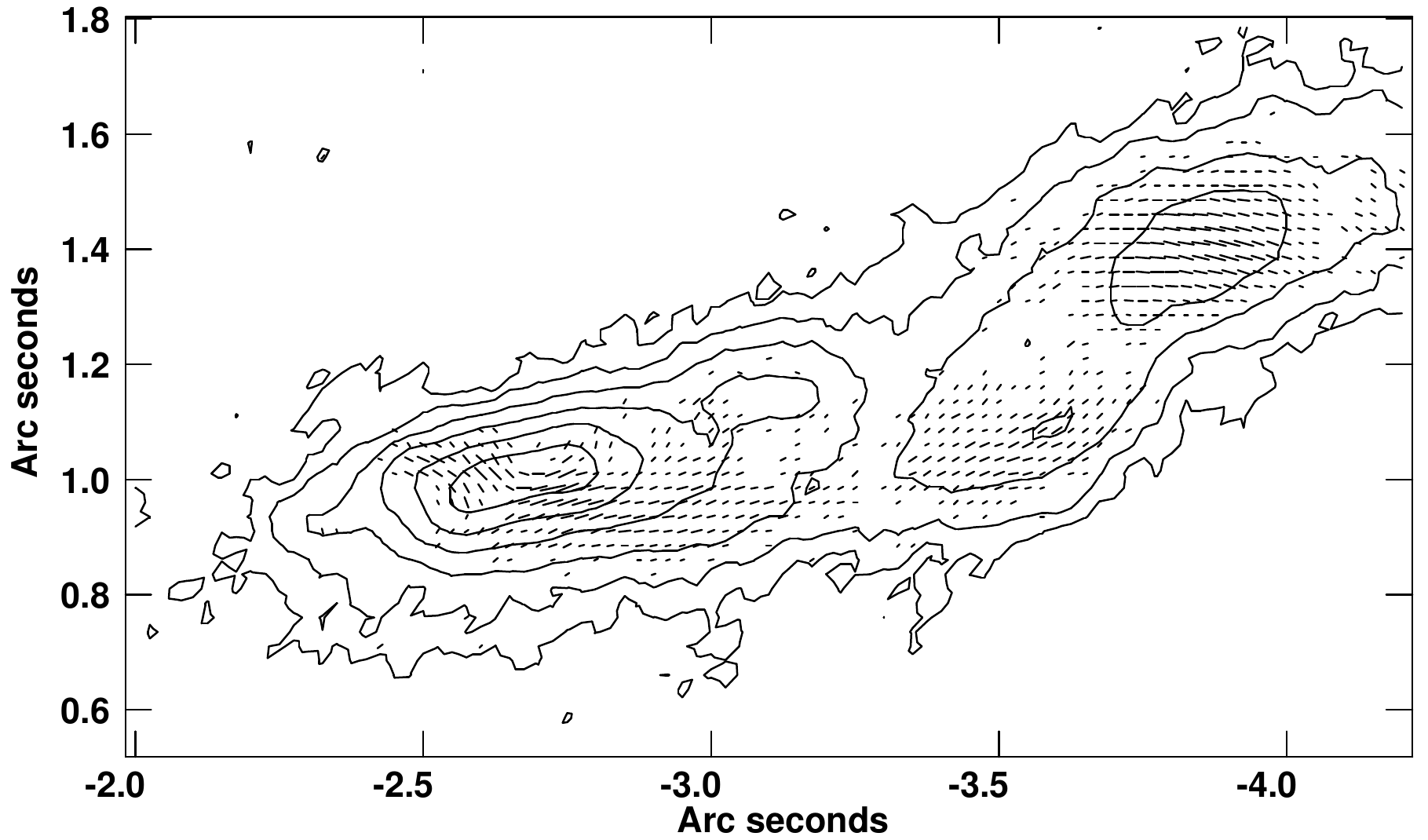}
\end{tabular}
 \caption[Radio and optical flux and polarization images of inner jet, knot D]{Knot D. Comparison of the radio and optical polarimetry. Top: 22 GHz, Bottom: F606W; Left panel: Fractional polarization, Right panel: Polarized flux. The images show combined epochs between 2002 and 2008 for both the wavebands. The contours represent the flux overlaid by the magnetic field polarization angle vectors. The contour levels same as in Fig.~\ref{fig:nuc-hst1} with flux level at 0.5 mJy beam$^{-1}$ in radio and 8e-3 $\mu$Jy in optical. The length of the vectors represent the amount of percentage polarization in the region.}
  \label{fig:knotD}
\end{figure*}

\subsubsection{\textit{Knot D}}\label{sec:D}
\indent Another interesting region in the inner jet, located approximately between 2.$''$0 and 4.$''$0 from the nucleus, is knot D (Fig.~\ref{fig:knotD} top left). Knot D shows much more extended structure than knot HST-1 and shows more complex flux and polarization features. The knot is typically divided into three regions, D-East,  D-Middle, and D-West. We further identify sub-components of these three as labelled in Fig.~\ref{fig:knotD} (top left), based on the different flux and polarization morphology in each region, although some of these regions are below the 3$\sigma$ significance or are depolarized as seen in the polarized flux image on top right. We discuss below only the regions which are well above the 3 $\sigma$ significance.\\
\indent Knot D is highly polarized, ranging from 25\% to 50\%, at all observed radio wavelengths. The upstream sub-component knot D-East is the brightest and most extended among three sub-components. The upstream end of knot D-East, labelled as $\alpha$ and located at (2.4, 0.82) arcsec from the nucleus, shows low polarization in the fractional polarization image on top left but shows polarization well below 3$\sigma$ in the polarized flux image on top right and hence we cannot comment much on its polarization structure. Downstream of it is $\beta$ at (2.5, 0.9) arcsec from the nucleus, which shows polarized structure above 3$\sigma$. The MFPA in this region is observed to turn clockwise at the northern edge while it turns counter-clockwise at the southern edge. The polarization increase is also higher at the southern edge. Next is $\gamma$, at (2.7, 0.95) arcsec from the nucleus, the location of the flux maximum, where the MFPA once again becomes parallel to the axis. Here the polarization reaches to about 30\%. The flux maxima is about $0.''25$ upstream of $\gamma$. The downstream end of knot D-East is $\delta$ at (2.9, 1.0) arcsec from the nucleus, where the MFPA becomes complex and is seen to turn upward in the upper half and downward is the lower half of the knot, as shown by the red arrows in radio image. The northern and southern edges still have higher polarization as compared to the rest of the knot D-East. \\
\indent Knot D-Middle is shifted southward in the jet. This gives the impression of a small bend in the jet. Based on the polarization morphology we identify 3 distinct regions in knot D-Middle, namely, $\alpha$, $\beta$ and $\gamma$, among these $\gamma$ has the lowest polarization and is well below 3$\sigma$ level in polarized flux, while other two show significant polarization. D-Middle is also the region of highest polarization of all three. At the upstream end of knot D west, in $\alpha$ at (3.2, 1.1) arcsec from the nucleus, the MFPA is mostly east-west and the polarization is close to 40\%. Beyond this, in $\beta$ at (3.4, 1.18) arcsec from the nucleus, we see the flux maximum and the MFPA turns counter-clockwise by more than 40$^{\circ}$. The polarization starts decreasing and reaches minimum (well below 3$\sigma$) at the flux maximum of $\gamma$, at (3.55, 1.2) arcsec from the nucleus.\\
\indent D-West is also distinguished into four regions: $\alpha$ which is further divided into $\alpha$N and $\alpha$S (northern and southern edge regions), and $\beta$ and $\gamma$, which, the last two are below 3$\sigma$ significance of polarization.  D-W$\alpha$ form $\alpha$N and $\alpha$S, located at (3.5, 1.45) and (3.7, 1.1) arcsec, display a unique polarization morphology. The MFPA at $\alpha$S lies perpendicular to the local flux contours while in $\alpha$N, it is seen to be parallel to the local flux contours. The MFPA starts to converge back to the center and show significant decrease in polarization or a region of depolarization in D-W$\beta$, at (3.8, 1.35) arcsec from the nucleus, where the flux maximum is.\\
\indent Throughout the knot D complex, the flux and polarization maxima do not coincide. The local polarization maxima are observed to lie either upstream or downstream of the flux maxima by about $0.''2-0.''5$. The polarization structure in all three sub-components appear to be wrapped around the jet axis as shown by the red arrows in the Fig.~\ref{fig:knotD} and is consistent with the general trend along the jet (see \S3.1). \\
\indent Fig~\ref{fig:knotD} bottom shows the optical flux and polarization morphology of knot D. Similar to the radio, optical knot D can be divided into D-East, D-Middle and D-West regions {\bf and the same sub-components within each of them } i.e. $\beta$ and $\gamma$ in D-East, $\beta$ in D-Middle and $\beta$ in D-West. We clearly see the flux maxima of knots D-East and West and well resolved while D-Middle is relatively fainter similar to in radio. We notice that the optical flux maxima of all the knots is shifted slightly upstream as compared to the radio flux maxima. The flux maximum of knot D-Middle is seen to be shifted downward from the jet axis, which we also observe in radio.\\
\indent Knot D has overall higher fractional polarization in the radio $\sim$60\%), which is close to the upper limit of synchrotron emission polarization. D-East's flux maximum is the region of lowest fractional polarization in the group, a similar trend as seen in optical by P99. The polarization is lower along the center of the knot region while it increases near the edges. The MFPA is parallel in D-East and are observed to rotate clockwise on the southern edge and counterclockwise on the northern edge of the jet and become almost perpendicular to the jet flow. Moving further downstream, in the knot D-Middle, the MFPA stay mostly parallel; however, polarization is higher than D-East. In the knot D-West, most of the MFPA become perpendicular to the jet flow direction near and around the flux maxima, whereas their orientation is random near the edges.

\subsection{Intermediate jet}

\begin{figure*}
\begin{center}
\begin{tabular}{@{}cc@{}}
  \includegraphics[width=0.51\textwidth]{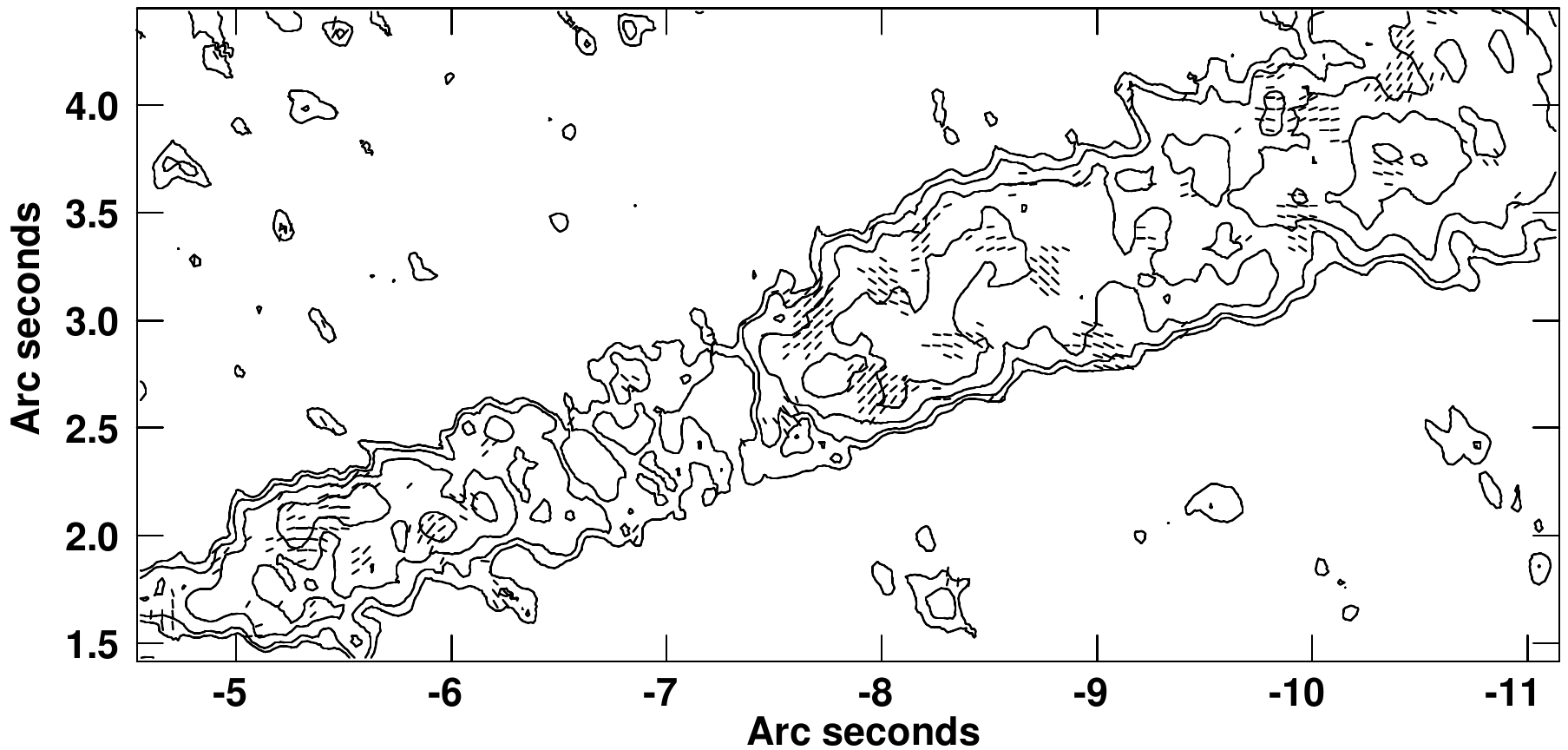} \\
  \includegraphics[width=0.51\textwidth]{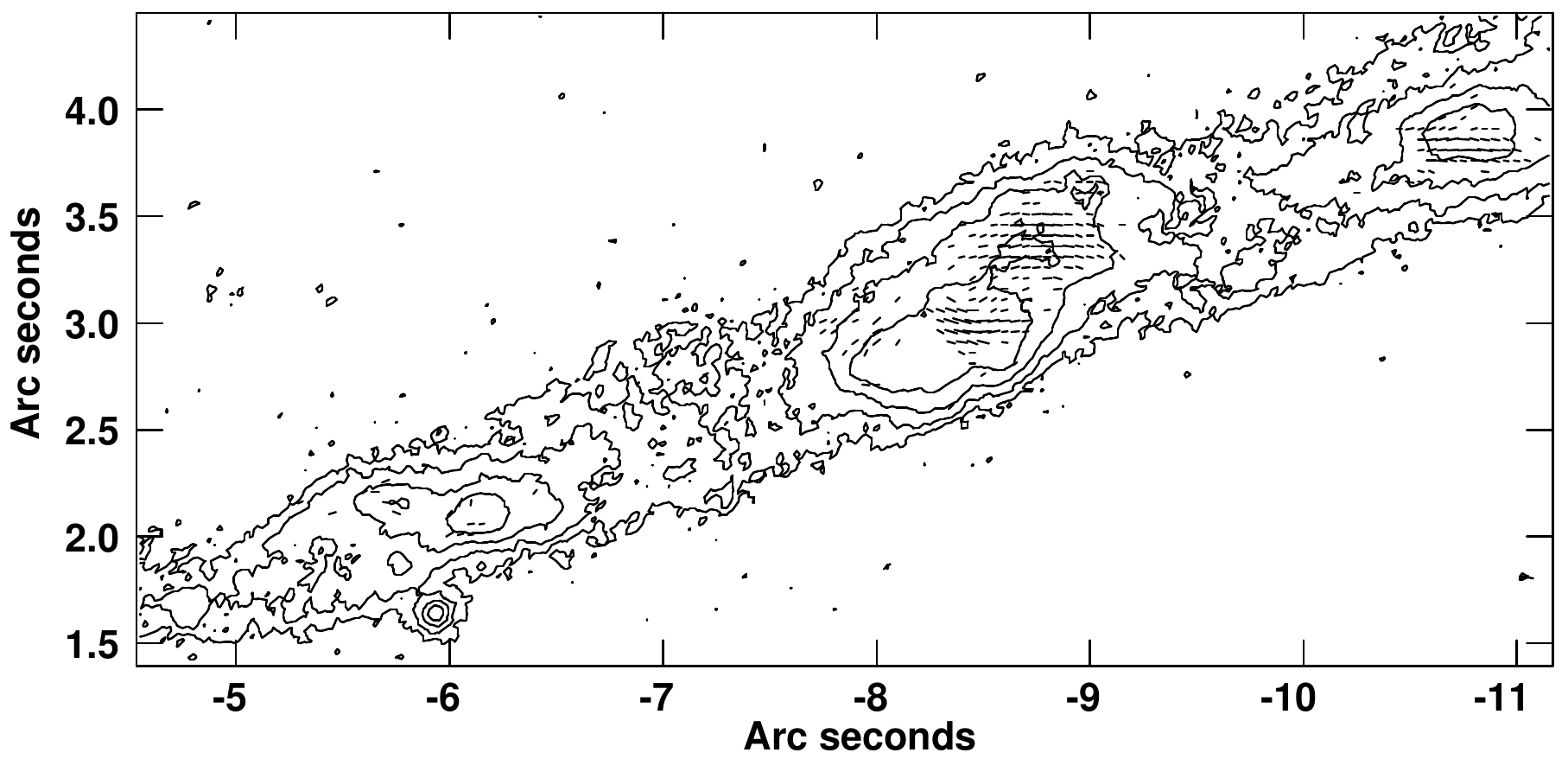}
\end{tabular}
 \caption[Radio and optical flux and polarization images of intermediate jet, knots E, F and I.]{Knots E, F and I. Comparison of the radio and optical polarimetry. Top: 22 GHz, Bottom: F606W; The images show polarized flux image of combined epochs between 2002 and 2008 for both the wavebands. The contours represent the flux overlaid by the magnetic field polarization angle vectors. The contour levels same as in Fig.~\ref{fig:nuc-hst1} with flux level at 0.2 mJy beam$^{-1}$ in radio and 8e-3 $\mu$Jy in optical. The length of the vectors represent the amount of percentage polarization in the region.}
  \label{fig:knot-efi}
  \end{center}
\end{figure*}

\indent Knots E, F and I lie between $\sim$5.$''$0 and 11.$''$0 from the nucleus of M87. Fig.~\ref{fig:knot-efi} show the polarized flux contour plots of this region. The knots show complex flux and polarization features. The signal to noise ratio is not sufficient to resolve the individual sub-components as we see in case of inner jet, however. Most of the regions of these knots show very low polarization (well below 3$\sigma$) or regions of depolarization as can be seen from the figure, hence we do not show the fractional polarization plots here. As a result we cannot comment on their polarization structure, however we describe their flux structures below.

\subsubsection{Knot E}\label{sec:E}
\indent Knot E is the most compact knot among the three. The radio morphology of knot E does not seem to have a clear flux maximum.  What we see instead is a complicated flux distribution with many small local maxima that appear to be situated near the jet edges. Note, however, that due to the faintness of these knots, our signal to noise is the lower in this region than the other brighter regions, although it is significantly higher than in P99. \\
\indent Unlike in the radio, the optical morphology of knot E shows a clear maximum, located at approximately (6.1, 2.2) arcsec from the nucleus, superposed on a region of increased surface brightness that appears to extend diagonally from northeast to southwest.  The optical flux maximum is significantly downstream of the locations where knot E appears to be brightest in the radio, although it nearly corresponds with the local radio flux maxima at (6.0, 2.0) arcsec and (6.1, 2.2) arcsec from the nucleus.

\subsubsection{Knot F}\label{sec:F}
\indent Knot F is the brightest in the intermediate part of the M87 jet. It displays a complex flux structure with a few flux maxima throughout. \\
\indent The optical flux structure of knot F is quite different from what we see in the radio.  A broad optical flux maximum region extends from (8.0, 2.8) arcsec from the nucleus to about (8.8, 3.0) arcsec from the nucleus. In P99, this region was seen to separate into two flux maxima.  This is less apparent in the radio flux contour maps of Fig.~\ref{fig:knot-efi} bottom, possibly due to the temporal evolution of the knot, but similar features can still be seen on Fig. 7 of P99. We do not see distinct flux features in optical as described in radio above, but there are a few broad correspondences between the two flux regions in optical, at (8.1, 2.8) arcsec and (8.5, 3.0) arcsec from the nucleus, and regions identified as $\beta$ and $\gamma$ in radio. The region of second highest flux in optical, at (9.0, 3.4) arsec from the nucleus, correspond to the region $\delta$ in radio.\\
\indent Similar to the radio, the optical polarization structure is well below 3$\sigma$ or is depolarized, except the regions of $\gamma$ and $\delta$, where we see very low polarization. The MFPA vectors in this region are mostly oblique to the jet direction. The polarization is too small to say anything affirmatively. We can also see that the polarization maxima in knot F is shifted downstream slightly, which possibly is an indication of lack of spatial correlation between flux and polarization.

\subsubsection{Knot I}\label{sec:I}
\indent Fig.~\ref{fig:knot-efi} top shows radio flux and polarization structure of knot I. Knot I shows a clearly resolved radio flux maxima at $\sim$(10.4, 3.7) arcsec from the nucleus, along with a few secondary flux maxima region close to the upper edge of the it. Knot I is one of the lowest signal to noise regions along the jet. As can be seen in the polarized flux image in radio, top right, most of the knot is well below 3$\sigma$ significance. The northern part of the knot show very low polarization, however the information is not sufficient to comment on the polarization structure. \\
\indent The flux morphology of optical knot I is very different than its radio counterpart. We see one distinct flux maximum at (10.9, 3.9) arcsec from the nucleus, the same place as in radio. Similar to the radio, most of the optical knot I show very low polarization or depolarization (Fig.~\ref{fig:knot-efi}, bottom right). Although, we can see that the polarization is seen to be highest at the flux maxima in optical. The vectors are oriented in the east-west direction at the maxima. The rest of the regions do not show much information about the polarization.

\subsection{Outer jet}
\begin{figure*}
    \begin{center}
        \begin{tabular}{@{}cc@{}}
            \includegraphics[width=0.5\textwidth]{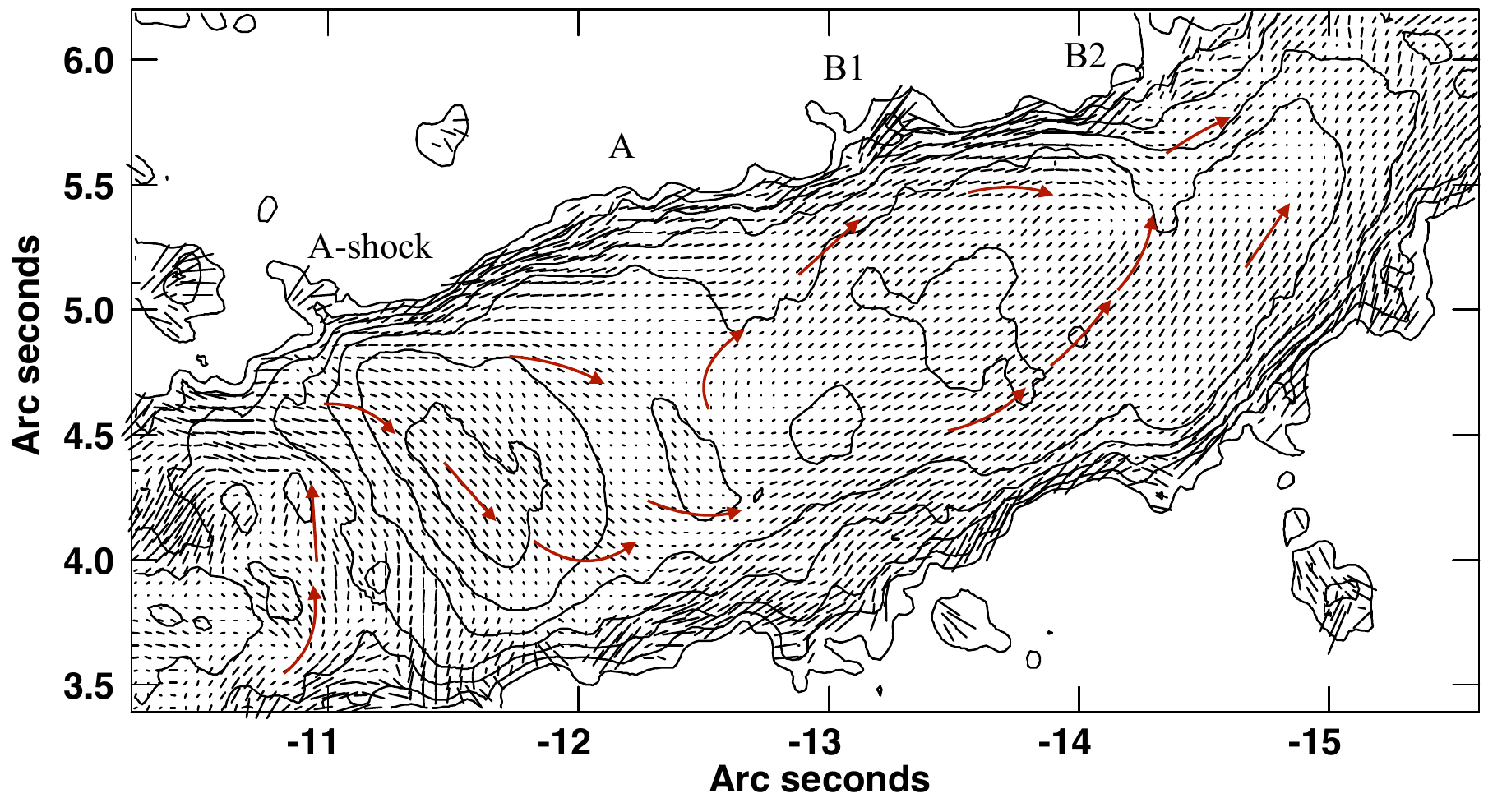}
            \includegraphics[width=0.5\textwidth]{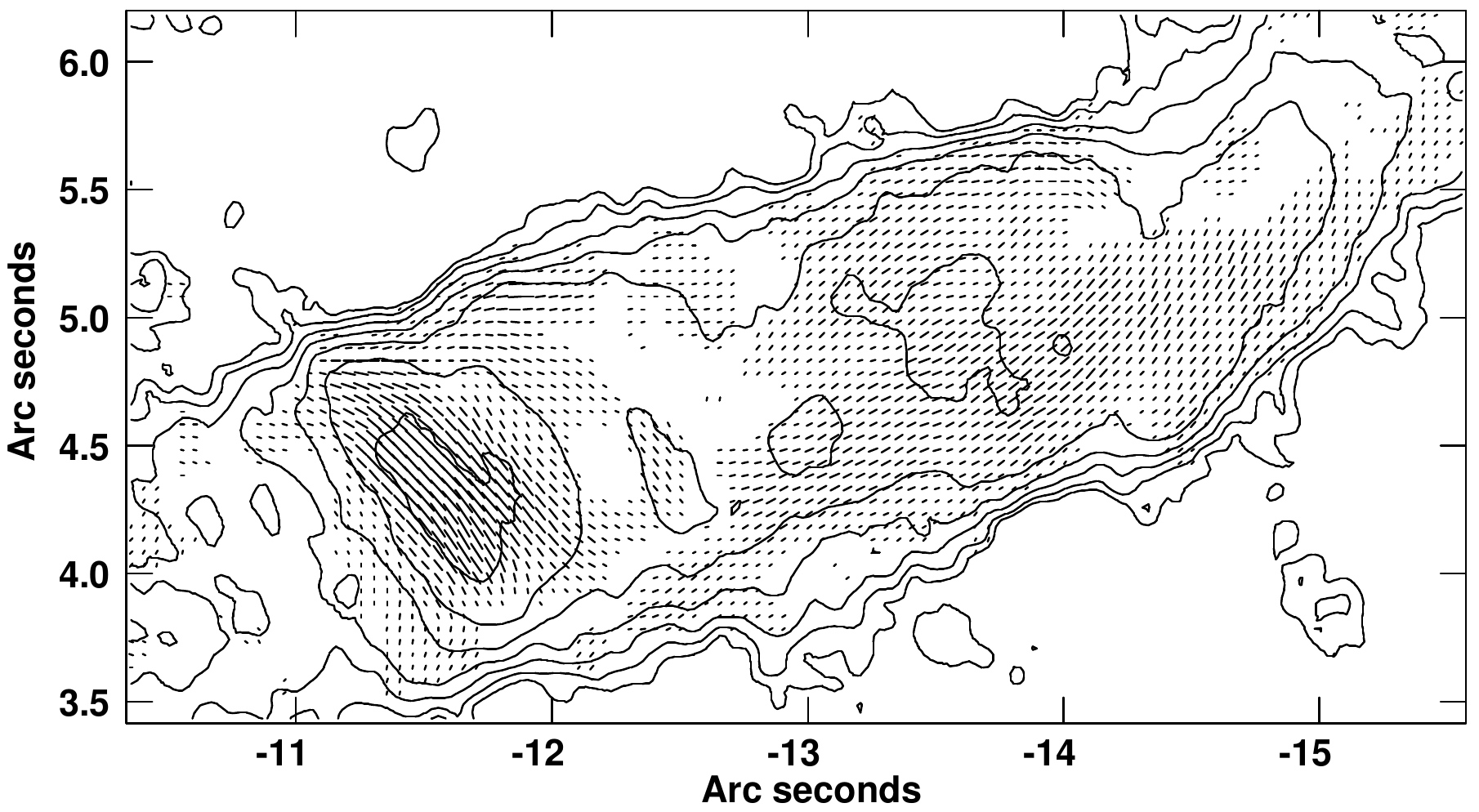}  \\
            \includegraphics[width=0.5\textwidth]{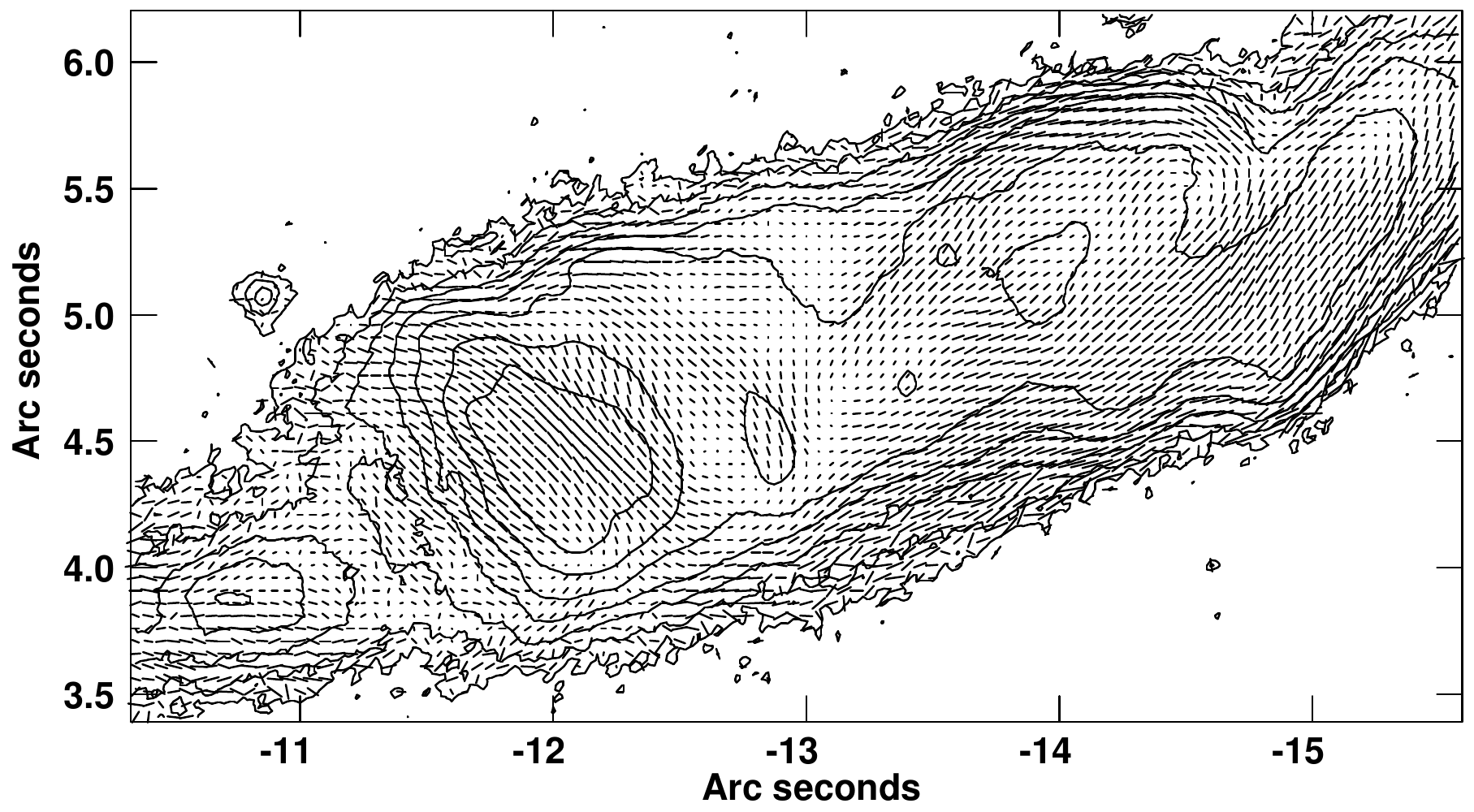}
            \includegraphics[width=0.5\textwidth]{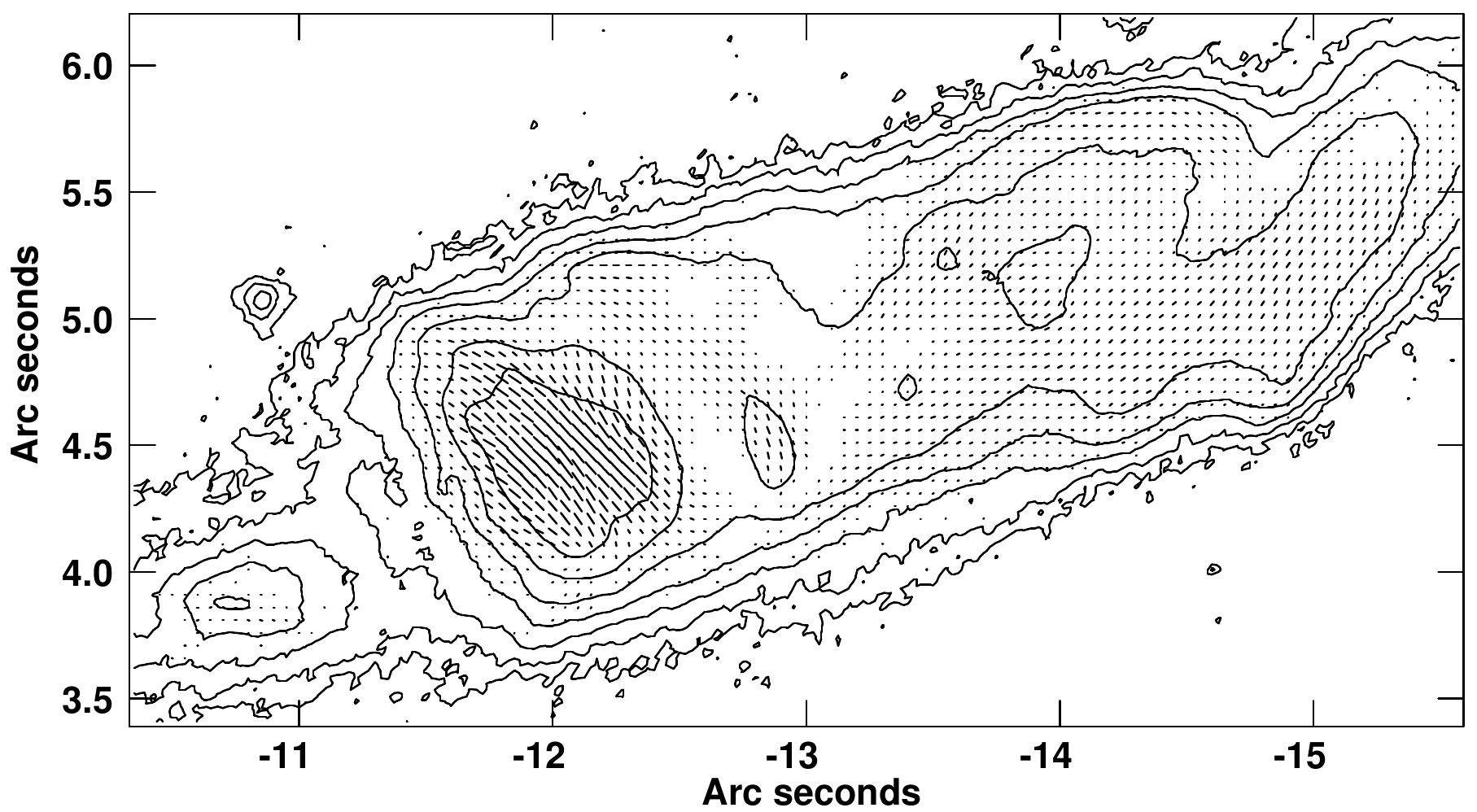}
        \end{tabular}
        \caption[Radio and optical flux and polarization images of outer jet, knots A and B.]{Knots A and B. Comparison of the radio and optical polarimetry. Top: 22 GHz, Bottom: F606W; Left panel: Fractional polarization, Right panel: Polarized flux. The images show combined epochs between 2002 and 2008 for both the wavebands. The contours represent the flux overlaid by the magnetic field polarization angle vectors. The contour levels same as in Fig.~\ref{fig:nuc-hst1} with flux level at 0.2 mJy beam$^{-1}$ in radio and 8e-3 $\mu$Jy in optical. The length of the vectors represent the amount of percentage polarization in the region.} \label{fig:knot-ab}
    \end{center}
\end{figure*}

\indent The outer jet comprises knots A, B, C and G, beyond which the jet disrupts and feeds matter into the eastern inner radio lobe. In figures \ref{fig:knot-ab} and \ref{fig:knot-cg}, the total flux shown on the left and the polarized flux is shown on the right (radio on the top and optical at the bottom), both overlaid by the MFPA vectors. \\
\indent The overall morphology of the outer knots, A, B, C, and G are quite different than rest of the jet. The difference in the thickness of the radio and optical jet is clearly evident from figures \ref{fig:knot-ab} and \ref{fig:knot-cg}. The radio jet is much thicker and disorganized, while the optical jet shows much more defined structure, near the bend as well as in the individual knots. This structure may suggest the presence of a layered surface, emitting different energy electrons from different physical regions as modelled in P99. This may be caused due to the internal magnetic field structure changing in the outer jet. P99 report a similar difference in the radio and optical morphology of inner and outer jet knots. \citet{1989ApJ...340..698O} and \citet{1996ApJ...467..597B} suggested this could be due to the Kelvin-Helmholtz instabilities causing the shocks within the jet and disrupting the magnetic field structure. We discuss the features in each of these knots in detail below.

\subsubsection{\textit{Knot A}}\label{sec:knotA}
\indent Fig.~\ref{fig:knot-ab} shows the region of the knot A+B complex. The left side panels show flux contours overlaid by MFPA vectors whereas on the right, we plot the polarized flux (signficant at 3$\sigma$) with the MFPA vectors. This extended flux region is the second brightest, after the nucleus, in the jet of M87, located between $\sim$11.$''$0-16.$''$0. At this location in the jet, the structure of the jet starts to change drastically, which is evident by the flux and MFPA vector morphology. The knot A+B complex has high signal to noise and shows fine details in each at 22 GHz. \\
\indent The upstream broad flux region is the brightest in knot A, called A-shock in the literature (P99) located at (11.7, 4.25) arsec from the nucleus. Downstream of which we can identify two flux maxima in knot A close to its southern edge located at (12.5, 4.4) and (13.0, 4.5) arcsec from the nucleus, respectively, which are better resolved as compared to P99. \\
\indent In radio knot A-shock, the MFPA vectors (top, left panel) are observed to be nearly perpendicular to the direction of jet. The vectors gradually become parallel to the jet direction near the edges. Knot A is highly polarized, affirming a highly ordered magnetic field in this region. The polarization vectors are found to rotate in a systematic pattern all through the knot A and B complex. They become parallel to the jet direction in the inter-knot region between the two and are shortened in length, implying a decreased polarization. Knot A-shock is the highest polarized region in the A+B complex with a fractional polarization of $\sim$40\%. The maximum of fractional polarization coincides with the flux maximum of A-shock. These two do not coincide at the flux maximum of knot A, however; the polarization vectors show an intriguing circular rotating pattern here. Following the general trend along the jet, the polarization maximum of knot A lies $\sim0.''5$ downstream of its flux maximum, however, it is slightly less polarized, at 20\%, as compared to the knot A-shock. This knot is also a classic example of apparent helical structure of the jet. We show red arrows on the Fig.~\ref{fig:knot-ab} (top, left) as a guide to the eye to trace the peculiar sinusoidal pattern in this region.\\
\indent Fig~\ref{fig:knot-ab} (bottom) shows the optical flux and polarization morphology of knot A. The structure in the optical is observed to be similar and equally complex as in radio. Knot A is the brightest optical knot of the outer jet. Knot A-shock displays a broad flux morphology, similar to the radio. The flux maximum of knot A-shock is located slightly downstream, at (12.0, 4.4) arcsec from the nucleus, as compared to the radio flux maximum, at (11.7, 4.25) arcsec from the nucleus. We see another flux maximum downstream of A-shock, at (12.9, 4.5) arcsec from the nucleus, which corresponds to the feature located at (12.5, 4.4) arcsec from the nucleus in the radio. The other flux maximum, which is seen in radio image at (13.0, 4.5) arcsec from the nucleus, is not well resolved in optical. The overall flux morphology in optical is slightly narrower that in radio, consistent with the spine-sheath model of \citet{2007ApJ...668L..27K}.\\
\indent The optical MFPA at A-shock is predominantly perpendicular to the jet flow direction and the fractional polarization is $\sim$40\%. The polarization maximum of A-shock just barely overlaps the flux maximum. Due to the broad A-shock feature, polarization maximum seems to have shifted slightly downstream, similar to the radio. The MFPA vectors are observed to rotate clockwise in moving away from the center and a slight increase in their size representing a relatively higher fractional polarization ($\textgreater$50\%) near the edges. Moving downstream from knot A-shock, the MFPA is seen to rotate clockwise and fractional polarization is significantly reduced to $\sim$20\% in the knot A region (at $\sim$12.$''$7 out), while the MFPA along the edge become parallel to the jet flow. The inter-knot region between A and B1 (the region upstream of knot B) is relatively less polarized and in some places the polarization reaches close to zero, as seen by the gaps in the polarization map.\\
\indent The figures on the right show the polarized flux in the knot A (top, radio; bottom, optical). The flux features seen in these images is very similar to the total flux images on the left. The radio polarized flux shows the increased polarization at the location of flux maxima. However, we do not see the higher polarization seen along the edges in the total flux image, which probably arises in total flux due to the shearing of field lines along the jet surface. We see a region of depolarization or which has polarization well below 3$\sigma$ near the flux maxima of knot A in both the polarized flux images. The two regions, in radio and optical, are slightly different in location, pointing toward the evidence of different origins of radio and optical electrons and the stratified jet model of P99.

\subsubsection{\textit{Knot B}}\label{sec:knotB}
\indent The image of knot B displays two sub-components, namely, knot B1 and B2. Knot B1 shows a clear flux maximum at (13.5, 4.9) arcsec from the nucleus and is brighter and broader of the two sub-components. Knot B2 shows more diffuse flux morphology spread between 14.$''$0 and 15.$''$0 from the nucleus. The jet displays a small bend at $\sim$14.$''$25 from the nucleus, where it bends northward through a small angle before forming knot C. \\
\indent The radio flux and polarization does not follow each other throughout the knot. Both the sub-components are moderately polarized in the radio to about 20-30\%. The region of high polarization is found to be $\sim 0.''5$ downstream of radio flux maximum of knot B1 in our maps. The jet starts bending toward north at knot B2. The polarization vectors are lying along the direction of the jet axis and are observed to turn along the direction of jet at the bend. At the downstream end of knot B2, polarization vectors display a very peculiar structure. The vectors tend to turn counterclockwise at about 14.$''$8 from the nucleus, where we also observe that the jet starts to bend upward and forms into the knot C. The vectors seem to form a circular pattern at this point and the polarization is minimum, which is very similar to what is observed in knot A. This is consistent in the all other radio wavelengths as well as in the optical images. The degree of polarization is observed to be high near the edges throughout knot B as is the case with almost all the knots in the jet. Knots B1 and B2 also display an apparent helical structure, also seen in knot A and described in the \S3.1.\\
\indent The optical flux morphology of knot B is very similar to the radio. We see two bright regions knots B1 and B2 located at $\sim$(13.9, 5.2) and (15.0, 5.1) arcsec from the nucleus. The locations of these two regions are shifted slightly downstream in optical as compared to the radio. Knot B1 is brighter than B2 in optical which is much diffuse and compact. The optical flux region is compact as compared to the radio. It also displays a small bend at approximately 14.$''$5 from the nucleus. This bend is much more prominent and is shifted downstream as compared to the bend in radio.\\
\indent The optical knots B1 and B2 are another high polarization regions. The optical MFPA direction is mostly parallel to the jet flow direction, however, it is observed to rotate counterclockwise in moving further out. In knot B1, polarization is minimum at the flux maximum, while in knot B2, the polarization is high where the flux is high. The polarization vectors are seen to be crowded near the edge of the two knots and are majorally parallel to the local direction of the jet. The optical jet shows a similar bend just downstream of knot B2 at $\sim$14.$''$8 out from the nucleus, as seen in the radio jet. At this point the jet bends toward the north and forms knot C at $\sim$16.$''$5. In the small region just upstream of this bend, the vectors are orthogonal to the jet and the vectors form a small circular pattern. \\
\indent The overall polarized flux morphology (Fig.~\ref{fig:knot-ab}, right) of knot B is similar to that of total flux, although we see a few regions where the polarization is well below 3$\sigma$ or the region is depolarized. Especially in optical, close to the northern end of B2, we see a broad region of depolarization which lies close to the bend in the jet. Beyond this, up to the flux maxima of knot C1, we see the polarization is very low. Also the edges of B1 and B2 show very low or no polarization on both the polarized flux images. These features are slightly different in radio and optical morphology again pointing toward the stratified jet model of P99.
\indent The overall structure of the polarization vectors in the knot A and B complex is unique. The peculiar polarization vector structures seen in the knot A+B complex are consistent with morphology observed in the optical images. P99 discuss the helical pattern traced by MFPA vectors in their radio and optical images from 1994-95 observations suggesting the presence of a magnetic field in the form of a tightly would helix precessing outward from the central engine. We will discuss the presence of a helical magnetic field and/or structure in a follow-up paper on polarization structure modeling.

\subsubsection{\textit{Knot C}}\label{sec:knotC}

\begin{figure*}
\begin{center}
   \begin{tabular}{@{}cc@{}}
  \includegraphics[width=0.5\textwidth]{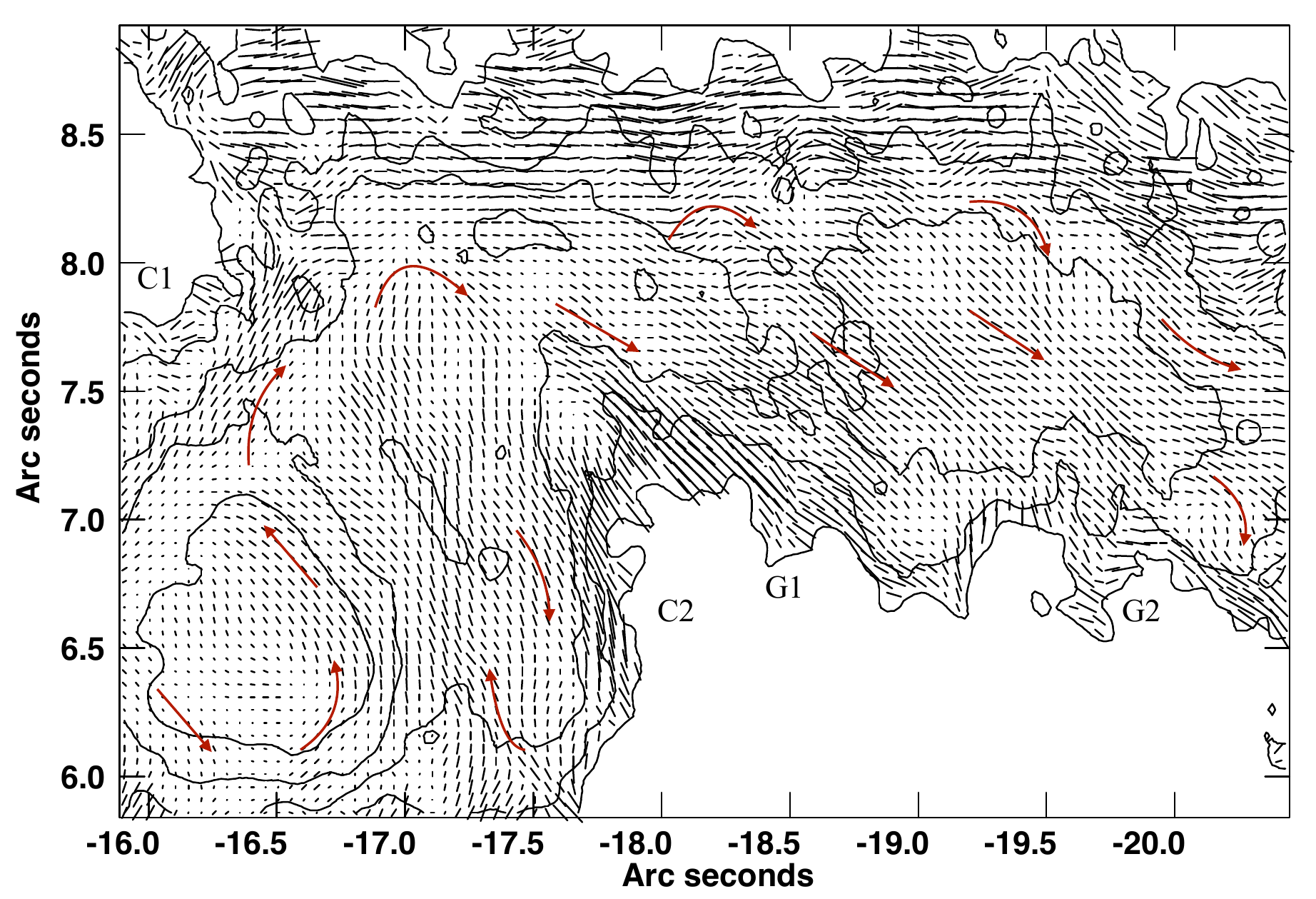}
  \includegraphics[width=0.5\textwidth]{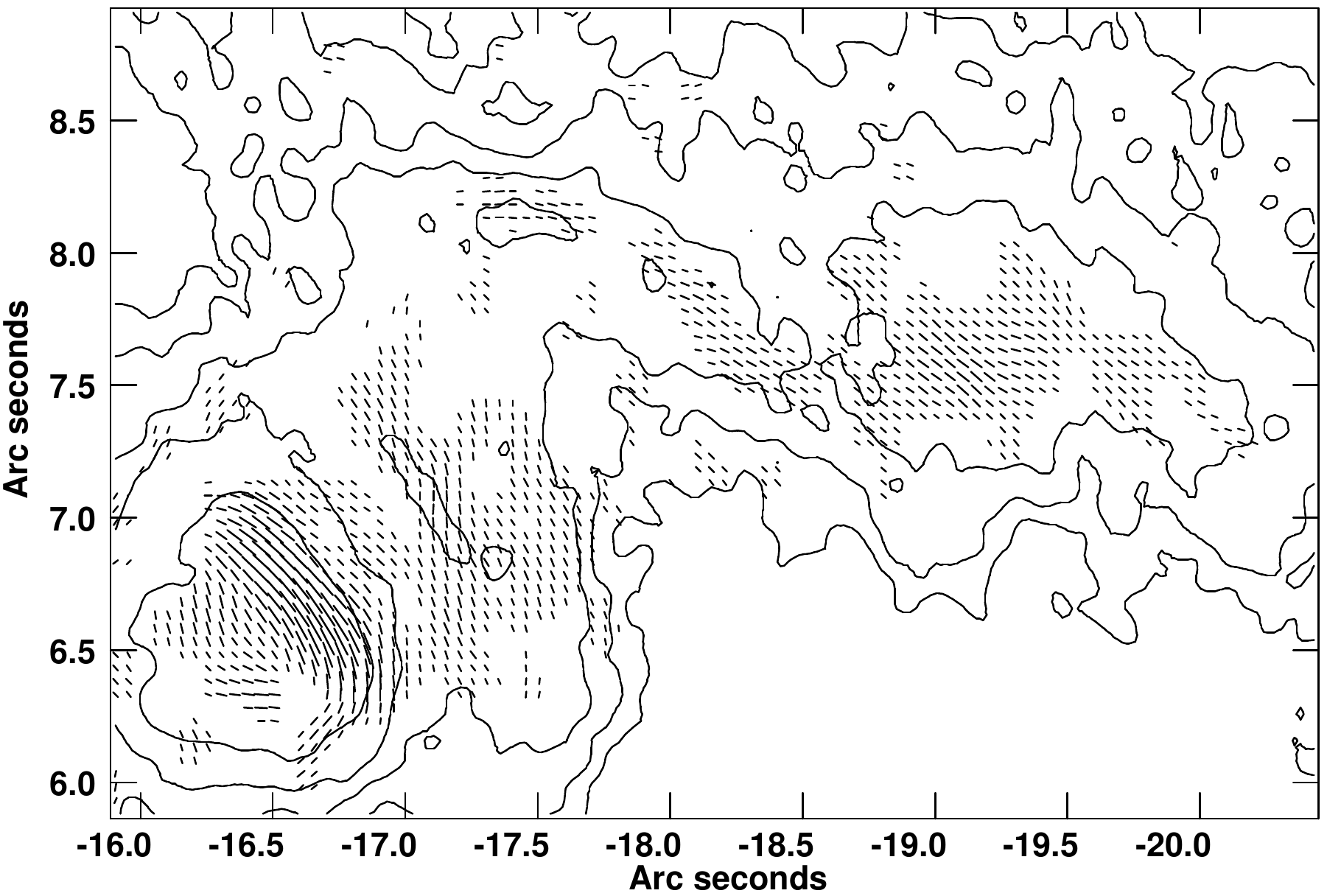} \\
  \includegraphics[width=0.5\textwidth]{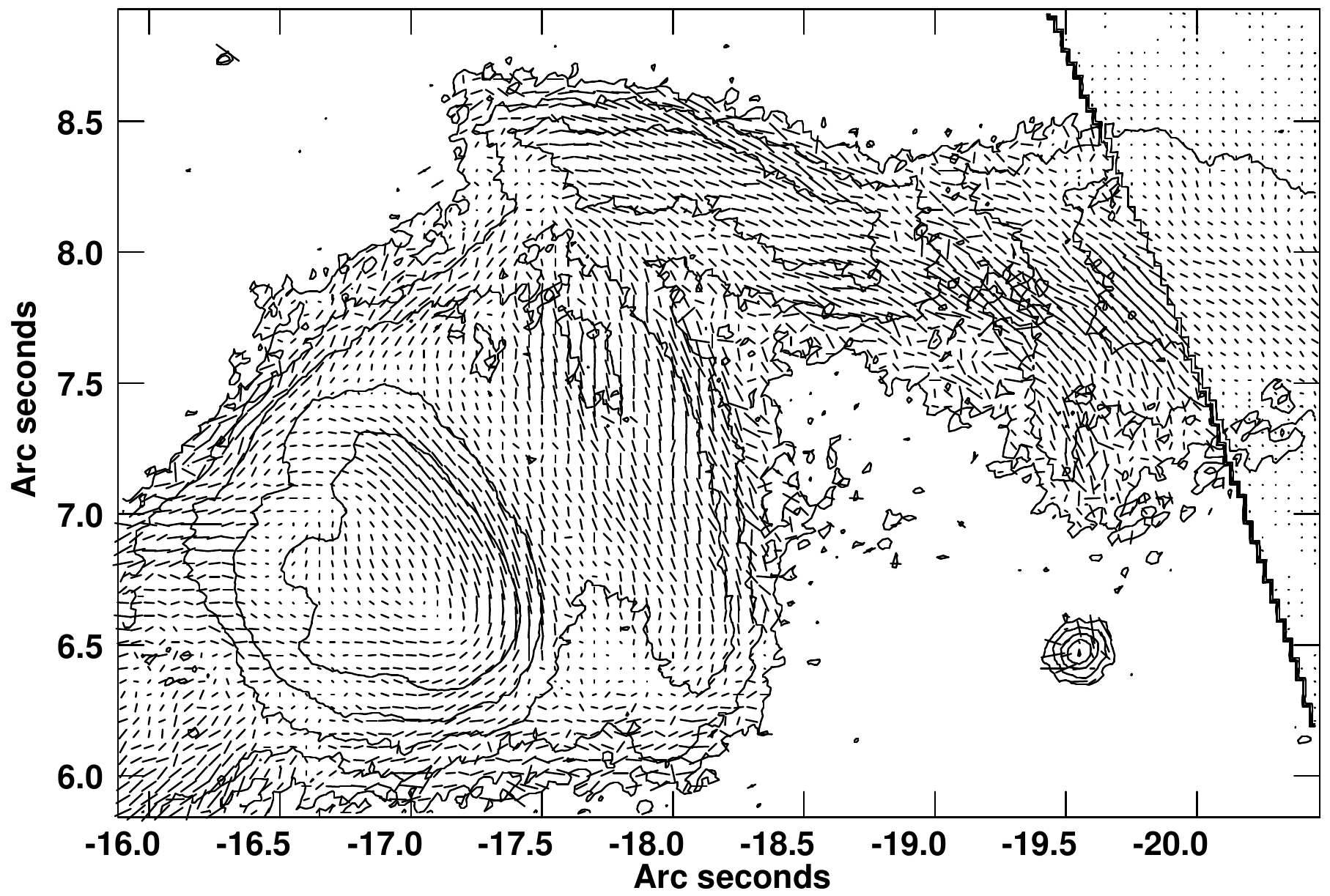}
  \includegraphics[width=0.5\textwidth]{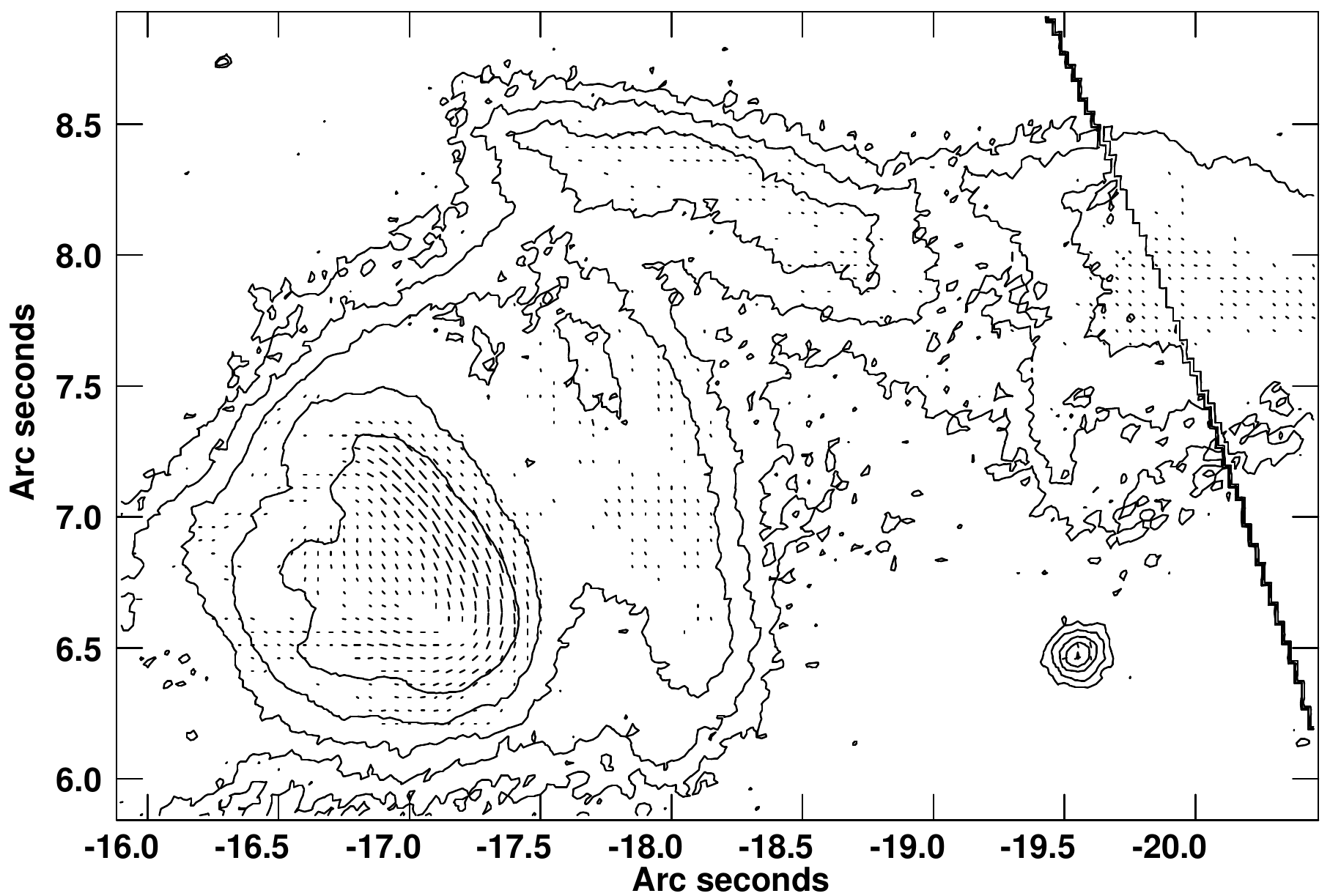}
\end{tabular}
 \caption[Radio and optical flux and polarization images of outer jet, knots C and G.]{Knots C and G. Comparison of the radio and optical  polarimetry. Top: 22 GHz, Bottom: F606W; Left panel: Fractional polarization, Right panel: Polarized flux. The images show combined epochs between 2002 and 2008 for both the wavebands. The contours represent the flux overlaid by the magnetic field polarization angle vectors. The contour levels same as in Fig.~\ref{fig:nuc-hst1} with flux level at 0.3 mJy beam$^{-1}$ in radio and 8e-3 $\mu$Jy in optical. The length of the vectors represent the amount of percentage polarization in the region. The feature covering the upper right hand corner of the lower panel is the outer limit of the subtracted galaxy model.}
  \label{fig:knot-cg}
  \end{center}
\end{figure*}

\indent Knot C is one of the outermost bright knots in the M87 jet. Fig~\ref{fig:knot-cg} (top,left) shows the flux and polarization structure of knot C. Similar to the knot A and B complex, this outermost region is significantly different than the inner and intermediate knots. The extended structure of knot C is distinguished between knots C1 centered at (16.5,6.5) arcsec and knot C2 centered at (17.5,8.2) arcsec from the nucleus, respectively. Knots C1 and C2 display broad radio flux maxima, with a few local flux maximum spread over the length and width of the knots. The jet displays a large bend near these knots. One bend is seen upstream of knot C1, in the interknot region between knots B2 and C1 at about 15.$''$0 from the nucleus where the jet turns northward through about 45-50$^{\circ}$ off the jet axis. The other bend is between knots C2 and the downstream knot G1, at about 18.$''$0 from the nucleus, where the jet turns westward through close to 90$^{\circ}$.\\
\indent The upstream end of knot C1 is less polarized than the downstream end ($\textless$20\%). The direction of polarization vectors at the flux maximum of C1 is predominantly parallel to the jet direction whereas just downstream of it, the vectors become perpendicular and the fractional polarization is increased to $\sim$30\%. The knot C2 is another high polarization region in the outer jet. Although the fractional polarization is between 20-30\%, the vectors lie in the direction of jet. The polarization is maximum at the edges of the knot C1 and C2 complex and mostly oblique to the direction of jet. The region of large bend between these two knots is another peculiar region in the outer jet. The polarization vectors here are in the direction of the jet, as a results they happen to lie almost perpendicular to each other in the upstream and downstream regions of the knot C2. \\
\indent Fig~\ref{fig:knot-cg} (bottom, left) shows the optical flux and polarization morphology of knot C. The optical flux morphology of the knot is very similar the radio. The optical knot C can also be divided in two parts, C1 and C2. Both of them have a broad but defined flux region. The optical knot is narrower than radio, especially knot C2. The large bend in the jet, between knots C2 and G1 (see the following section for description of knot G1) is much more prominent and clear in the optical. Also the inter-knot region between knots C2 and G1 is much narrower and the transition is more defined in optical than in radio.\\
\indent The optical polarization morphology of knot C is much simpler as compared to the radio (Fig~\ref{fig:knot-cg} (bottom left)). The inter-knot region between knot B2 and C is $\textless$20\% polarized. The MFPA vectors are randomly oriented in this part of the jet. The sub-components C1 and C2 show quite similar morphology as in radio. Both knots are moderately polarized and vectors at the upstream edge are parallel to the jet direction while the vectors at the downstream edge become perpendicular to the local flow direction. Between 17.5$''$ and 18.5$''$ out from the nucleus, at the upstream edge of the knot G, the jet displays a similar large eastward bend as seen in radio. The polarization vectors at the bend are randomly oriented and, also, the fractional polarization is found to decrease to about 20\%.\\
\indent The polarized flux morphology (Fig.~\ref{fig:knot-cg}, right panels) of knot C shows broad regions of low or no polarization. Although, radio flux and polarization maximum of C1 coincide, those in C2 do not show any correlation. Knot C1 has much higher polarization than C2. The optical flux and polarization of C1 coincide, however; the region downstream of C1 shows a very low polarization. The region closer to the jet axis has polarization well below 3$\sigma$ or is region of depolarization. The optical polarized flux and polarization is slightly different than the radio, a possible evidence of the different origins of the radio and optical electrons.

\subsubsection{\textit{Knot G}}
\indent Beyond knot C the jet bends eastward through a large angle forming the outermost knot in the jet, G. This knot has two broad sub-components, G1 and G2 located at (18.5, 8.0) arcsec and (19.7, 7.6) arcsec from the nucleus, respectively. Both G1 and G2 are more diffuse as compared to any other knot in the jet. \\
\indent These two knots are  $\sim40\%$ polarized in radio. The upstream end of G1 is the least polarized in the knot, however its polarization increases to more than 30$\%$ at approximately 0.$''$5 downstream of flux maximum of G1. The MFPA vector structure at the polarization maximum is unique. It displays a large counterclockwise rotation in MFPA vectors in the upper part while it is seen to rotate clockwise in the lower part. The two regions of MFPA merge with each other toward the downstream end of G1 and become parallel to the local jet flow direction in the center of the jet in G2. The MFPA vectors near the edges are still random in orientation and display a slightly reduced polarization off the center on north and south. The polarization is uniformly higher near the edges of G1 and G2. The MFPA vectors continue to stay parallel all the way up to 20.$''$0 out from the nucleus. The inter-knot region between C and G features highest polarization apart from the edges. The unique circular structure of the MFPA vector may suggest a presence of shock in the jet, compressing the local magnetic field and hence polarization vectors.\\
\indent Fig~\ref{fig:knot-cg} (bottom) shows the optical morphology of the knots G1 and G2.  The MFPA and flux morphology is much more uniform in optical than in radio. Similar to the radio, we can identify knot G1 flux region. The galaxy subtraction stopped at approximately at the location of knot G2 ($\sim$20.$''$0 out from the nucleus). This affects the region of G2 and we do not see much of it. The part of knot G1 shows a few dispersed flux maxima. The optical knot G is much narrow in width as compared to the radio, which is consistent with the general trend along the jet and consistent with the spine-sheath model of \citet{2007ApJ...668L..27K}. \\
\indent The optical polarization of knot G is of the same order as the radio. The fractional polarization of G1 is close to 40\% and is the highest in knot G. The MFPA stays mostly oblique to the local jet flow direction and we do not see as much complexity of polarization structure in optical as in radio. In the reduction of optical data, the galaxy subtraction was applied to correct for the galaxy flux affecting the flux of the jet. We cannot comment much about the polarization structure in G2, except that the upstream end of the knot has lower polarization.\\
\indent Knot G is mostly depolarized in polarized flux images (Fig.~\ref{fig:knot-cg}, right panel). We do not see a clear polarization maxima in G1 in either of the bands. Although radio shows a bit more polarization as compared to optical, none of it is significant enough to comment on the any correlation to the flux features. \\

\section{Variability Study} \label{sec:variability}
\indent \citet{2011ApJ...743..119P} published results of optical polarization and spectral variability study of the nucleus and HST-1. We use their results along with our higher resolution radio data and do a comparative study of flux and polarization variability in the following section.


\begin{figure*}
    \begin{center}
        \begin{tabular}{@{}cc@{}}
            \includegraphics[width=0.5\textwidth]{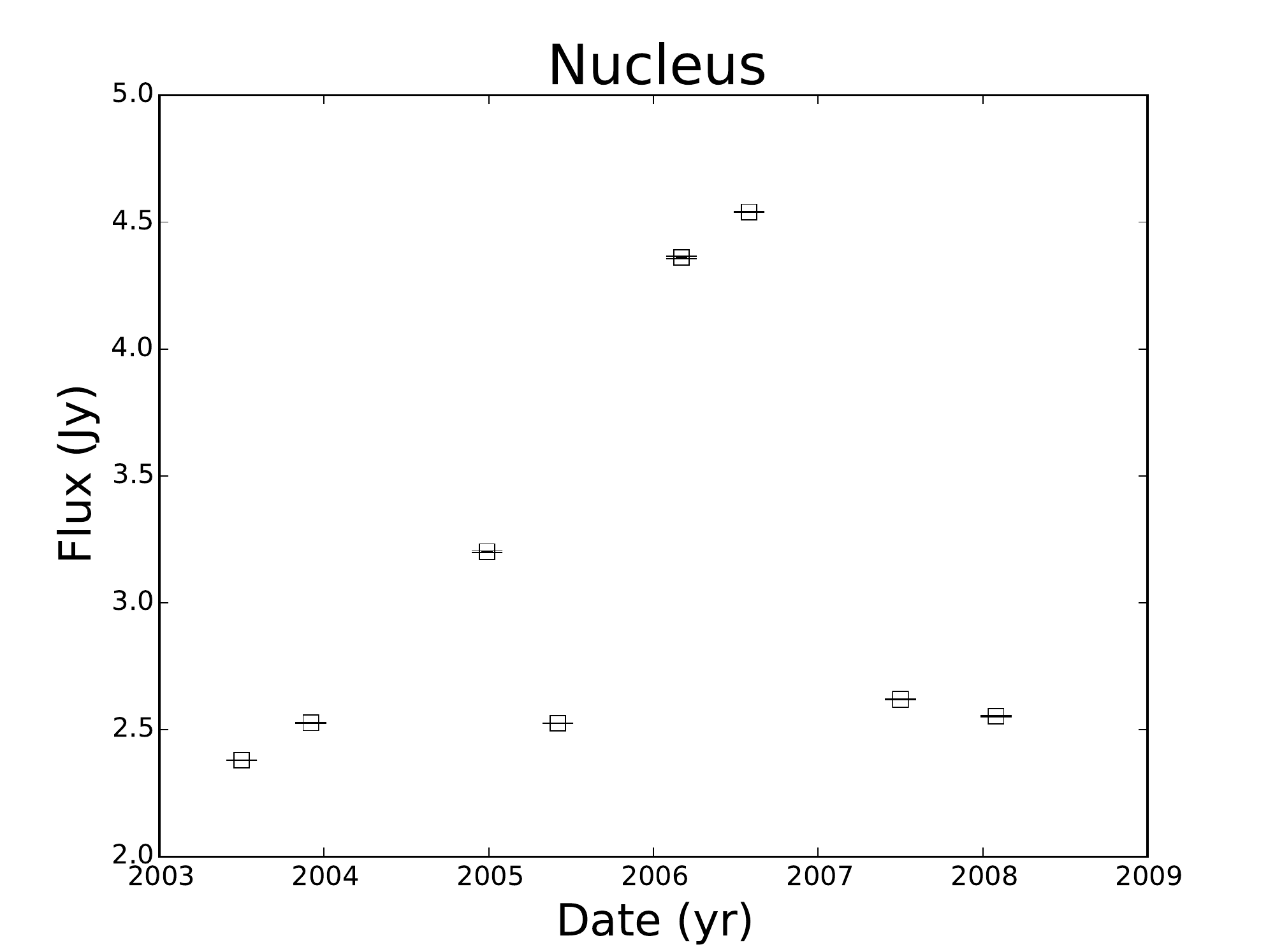}
            \includegraphics[width=0.5\textwidth]{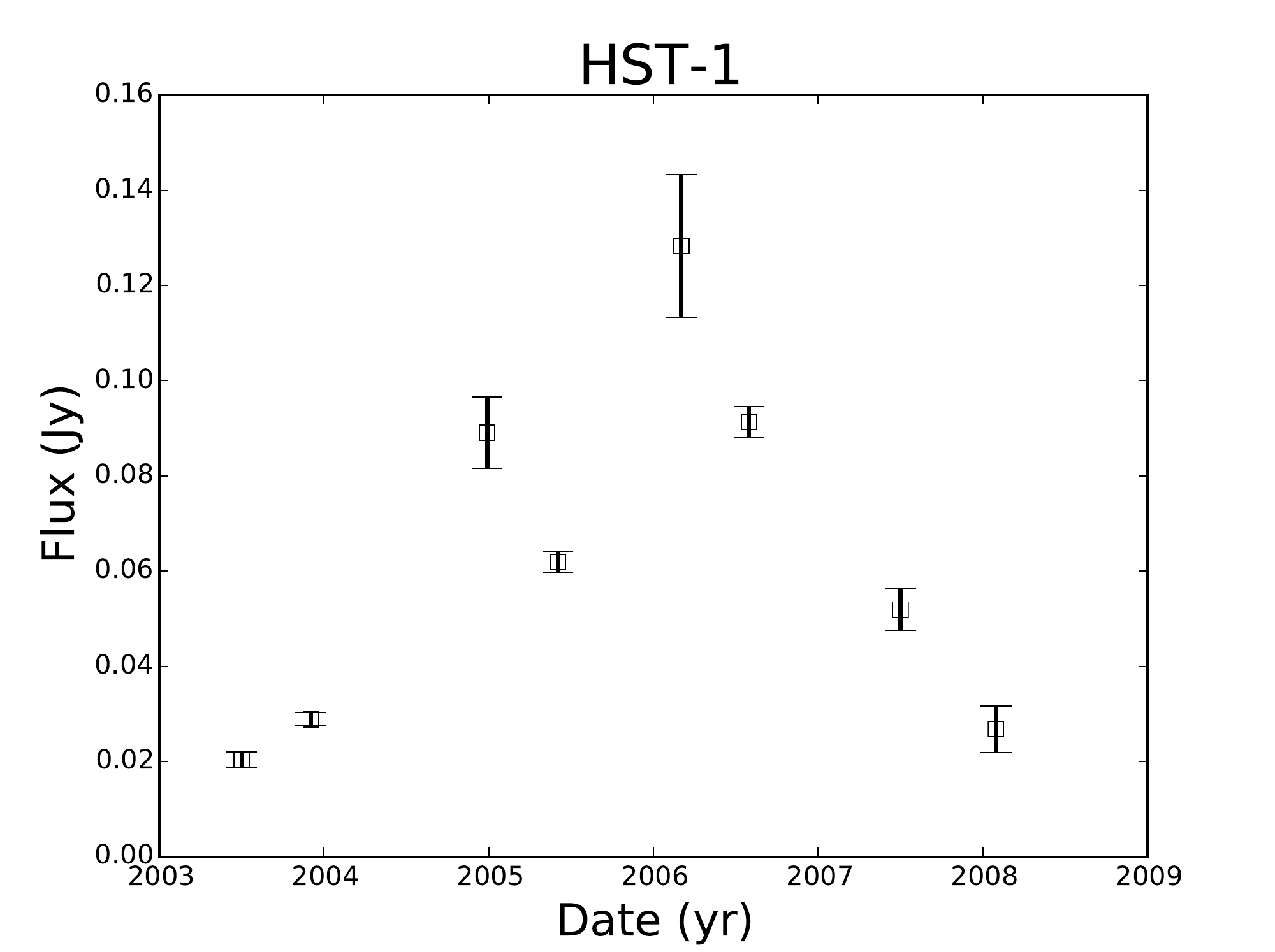}\\
            \includegraphics[width=0.5\textwidth]{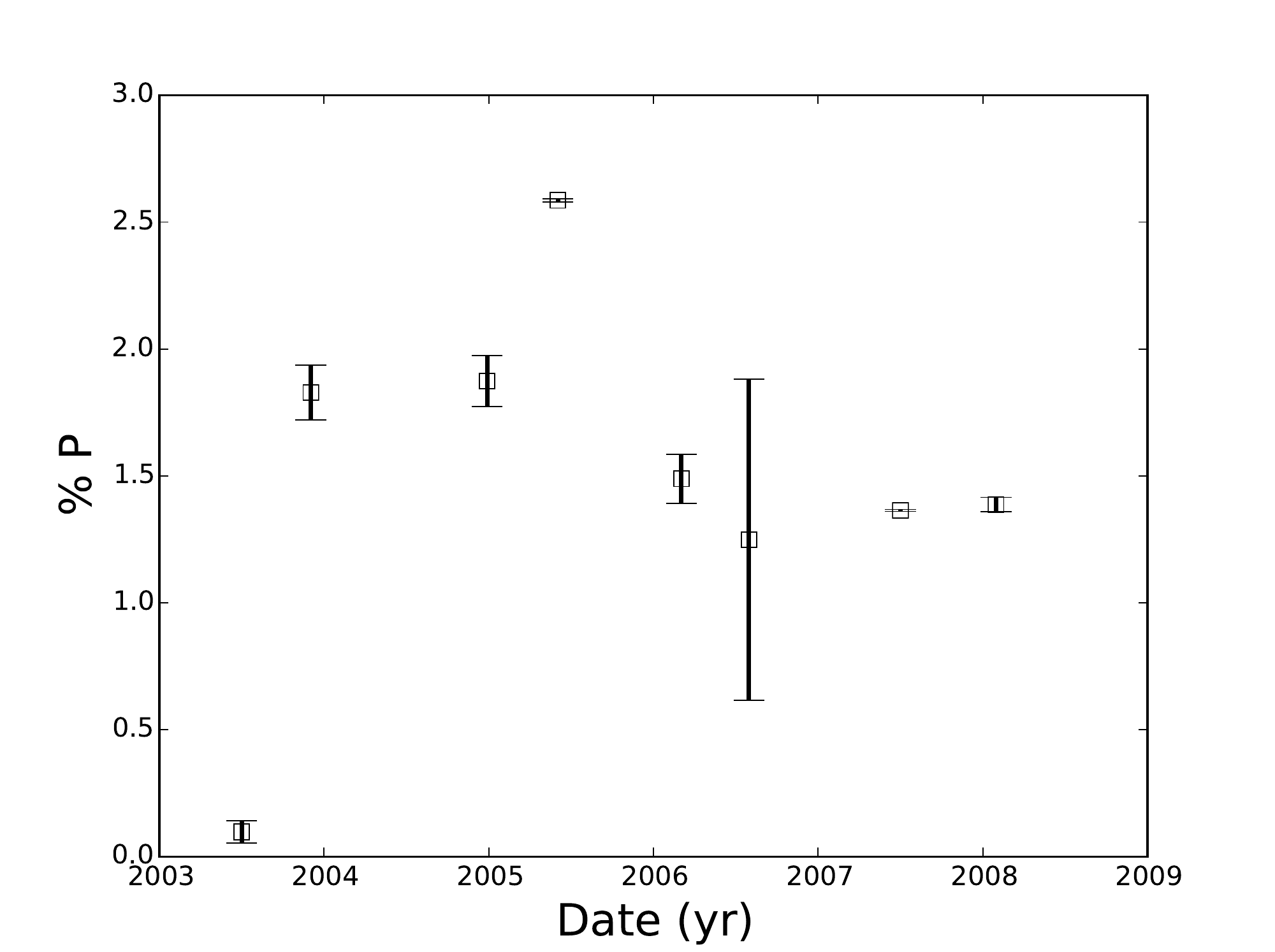}
            \includegraphics[width=0.5\textwidth]{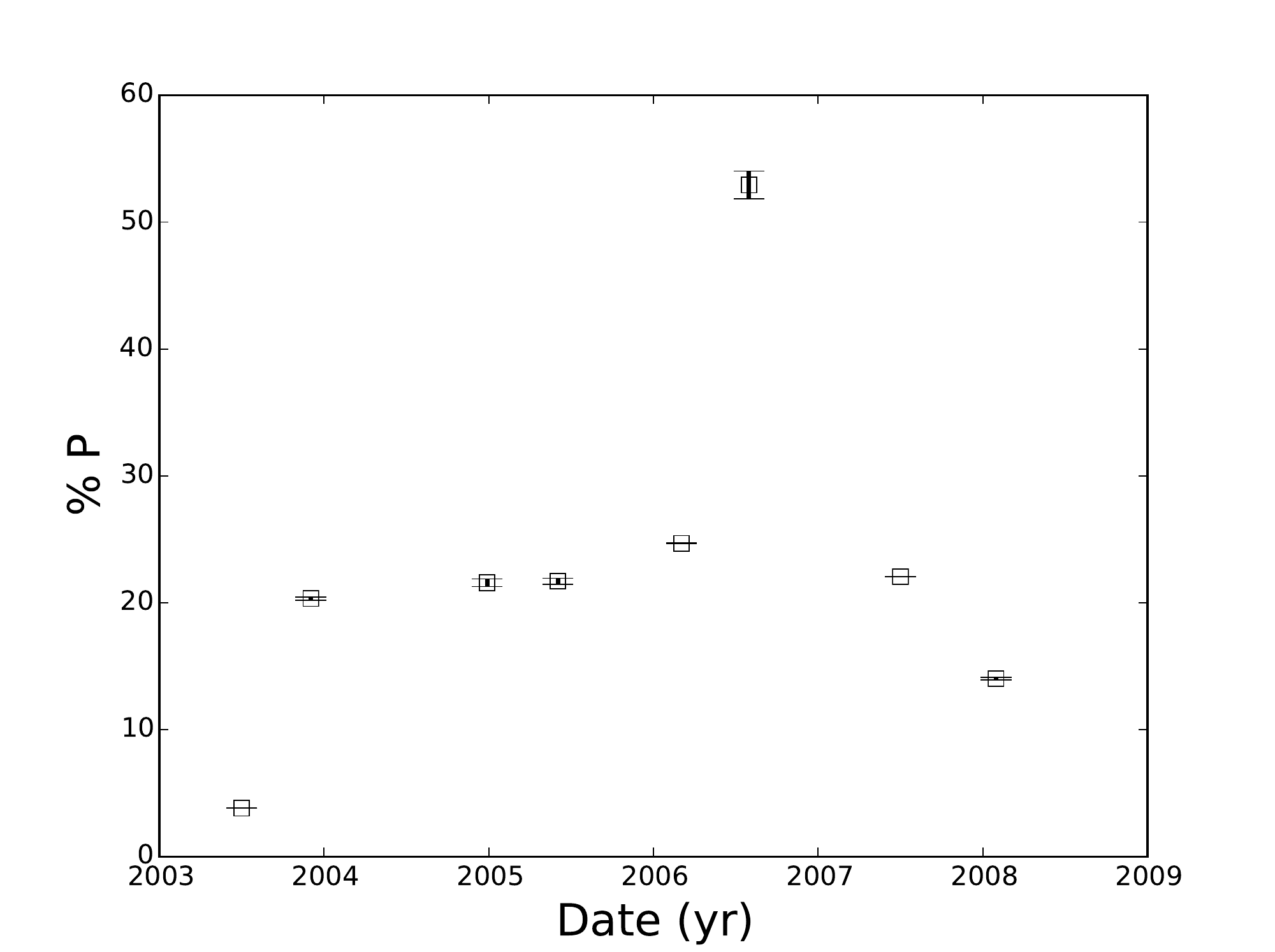}\\
            \includegraphics[width=0.5\textwidth]{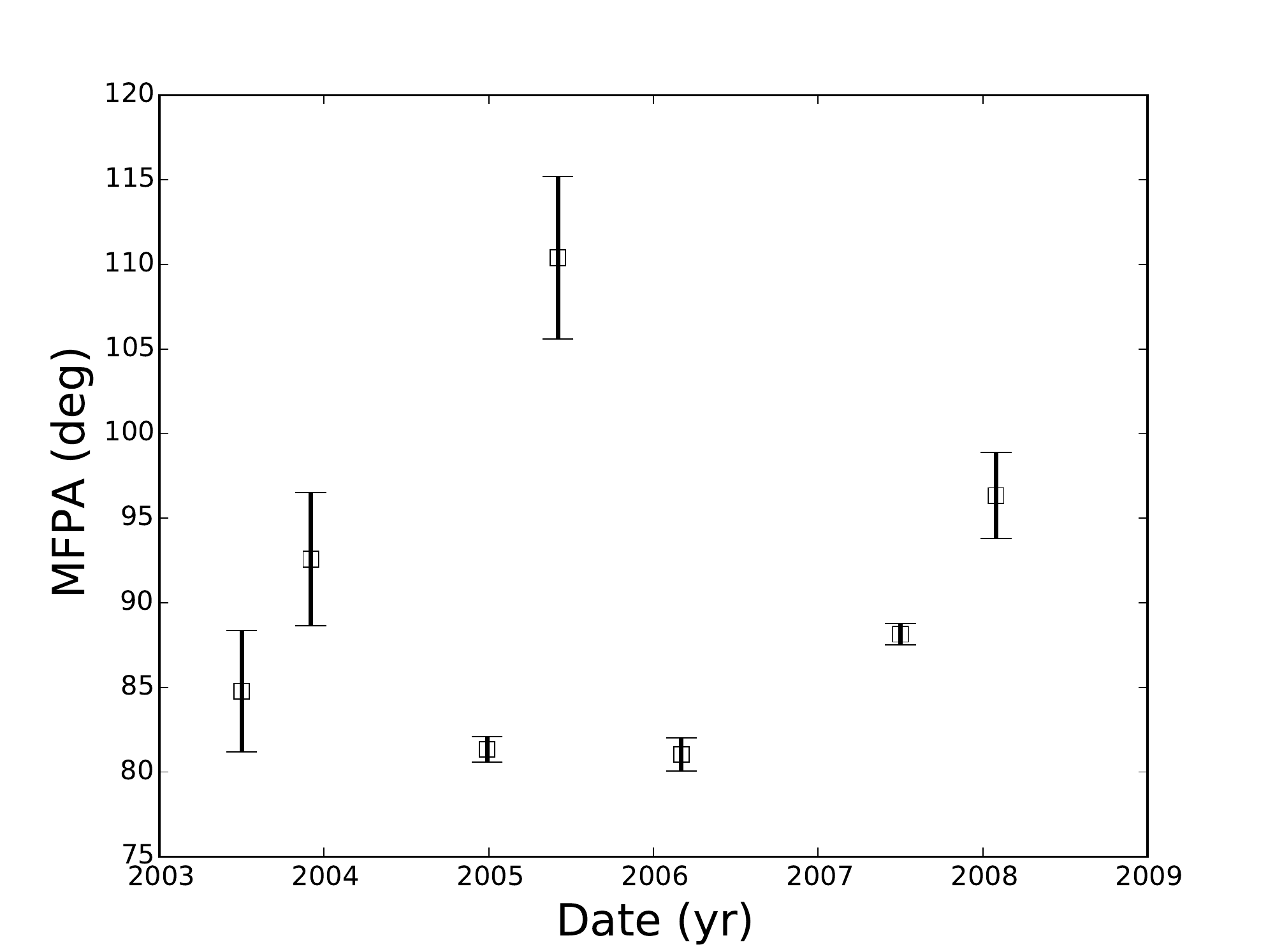}
            \includegraphics[width=0.5\textwidth]{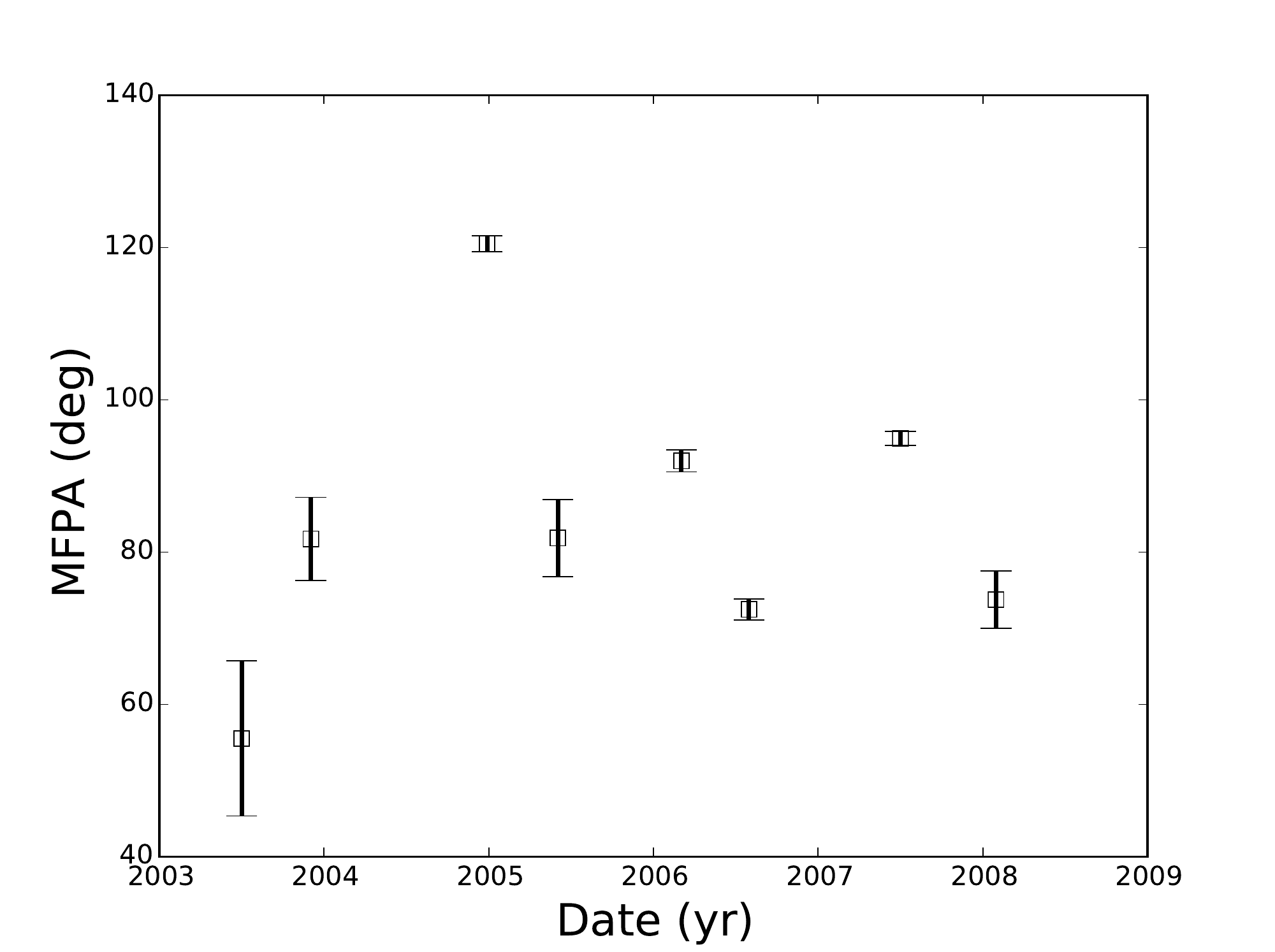}
        \end{tabular}
        \caption{Variability of total flux (i.e. Stokes I), fractional polarization and MFPA of the nucleus (left panel) and HST-1 (right panel). See the \S4.1 for the description.} \label{fig:7}
    \end{center}
\end{figure*}

\subsection{Flux and Polarization variability}

\begin{figure*}
    \centering
        \newpage
            \begin{tabular}{@{}cc@{}}
               \includegraphics[width=0.5\textwidth]{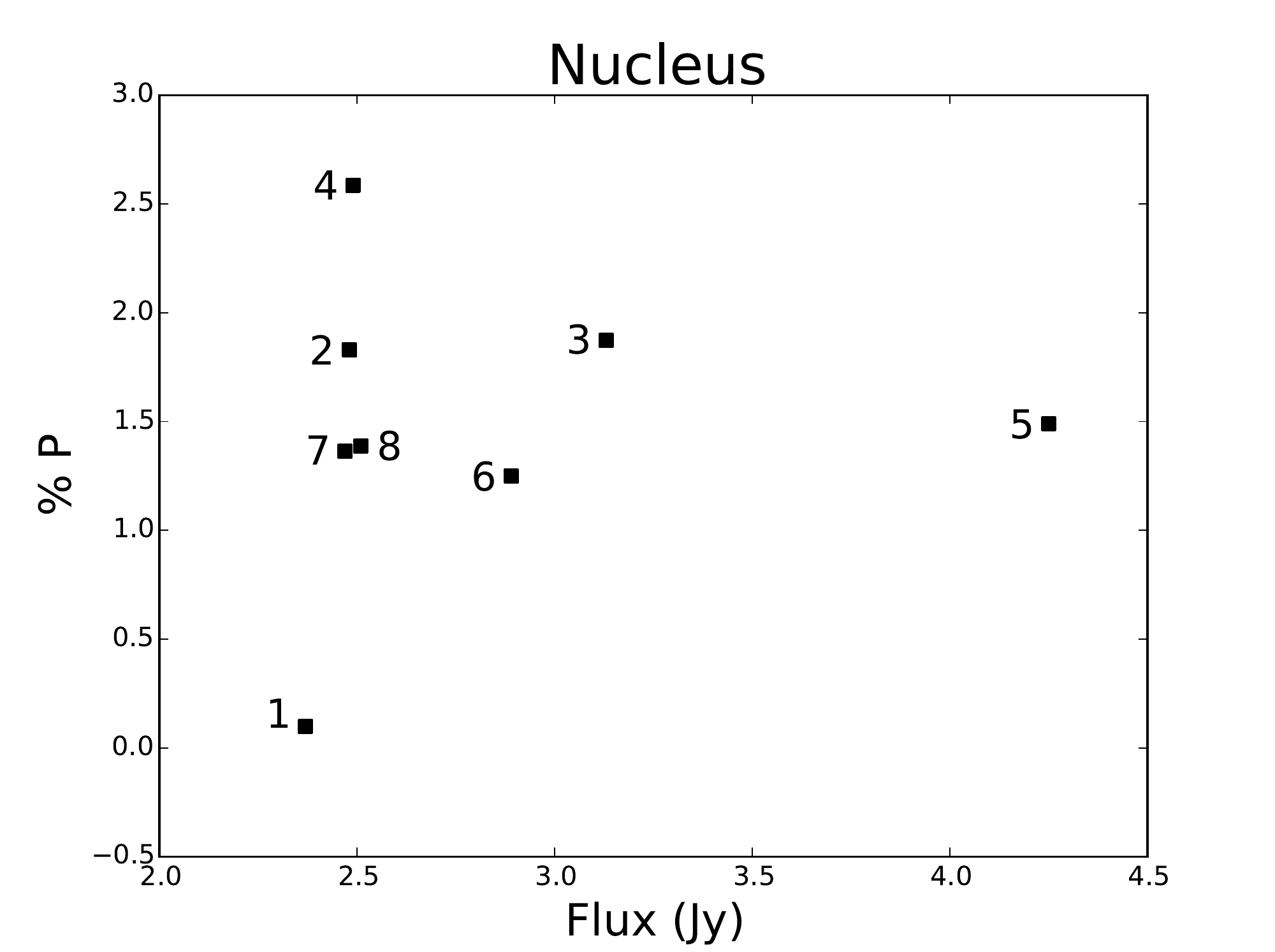} 
               \includegraphics[width=0.5\textwidth]{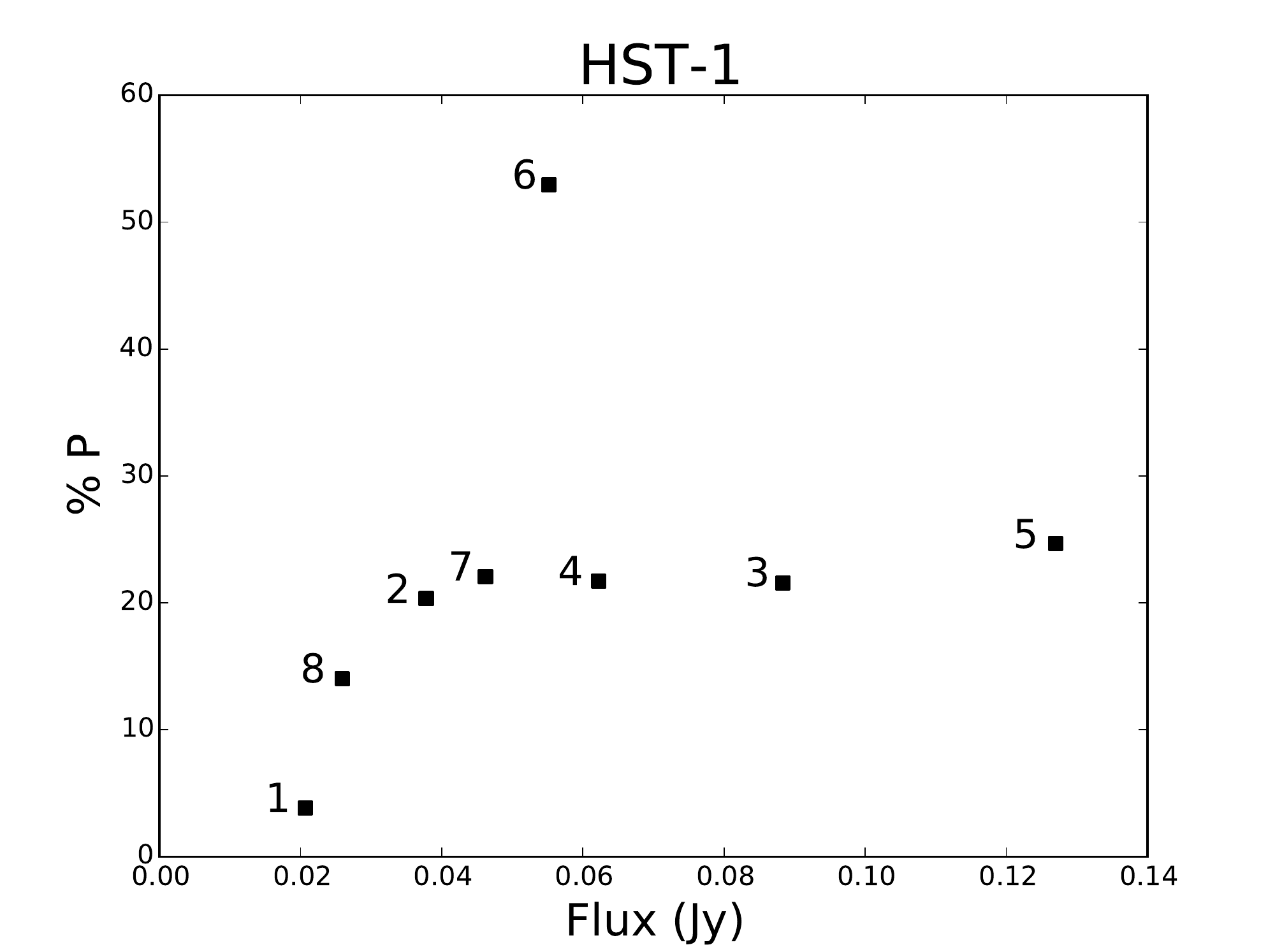} 
            \end{tabular}
        \caption{Polarization versus Stokes I flux (Jy) for the nucleus (left) and for HST-1 (right).}\label{fig:8}
\end{figure*}

\indent The total radio flux variations of the nucleus and HST-1 along with the large flare in HST-1 around 2005 are similar to the optical variability of \citet{2011ApJ...743..119P} (their figures 1 and 2) and X-ray variability of \citet{2009ApJ...699..305H} (their figure 9). The optical-UV data also show two small flares in the nucleus just before and after the flare in HST-1 \citep{2011ApJ...743..119P}. The majority of optical observations were taken during 2004-2006 when knot HST-1 was already bright and hence full dynamic range of the variability of HST-1 was not observed in optical. On the other hand - radio data were taken during 2003-2008 and spans over the time domain of HST-1's flare.\\
\indent Both HST-1 and the nucleus show much similar behavior in polarization and MFPA variability in radio as can be seen in middle and lower panels of Fig.~\ref{fig:7}, respectively. Polarization behavior of the nucleus is much consistent with values between 1-4\% as compared to the larger variations observed in optical which were ~1-13\%. During this time, the optical MFPA changed as much as $90^{\circ}$ whereas in radio it stays more stable between 80-90$^{\circ}$. HST-1 is much more highly polarized than the nucleus in both optical and radio. In the optical, its polarization ranged between 20\%-45\% with little variability, compared to close 20\% in radio with small variability, and was strongly correlated with flux with nearly constant EVPA $\approx-62^{\circ}$. \\
\indent Similar to the nucleus, MFPA of HST-1 stayed close to 80$^{\circ}$ on an average. This value differs from the optical MFPA ($\approx$ 30$^{\circ}$)\citep{2011ApJ...743..119P} suggesting these particles may represent very different populations in space and may indicate optical emission during the flare being much more dominated by the flaring component. Polarization behavior of HST-1 at the epoch June 2007 is significantly higher than at the other epochs. This epoch is very close to the second flare in the nucleus as well as a smaller second flare in HST-1. The immediate next epoch observed after June 2007, was 7 months later in Jan 2008, which has fractional polarization values consistent with the rest of the epochs.\\
\indent In Fig.~\ref{fig:8} we have plotted fractional polarization versus flux at 22 GHz, on the left for the nucleus and on the right for HST-1. Neither the nucleus nor HST-1 shows any evidence of correlation between polarization and flux. During the earlier epochs (points 1 through 5), the polarization of HST-1 increases linearly with the increase in its flux, however; in last two epochs (points 6 and 7), the polarization do not seem to have any correlation with the flux value. On the other hand, the nucleus shows a very different behaviour during same period. The nuclear polarization does not have any correlation with its flux, whatsoever and changed randomly. \citet{2011ApJ...743..119P} explain the correlation between flux and polarization in terms of a looping. They indicate this looping as 'hard lags' (clockwise looping) and 'soft lags' (counter-clockwise looping) in their figures 3 (for the nucleus) and 4 (for HST-1). They explain the correlation with the help of connection between acceleration and cooling timescales controlling the spectral evolutions of radiating particles.

\begin{figure*}
    \begin{center}
        \begin{tabular}{@{}cc@{}}
            \includegraphics[width=0.5\textwidth]{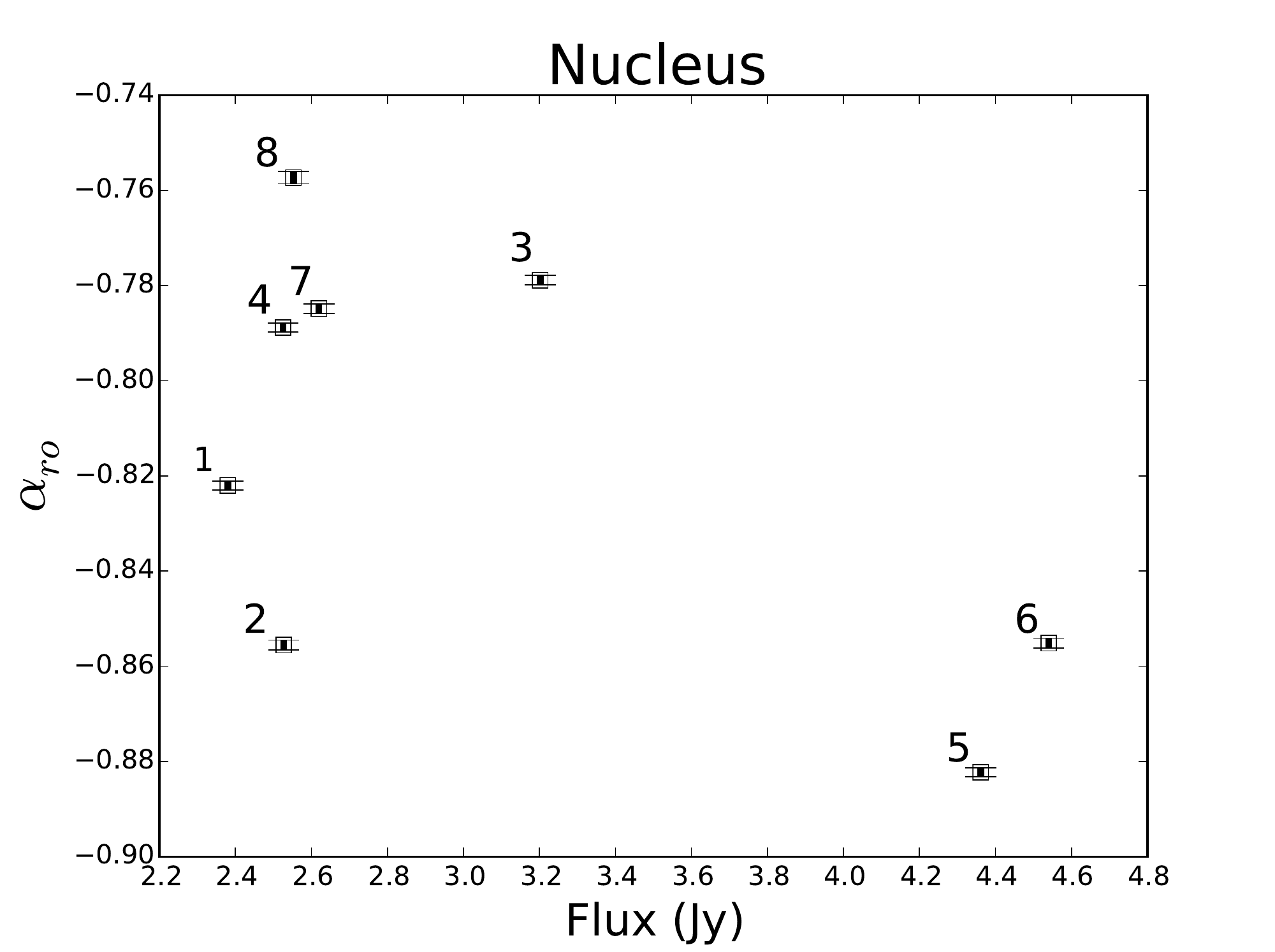}
            \includegraphics[width=0.5\textwidth]{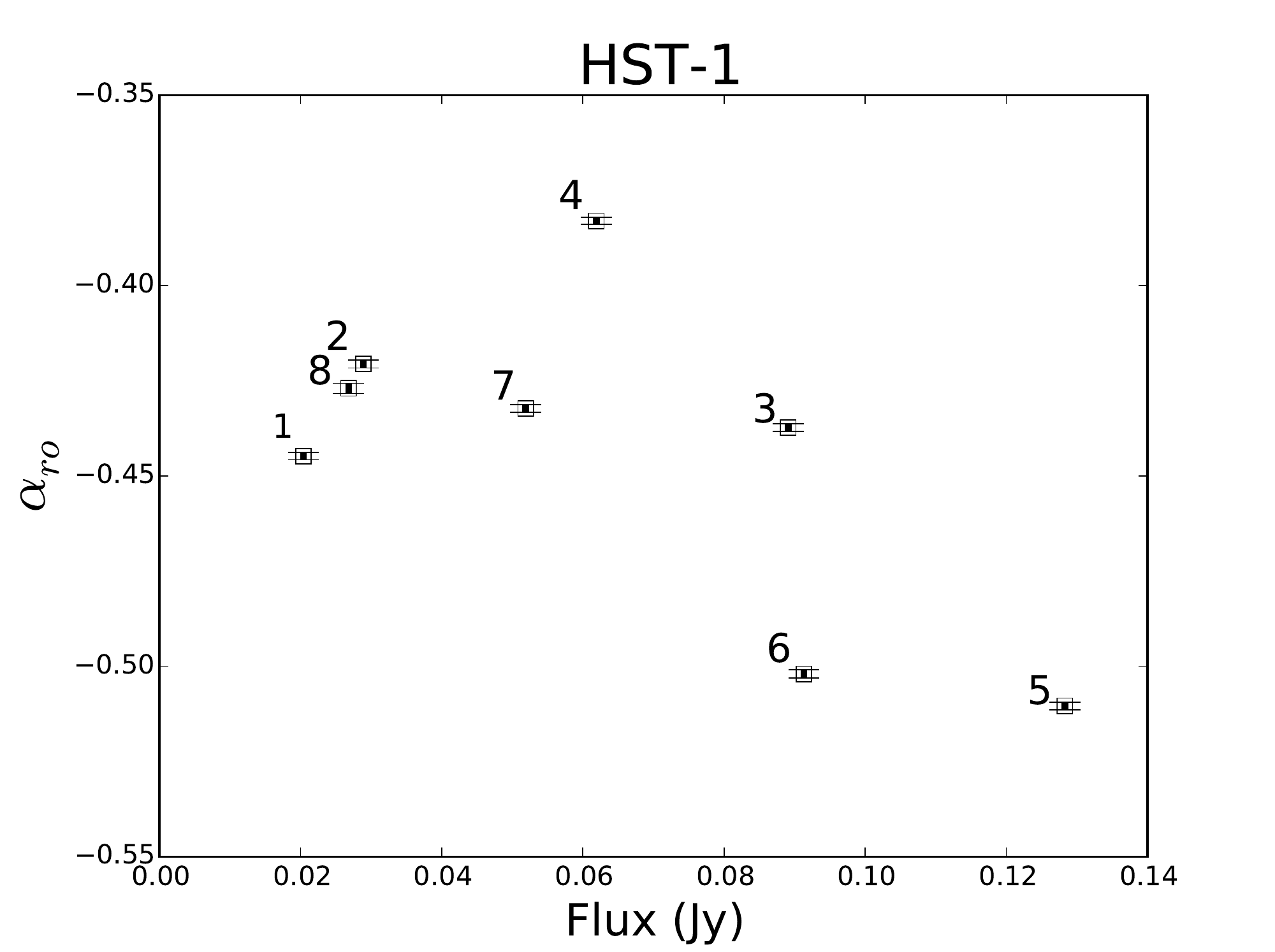}\\
            \includegraphics[width=0.5\textwidth]{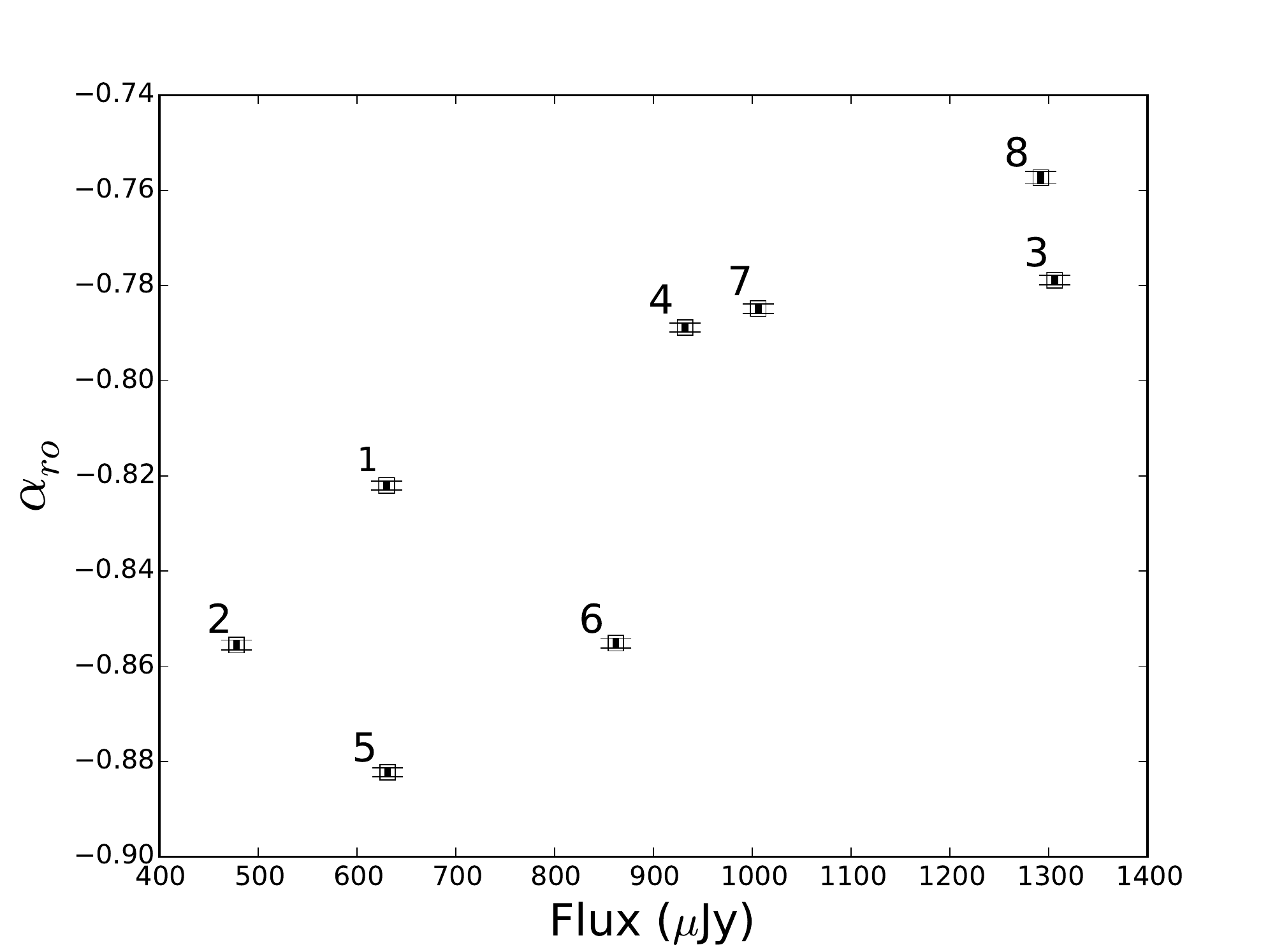}
            \includegraphics[width=0.5\textwidth]{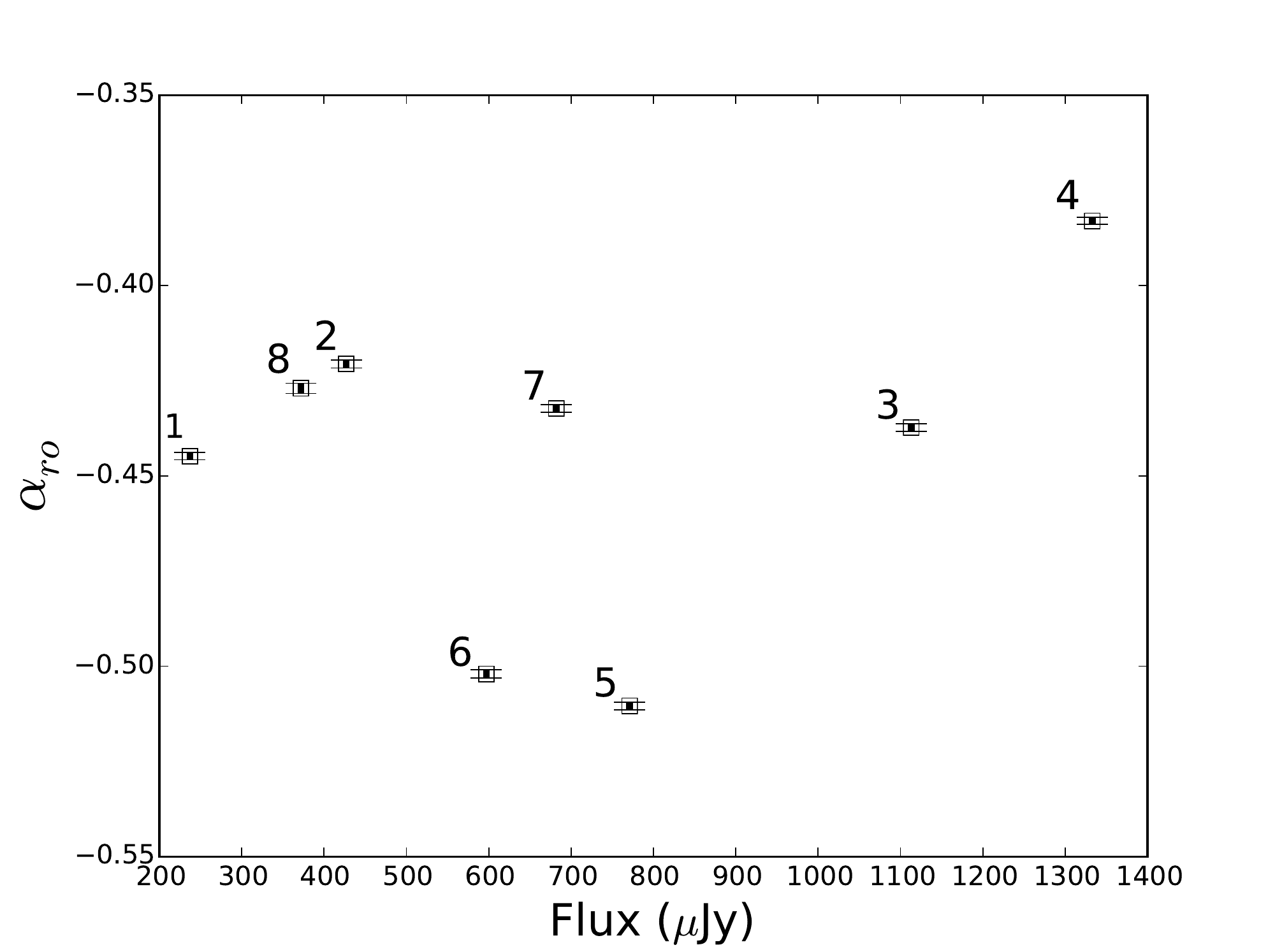}
        \end{tabular}
        \caption{Spectral index variability versus Stokes I flux for the nucleus and HST-1. Top - $\alpha_{ro}$ versus radio flux at 22 GHz; Bottom - $\alpha_{ro}$ versus optical flux at F606W.}\label{fig:9}
    \end{center}
\end{figure*}
\begin{figure}
    \centering
        \includegraphics[width=0.5\textwidth]{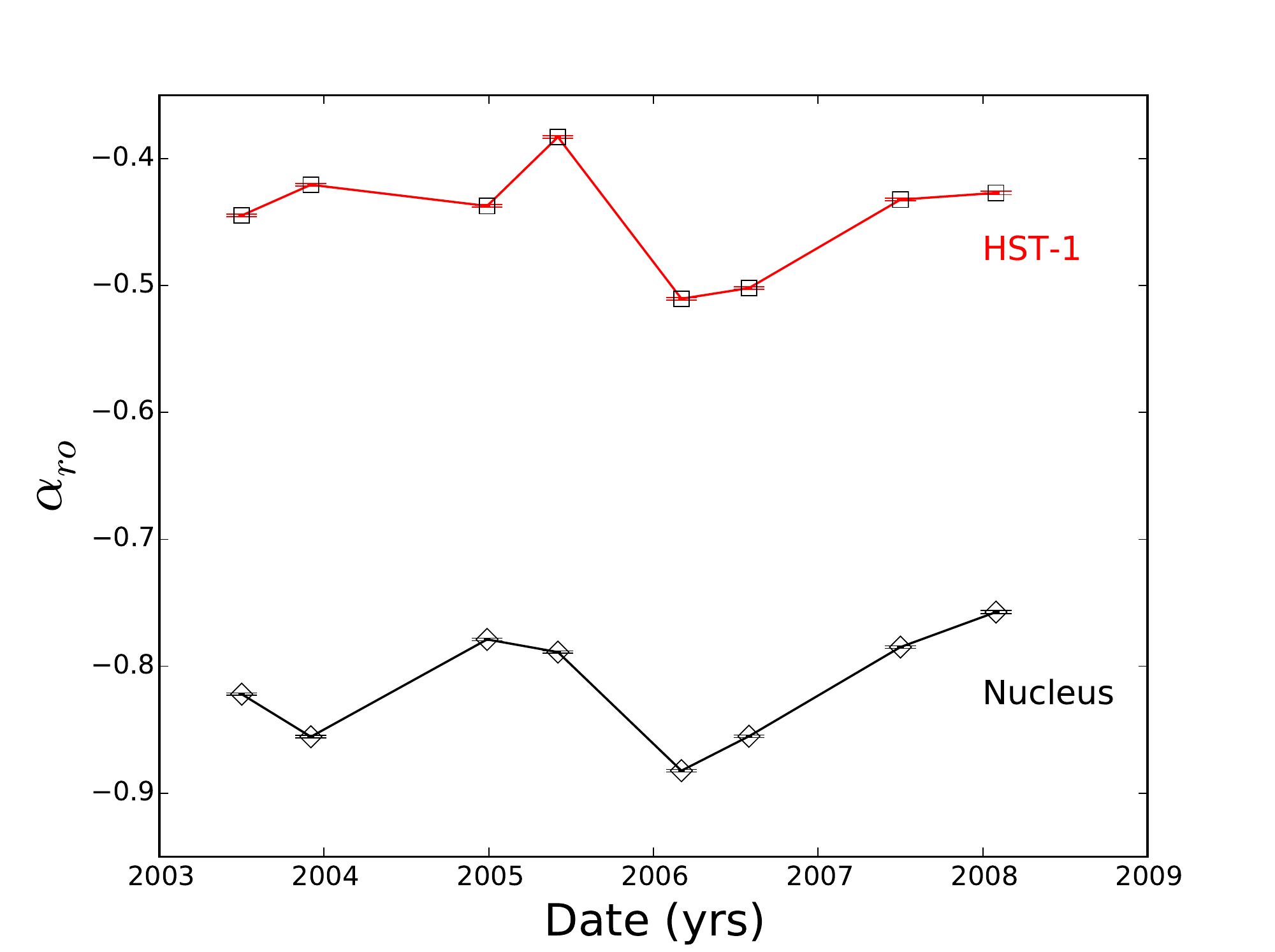}
\caption{Spectral index variability between radio and optical with time. The diamonds represent $\alpha_{ro}$ for the nucleus, while the squares represent the same for HST-1.} \label{fig:10}
\end{figure}

\subsection{Comparison with Optical SED}
\indent We plot radio-optical spectral indices using flux values at 22 GHz and F606W ($\approx$ 4.95$\ast$10$^{5}$ GHz). We compare them to the optical SED of \citet{2011ApJ...743..119P}, figures 3 and 4. Fig.~\ref{fig:9} (top) shows the evolution of spectral index $\alpha_{ro}$ versus total radio flux for the nucleus (on left) and HST-1 (on right). We assumed the relation between the spectral index $\alpha$ and flux as $S_{\nu}$ $\propto$ $\nu^{\alpha}$. As can be seen, there is no direct evidence of correlation between spectral index and flux. For the nucleus, the spectral index evolves independent of total flux. At the beginning of the observations, up to epoch 5, it does not show any correlation with the flux whatsoever, and oscillates between -0.85 and -0.75; however, epoch 5 onward, it shows a monotonous increase from low $\sim$ -0.9 to a high of $\sim$ -0.75. \\
\indent In case of knot HST-1, $\alpha_{ro}$ shows a similar behavior. The spectral index stayed close to -0.45 before a sharp increase in it around 2005 flare when the spectral index increased to $\sim$ -0.35. At the epoch 5, we see a sudden drop in the spectral index however beyond this, we see a monotonous increase in its value until the last epoch 8. During all this time, the spectral index of HST-1 varied between $\sim$ -0.5 to -0.35.
\indent Bottom two plots of Fig.~\ref{fig:9} show the evolution of $\alpha_{ro}$ versus the optical flux. As can be seen, there is no clear correlation of spectral index with the optical flux either. \citet{2011ApJ...743..119P} observed a ``looping" behavior during the maximum of HST-1's flaring in 2005 in their plot of $\alpha_{UV-O}$ versus the optical flux plot (their Fig.4); however we do not see any such trend in $\alpha_{ro}$ plot of HST-1 or the nucleus. This may point toward the possibility that the radio and optical electrons may originate in different regions within the jet.\\
\indent Fig.~\ref{fig:10} summarizes the spectral index variability of the nucleus and HST-1 in a single plot over the period of observations. The spectral index of HST-1 shows larger variations as compared to the nucleus, especially close to its flaring in 2005. The $\alpha_{ro}$ of HST-1 varied by about 30\% between -0.38 in mid 2005 to -0.51 in early 2006, whereas that of nucleus shows variations close to 15\% in the same time.

\section{Discussion} \label{sec:discussion}
\indent The jet of M87 is highly complex, presenting a variety of features both from a morphological and
spectral point of view as well as from the polarimetric point of view.  Our results show a number of regions
with increased polarization where the magnetic field vectors rotate by $\sim$ 90$^\circ$, in particular in knots
HST-1, D-East, A and the flux maximum region of knot C.  We also see many features with fascinating,
apparently helical undulations in the magnetic field vectors, particularly in the downstream regions of nearly
every bright knot.  There are also differences between the optical and radio polarization map.  All of these have
been discussed at length in \S\ref{sec:compare_pol}.  Some of these features were seen
before in P99; however, many of these features are seen here for the first
time, thanks to the increased angular resolution of these data. At the same time, it is worth noting that these images represent an epoch approximately 10 years later than the P99 study, and so it is possible that some of the differences between those images and the
ones presented here may be the result of temporal evolution.\\
\indent All of these features give us an increasingly full picture of the complex physics occurring within the M87 jet.
In particular, since polarization images contain information about the local ordering and direction of the magnetic
field, they represent a key tool that can help us relate jet structure to jet dynamics, particle acceleration,
high-energy emission processes. In this section, we attempt to bridge the gap between the polarization morphology and physics.  We divide our
discussion into two parts: in \S\ref{sec:shocks} we concentrate on the shock-like features, while in
\S\ref{sec:helical} we concentrate on helical undulations within the jet. This also gives us the opportunity to
discuss in \S\ref{sec:concl} several models for jet structure, including the spine-sheath models, where a
faster, possibly more energetic particle flux is seen in the jet interior than at its edges.  This model family includes
the energetic stratification proposed by P99, as well as later works, such as \citet{2007ApJ...668L..27K}.

\subsection{Shocks and Shock-like Features} \label{sec:shocks}
\indent Several regions of the M87 jet have been labelled as shocks by a wide variety of authors, including, for example,
\citet{1989ApJ...340..698O, 1996ApJ...467..597B, 1999AJ....117.2185P, 2006MNRAS.367..851C, 2012A&A...543A.115N}.
While most previous works have considered perpendicular shock features, in which the magnetic field is strongly
compressed perpendicular to the jet flow, such features can also be oblique or conical (e.g., \citet{1990ApJ...350..536C,
1996ApJ...467..597B}).  In addition, if the overall jet magnetic field structure has a helical element, the effect on the polarization morphology by any of these features can be considerably more complicated.  The high resolution of these images allows us to have
a nuanced discussion of these issues.\\
\indent The two strongest shock features are knots HST-1 and A.  Knot HST-1 is believed to be the location of a recollimation
shock, as detailed in a number of previous papers, including \citet{2006MNRAS.367..851C, 2009ApJ...699.1274B,
2010ApJ...721.1783N, 2012ApJ...745L..28A}.  Recollimation shocks are believed to occur in a wide variety of jet systems as a result of interactions between the jet and the
surrounding interstellar medium (ISM) \citep{2015ApJ...809...38M} near the Bondi radius, where the local ISM becomes dominated by the influence of the
nuclear supermassive black hole. This is observed not only in AGNs, but also X-ray binaries and protostellar objects. As a result, M87 is not the only AGN where recent work has noted the presence of
a recollimation shock.  Other examples include 3C 120 \citep{2012ApJ...752...92A}, BL Lac \citep{2014ApJ...787..151C},
CTA 102 \citep{2015AN....336..447F}, and 1803+784 \citep{2013ApJ...772...14C}.  However, of these objects, M87
lies by far the closest to us, thus giving us a unique opportunity to study these structures, believed to be universal, in detail.\\
\indent Models for generating relativistic jets from magnetized accretion disks almost unanimously consider the magnetic
field geometry of the nuclear jet to have a dynamic, helical morphology (e.g., \citet{2015MNRAS.453.3213S,
2015ASSL..414...45T, 2014MNRAS.440.2185D, 2012JPhCS.372a2040T}.  The compression of the flow in a recollimation
shock should change the morphology of the magnetic field structure.  In M87, we see a combination of features in the nuclear and HST-1 regions (Figure \ref{fig:nuc-hst1}). In particular, we see helical undulations in the polarization PA in the
region between the nucleus and HST-1 (features Nuc-$\alpha, \beta, \gamma$ in the radio image).  Within the flux
maximum of HST-1, we see a significant increase in polarization, particularly in the optical, as well as a nearly
perpendicular magnetic field direction.  The radio 22 GHz image, while showing the increased polarization at the flux
maximum, does not show a perpendicular magnetic field direction.  However, this may be due to an increased rotation
measure at HST-1, as discussed by \cite{2011MNRAS.416L.109C}, who used the same VLA data as we did, but
concentrated only on the nuclear region of M87.  Their images (see their Fig 1) show a nearly perpendicular
magnetic field orientation at the highest radio frequency (43 GHz), which would have the lowest Faraday rotation.
Downstream of HST-1's flux maximum, the radio image once again shows helical undulations in the magnetic field
vectors (features HST-1 $\alpha, \beta, \gamma,\delta, \epsilon$).  These features are consistent with the compression
of a helical magnetic field (seen both upstream \& downstream of HST-1) in the recollimation shock, where the
perpendicular component would become dominant.  To explain the higher degrees of perpendicular polarization, \citet{2006MNRAS.367..851C} assumed that, there has to a combination of chaotic and poloidal magnetic field present in the jet at this location, a notion that is consistent with our data.\\
\indent Knot A, shown in Fig. \ref{fig:knot-ab} and the surrounding region is also highly interesting.
\cite{1996ApJ...467..597B} analyzed it as the site of a strong, highly oblique shock feature, which Nakamura and collaborators \citep{2014ApJ...785..152N, 2010ApJ...721.1783N, 2004ApJ...617..123N} have described it and knot C as a pair of fast-mode shocks.  Our data throw interesting light on this.
In particular, knot A, which previous optical imaging data \cite{1996ApJ...473..254S} has revealed to have a highly complex internal structure, is now seen to have a complex polarization structure as well. The maximum optical polarization
occurs well upstream (by 0.''4) of the flux maximum.  This is also where the magnetic field direction is first seen to rotate.  This
rotation is seen in both the optical and radio, but the radio does not show increased polarization in this region.  Downstream of
this, the flux maximum region is seen to have increased polarization as well (and in fact this is where the maximum radio
polarization is located).  Surveying the polarization maps in this region, one sees four features, each with slightly different
magnetic field orientations, giving the impression of either magnetic ``sheets" or filaments that are wound with a slightly
different PA.  We also see reduced polarization downstream as well as  a couple of features with almost zero
polarization with nearly circularly symmetric magnetic field vectors that surround it.   This is consistent with a compression
of the overall helical magnetic field at knot A, but we find it implausible to think of A and C as a connected shock system as
discussed by \cite{2014ApJ...785..152N}.  Not only is there a large (400 pc) distance between these two knot complexes,
but also the very different  morphology of the magnetic field vectors argues strongly against it.  Note also that in knot A we
also see higher optical polarization along the jet edges (only statistically significant in knots A and B), further suggesting
interactions with the surrounding ISM.
A likely shock feature is also seen in knot C (and possibly G).  This region, shown in Fig. \ref{fig:knot-cg},
has a conical shape, and the polarization map
shows a nearly perpendicular magnetic field orientation and increased polarization at both the upstream and downstream
end of the flux maximum region, with a polarization minimum between and vectors along the knot edges that follow
the morphology of the flux contours.  This suggests a double (or triple, if G is included) shock structure that might
contribute to the breakup of the jet downstream of knot G, something that has been noted before in a variety of papers.
\citet{2016MNRAS.tmpL..44T} have pointed out that the kink instability that
is important in  FR I jets (see also below, \S\ref{sec:helical}), can lead to features like this on kiloparsec scales. Our
polarization map is consistent with this idea. \\
\indent In another study by \citet{2011MNRAS.416L.109C} claim high Faraday rotations in the inner jet of M87, i.e. in the nucleus and HST-1. They analyze the same data as ours, taken between 2003 and 2007 at 8, 15, and 22 GHz. Their studies show quite significant internal Faraday rotations in HST-1 at 8 GHz radio observations along with significant variations in the EVPA and fractional polarization during the period of observations. They claim that the variability in fractional polarization and observed helical undulations in the polarization structure of HST-1, most possibly was caused due to the internal Faraday rotation during the time of flare in HST-1. We do not find significant Faraday rotations in our observations, hence we can neither confirm nor deny their claims, although the observed differences in the MFPA vector orientations in the radio and optical may be explained with their results.

\subsection{Helical Features} \label{sec:helical}
\indent A number of regions of the jet have a helical morphology and magnetic field vectors. Some of these regions are
shock-like and/or connected with shock-like features, such as the features upstream and downstream of the flux
maximum of knot HST-1, or the region downstream of knot A (both discussed above), but other features, such
as knot D and knot B, do not appear to be strongly shock related.  Here, we will discuss first the
non-shock-related helical features, and then attempt to bring them and the helical features in the more shock-like
knots into a coherent picture.\\
\indent The apparent helical morphology of knot D is evident in both its flux morphology as well as its polarization vectors.
The knot gives the appearance of being a braided, filamentary structure. P99 described it as being shock-like, and
the differences observed by those authors between the optical and radio polarization vectors was one of the reasons
behind their suggested model of a stratified energetic structure.  As can be seen in Fig.~\ref{fig:nuc-hst1}, our increased
resolution throws a significant amount of new light on this issue, although it should also be mentioned that it is likely that the motion of some components (which are seen to be as fast as $\sim 5 c$, \cite{2013ApJ...774L..21M}), could
play a significant role, producing differences of as large as 0.''2 in the positions of the fastest components over 10 years.
The polarization vectors of knot D also give a strongly helical appearance.  In the brightest regions these appear to follow
the flux contours.  One exception to this is the flux maximum region of knot D-East, which is very low polarization in the
optical and shows some signs of cancellation in that band, and the upstream half of its flux maximum region in the radio.
This is significantly more detail than could be seen in P99, and in fact comparing our images to theirs we see that the low
polarization optical regions are larger in the more recent data.  This suggests that the polarized structures are moving
down the jet flow, and while cancellation may play a non-negligible role in the differences between the optical and radio,
it also seems clear that there are spectral differences as well, such as those described in the P99 model. Indeed, as discussed
in \cite{2013ApJ...774L..21M}, the fastest moving parts of the knot D complex are located in its eastern part, apparently
being ejected from a feature at its upstream end. That feature is at the approximate position of D-East$\alpha$ in Fig. \ref{fig:knotD}.
A second  exception is seen in the apparent double structure that appears to develop in D-Middle that is seen primarily in
the radio.  However, when compared to the total flux morphology that reenforces its apparent braided nature. \\
\indent Knot B's appearance is also striking.  The brightest regions (Fig. \ref{fig:knot-ab})
also appear to have a filamentary structure, but it is less clear that they are helical.  The inclination of these brightest
features as compared to knot A, its neighbor, is striking:  the brightest region extends for about 2'' and extends for nearly the
entire width of the flow, starting in its southern half at the upstream end and appearing to spread to the entire flow by knot
B's downstream end.  This region is fairly clearly demarcated by the polarization vectors, and sketched out by the red arrows
in Fig. \ref{fig:knot-ab}.  The upstream end of that region shows a polarization minimum in the radio, and multiple regions of low
polarization in the optical, combined with other regions of higher polarization but filamentary magnetic field features seen in
the radio. The differences are significant, and suggest that there are some spine/sheath issues to the jet structure in this
region as well.  The downstream end of knot B, historically called B2, shows radio polarization vectors that follow the flux
contours, but the optical polarization vectors show somewhat more structure, more clearly delineating the apparent change
in direction seen at knot B2 and also not including the apparent radio polarization minimum seen near the centroid of B2's
flux contours. The appearance of the vectors suggests that the latter is due to cancellation.  Also, as seen in knot A, the
edge regions of the knot have a somewhat higher fractional polarization than other parts.  This suggests a continuing interaction
with the ISM in knot B, but while it is clear that knots A and B are contiguous spatially, the changing inclination and complex
polarization morphology suggests that the two features are complex dynamically as well, not really consistent with single shock
complex. \\
\indent It is also worth pointing out that previous workers
have mentioned that knot E has a filamentary, helical structure (in particular \cite{2011ApJ...735...61H}.  While the
polarization vectors (Fig. \ref{fig:knot-efi}) in this region (particularly in the optical for knot F) are suggestive
in this regard, most parts of these knots fall below the 3$\sigma$ significance level both in the radio and optical.  The sole exceptions to this are the brightest parts of this region. We do see both perpendicular vectors near a flux maximum e.g. in knot A, but also regions where
cancellation is likely playing a role due to circularly symmetric vector patterns near the knot's upstream end e.g. in knot D-East and B2, where P99
noticed a polarization minimum.  In P99 we discussed a significant role for the Kelvin-Helmholtz instability in this region as well
as knots A and B, however; due to the low significance we do not discuss it further. A higher signal to
noise data are needed to study these in detail. \\
\indent In \S\ref{sec:shocks} above we discussed the prominent helical features seen upstream and downstream of the shock-like
components HST-1 and A.  These features are not dissimilar to the ones described here and point out that kinks may be
very important dynamically in jets on large scales. \citet{2016MNRAS.tmpL..44T} have pointed out that the kink instability that
is important in  FR I jets (see also \S\ref{sec:helical}).  Their model elucidates some of the issues brought out by the model
of \cite{2011ApJ...735...61H}, which attempted to model the entirety of the inner jet as a combination of braided helical
features. These changes can cause instabilities (like Kelvin-Helmholtz instability) which in turn can cause shearing of field lines
near the boundary between the jet surface and interstellar medium (ISM). Their  global 3D MHD simulations of low and
high power AGN jets and the instabilities produced within the jet as a result. Magnetic fields being the natural driving force
for launching the jets, can suppress the instabilities like Kelvin-Helmholtz and initiate the current-driven kink instabilities.
The kink instabilities, especially of the kink mode m=1, play important role in causing the jet to move sideways and
developing the helical motions, possibly as seen in our radio polarization images. The potential of the growth of these
kink mode depend on the Alfv{\'e}n wave travel time across the jet and also on the ambient density. The tightly
collimated jets, like in M87, are more susceptible to the kink instabilities. In such jets, the instabilities can rapidly form
and disrupt the jet and cause them to decelerate or stall.  Knot D, for example, may be an example of a kink developing
downstream of the knot HST-1 recollimation shock, while knot B may be an example of the continuing interaction of the
jet with the galactic ISM downstream of the main knot A shock region.  Interestingly, \citet{2013ApJ...774L..21M} showed the superluminal velocity vectors in outer jet, mainly in knots A and B, appear to line up in helical pattern.  We are in the process
of attempting to model the geometry of this region but this is work in progress. We will present the results of this work in a future paper.

\section{Conclusion}\label{sec:concl}
\indent The overall flux and polarization of the M87's jet shows striking differences as compared to the older observations of P99. We discussed what things are different in terms of resolution and the possibility of a few real sub-structures emerging on the sub-parsec scales near the nucleus and knot HST-1. As described in \S\ref{sec:shocks}, the structure is changing suddenly beyond the recollimation shock at knot HST-1, which compresses the local magnetic field \citep{2014ApJ...785..152N, 2006MNRAS.370..981S} and forces the field lines to become perpendicular to the flow. Further downstream, the interaction between a strongly magnetised relativistic plasma outflow and non-relativistic collimating magnetohydrodynamic winds can give rise to more shocks. As a result, the particles in these regions can be accelerated to relativistic speeds and move out from the knot forming new super-luminal sub-components seen in {\it VLBA} images \citep{2007ApJ...663L..65C}, which are likely to be responsible for the flaring behavior. A more recent study by \citet{2016MNRAS.tmpL..44T}, suggest that the presence of undulations in this region may have caused due to the successive compression and stretching of the local toroidal magnetic field resulting in the spinning of the magnetic field lines. While we see helical undulations in our data, we do not have enough resolution to comment if these components are moving out or are stationary features in the jet. \\
\indent In \S\ref{sec:trends} we described a few common features of the radio and optical jet that are observed in our images. We see that the optical jet is slightly narrower than the radio with the optical emission being more defined and concentrated closer to the center. This trend appear to be consistent with the previous model of a layered or stratified jet of P 99 (see their Fig. 7). They explain this in terms of origin of radio and optical electrons being very different. In this model, the more energetic optical electrons are probably located near the center whereas the lower energy radio electrons are from the outer layer of the jet. The differences in the flux morphology in the two bands also apparently indicate that the jet follow the ``spine-sheath'' model of \citep{2007ApJ...668L..27K}, in which they suggest (similar to P99) that the higher energy photons originate from the center of the jet while the lower energy photons originate from near the surface. \\
\indent Our results do not necessarily follow the stratified jet or ``spine sheath'' models. The similarities in the flux and polarization structure that we see as explained in \S~\ref{sec:trends}, are mainly due to the higher resolution of our data as compared to the previous data of P99. We see a lot more flux as well as polarization structure that was not seen in their images. As a result their model of stratified jet does not necessarily apply to each component in the jet. The newer flux details in the inner jet knots such as HST-1, D and F, were not see in old images of P99, as a result their model does not hold true in these regions. The flux and polarization structure in these regions show quite many similarities which does not support the stratified jet model. However,the stratified jet model can still hold in general for the outer jet components i.e. A, B, C and G, where we clearly see the differences in the radio and optical flux and polarization structure. \\
\indent Another striking difference is in the polarization morphology of the jet in two bands, especially in the inner jet, the nucleus, HST-1 and knot D. Our optical images show the predominant perpendicular MFPA features in the jet. The radio MFPA, on the other hand, stays mostly parallel to the jet direction. These differences can be explained either by arguing that the direction of local magnetic field is changing, or that the radio wavelengths are being Faraday rotated. At the location of perpendicular shock, the magnetic field lines can get squeezed and forced to turn in the direction perpendicular to the direction of jet plasma. The magnetic field lines may turn back to parallel downstream of the shock. This can cause the rapid changes in the in the directions of local magnetic field. If the shock lies in the interior of the jet i.e. closer to the jet axis, this may affect optical electrons only, lying closer to the axis of the jet, and not so much on the radio electrons closer to the surface of the jet.  P99, \citet{1996ApJ...467..597B, 1989ApJ...340..698O} suggested these changes can cause instabilities (like Kelvin-Helmholtz) which in turn can cause shearing of field lines near the boundary between the jet surface and interstellar medium (ISM). This may cause the increased polarization near the surface as observed in fractional polarization images in radio. \\
\indent In general, the flux as well as polarization structure in the inner jet and intermediate jet as well as the outer jet, show quite different characters in fractional polarization, which point toward the fact that structure of the magnetic field and its effects on the jet environments are completely different in each of these regions. These internal changes in the magnetic field can also affect the particle acceleration and emission mechanisms in respective regions, which we can clearly see from our radio and optical images. The effect of kink instability (as described in \S\ref{sec:helical}) on the kpc scales, away from the central engine, may play important role in defining the polarization structure in the outer jet and beyond. A more thorough followup observations of radio polarimetry and optical proper motions along the jet at higher resolution will help us gain further understanding of these processes. \\
\indent ESP and SSA acknowledge support from STScI through grants HST-GO-13759.003, 13676.003 and 13764.001. ESP and MG acknowledge support from NASA ASAP grant NNX15AE55G.


\bibliographystyle{apj}
\bibliography{polarimetry_paper}

\end{document}